\documentclass[reqno,a4paper,oneside,11pt]{amsart}

\usepackage[usenames,dvipsnames]{xcolor}
\usepackage[colorlinks=true,linkcolor=Blue,citecolor=Green]{hyperref}
\usepackage{amsmath,amssymb,epsf}
\usepackage{amsfonts,amsbsy,amscd,stmaryrd}
  \SetSymbolFont{stmry}{bold}{U}{stmry}{m}{n}
\usepackage{latexsym,amsopn}
\usepackage{amsthm}
\usepackage{esint}
\usepackage{mathrsfs}
\usepackage{slashed} 
\usepackage{enumitem}
\renewcommand{\baselinestretch}{1.05}
\usepackage[margin=1.3in,footskip=0.25in]{geometry}
\allowdisplaybreaks

\usepackage[utf8]{inputenc}
\usepackage[T1]{fontenc}   

\usepackage{graphicx}
\usepackage{multirow}
\usepackage{float}

\newcounter{smallarabics}
\newenvironment{arabicenumerate}
{\begin{list}{{\normalfont\textrm{(\arabic{smallarabics})}}}
  {\usecounter{smallarabics}\setlength{\itemindent}{0cm}
   \setlength{\leftmargin}{5ex}\setlength{\labelwidth}{4ex}
   \setlength{\topsep}{0.75\parsep}\setlength{\partopsep}{0ex}
   \setlength{\itemsep}{0ex}}}
{\end{list}}

\newcounter{smallroman}

\newcommand{\ben}{\begin{arabicenumerate}}  
\newcommand{\een}{\end{arabicenumerate}}

\def\init{\setcounter{equation}{0}}

\newtheorem{theorem}{Theorem}[section]

\newtheorem{proposition}[theorem]{Proposition}
\newtheorem{lemma}[theorem]{Lemma}
\newtheorem{corollary}[theorem]{Corollary}
\theoremstyle{definition}
\newtheorem{definition}[theorem]{Definition}
\newtheorem{remark}[theorem]{Remark}
\newtheorem{example}[theorem]{Example}

\newcommand{\beq}{\begin{equation}}
\newcommand{\eeq}{\end{equation}}
\newcommand{\bea}{\begin{aligned}}
\newcommand{\eea}{\end{aligned}}
\newcommand{\bear}{\begin{array}{rl}}
\newcommand{\eear}{\end{array}}
\newcommand{\bex}{\begin{example}}
\newcommand{\eex}{\end{example}}
\def\bel{\begin{lemma}}
\def\eel{\end{lemma}}
\def\bet{\begin{theoreme}}
\def\eet{\end{theoreme}}
\def\bed{\begin{definition}}
\def\eed{\end{definition}}
\def\ber{\begin{remark}}
\def\eer{\end{remark}}
\def\bep{\begin{proposition}}
\def\eep{\end{proposition}}
\newcommand{\qeds}{\qed\medskip}

\let\origmaketitle\maketitle
\def\maketitle{
  \begingroup
  \def\uppercasenonmath##1{} 
  \let\MakeUppercase\relax 
	\origmaketitle
  \endgroup
}

\def\rr{{\mathbb R}}
\def\zz{{\mathbb Z}}
\def\cc{{\mathbb C}}
\def\nn{{\mathbb N}}

\def\ss{{\mathbb S}}

\def\slim{{\rm s-}\lim}

\def\bar{\overline}

\def\cinf{C^\infty}
\def\proof{
\noindent{\bf Proof.}\ \ }

\usepackage{calrsfs}
\DeclareMathAlphabet{\pazocal}{OMS}{zplm}{m}{n}

\def\cY{{\pazocal Y}}

\def\cV{{\pazocal V}}
\def\cL{{\pazocal L}}
\def\cS{{\pazocal S}}

\def\cD{{\pazocal D}}

\def\cM{{\pazocal M}}
\def\cN{{\pazocal N}}

\def\cW{{\pazocal W}}

\def\cX{{\pazocal X}}

\def\cE{\pazocal{E}}

\def\cH{{\pazocal H}}

\def\sH{\mathcal{H}}

\def\sI{\mathcal{I}}

\def\CAR{{\rm CAR}}
\def\wf{{\rm WF}}

\def\i{{\rm i}}

\def\id{{\rm id}}

\let\Im\relax
\let\Re\relax

\DeclareMathOperator{\Ker}{Ker}
\DeclareMathOperator{\Ran}{Ran}
\DeclareMathOperator{\Im}{Im}
\DeclareMathOperator{\Re}{Re}

\DeclareMathOperator{\supp}{supp}
\DeclareMathOperator{\sing}{sing}
\DeclareMathOperator{\Char}{Char}

\def\p{\partial}

\def\14{\frac{1}{4}}

\def\12{\frac{1}{2}}

\def\e{{\rm e}}

\newcommand{\one}{\boldsymbol{1}}

\def\c{{\rm c}}

\def\sc{{\rm sc}}

\def\id{{\rm id}}

\def\coinf{C_{\rm c}^\infty}

\newcommand{\mat}[4]{\left(\begin{array}{cc}#1 &#2  \\ #3 &#4 \end{array}\right)}

\newcommand{\bra}{\langle} 
\newcommand{\ket}{\rangle}

\DeclareSymbolFont{boldoperators}{OT1}{cmr}{bx}{n}
\SetSymbolFont{boldoperators}{bold}{OT1}{cmr}{bx}{n}

\def\t{{\scriptscriptstyle\#}}

\makeatletter
\newcommand*{\defeq}{\mathrel{\rlap{%
                     \raisebox{0.34ex}{$\m@th\cdot$}}%
                     \raisebox{-0.4ex}{$\m@th\cdot$}}%
                     =}
                     
\makeatother
\makeatletter
\newcommand*{\eqdef}{=\mathrel{\rlap{%
                     \raisebox{0.34ex}{$\m@th\cdot$}}%
                     \raisebox{-0.4ex}{$\m@th\cdot$}}%
                     }
\makeatother

\def\Sol{{\rm Sol}}

\def\WF{{\rm WF}}

\def\Vol{{\rm vol}}

\def\dual{\!\cdot \!}
\def\kst{^{*}}
\def\stk{{}^{*}\!}
\newcommand{\col}[2]{\left(\begin{array}{c}#1 \\#2\end{array} \right)}
\def\DD{{\mathbb D}}
\def\GG{{\mathbb G}}
\newcommand{\LL}{%
            \mathrel{\raisebox{.0em}{%
            {\rotatebox[origin=c]{180}{$\mathbb{V}$}}}}}
\def\SS{{\mathbb S}}
\def\CAR{{\rm CAR}}
\def\tosim{\xrightarrow{\sim}}
\def\mo{\mathit{o}}
\def\mi{\imath}

\def\zero{{\mskip-4mu{\rm\textit{o}}}}

\def\dVol{\mathop{}\!d{\rm vol}}
\def\diff{\mathop{}\!d}
\def\End{\mathit{End}}
\def\Solit{\mathit{Sol}}
\def\bS{\mathbb{S}}

\def\MI{{\rm M}_{\rm I}}
\def\MII{{\rm M}_{\rm II}}
\def\MIp{{\rm M}_{\rm I'}}
\def\MK{{\rm M}}
\def\MIUII{{\rm M}_{{\rm I}\cup {\rm II}}}
\def\2Sol{{\rm Sol}_{{\rm L}^{2}}}
\def\largewedge{\mbox{\Large $\wedge$}}

\newcommand{\open}[1]{\mathopen{}\mathclose{\left]#1 \right[}}
\newcommand{\clopen}[1]{\mathopen{}\mathclose{\left[#1 \right[}}
\newcommand{\opencl}[1]{\mathopen{}\mathclose{\left]#1 \right]}}
\newcommand{\closed}[1]{\mathopen{}\mathclose{\left[#1 \right]}}

\begin{document}

\title[The Unruh state for massless fermions on Kerr spacetime and its Hadamard property]{\Large The Unruh state for massless fermions on Kerr spacetime and its Hadamard property}

\author{}
\address{Laboratoire de Math\'ematiques d'Orsay, Universit\'e Paris-Saclay, 91405 Orsay, France}
\email{christian.gerard@math.u-psud.fr}
\author{}
\address{Institut Fourier, Universit\'e Grenoble Alpes, 100 rue des Maths, 38610 Gi\`eres, France}
\email{dietrich.hafner@univ-grenoble-alpes.fr}
\author[]{\normalsize Christian \textsc{G\'erard}, Dietrich \textsc{H\"afner} \& Micha{\l} \textsc{Wrochna}}
\address{CY Cergy Paris Universit\'e, 2 av.~Adolphe Chauvin, 95302 Cergy-Pontoise, France}
\email{michal.wrochna@cyu.fr}
\thanks{\emph{Acknowledgments.} The authors are  grateful to Ko Sanders and Rainer Verch for stimulating discussions. Support from the grant ANR-16-CE40-0012-01 is gratefully acknowledged. D.H.~and M.W.~thank the MSRI in Berkeley and the Mittag--Leffler Institute in Djursholm for their hospitality during the thematic programs and workshops in 2019--20.  }

\begin{abstract} We give a rigorous definition of the Unruh state in the setting of massless Dirac fields on slowly rotating Kerr spacetimes. In the black hole exterior region, we show that it is asymptotically thermal at Hawking temperature on the past event horizon.  Furthermore, we demonstrate that in the union of the exterior and interior regions, the Unruh state is pure and Hadamard. The main ingredients are the Häfner--Nicolas  scattering theory, new microlocal estimates for characteristic Cauchy problems and criteria on the level of square-integrable solutions.
\end{abstract}

\maketitle
\section{Introduction and summary}

\subsection{Introduction} One of the major open problems in mathematical Quantum Field Theory on curved spacetimes is to 
determine the final quantum state arising from the collapse into a black hole and to describe its thermo\-dynamical properties.

In the last decade, valuable insight has been gained especially from simplified models in which the black hole is eternal and non-rotating. 

Most notably, years after Unruh's proposal for a distinguished state on Schwarzschild spacetime \cite{unruh} and subsequent developments, including e.g.~works by Candelas \cite{candelas} and Dimock--Kay \cite{DK1,DK2}, a rigorous definition was eventually provided by Dappiaggi--Moretti--Pinamonti \cite{DMP2}. The same authors gave also more clue to the physical relevance of the Unruh state by proving that it satisfies the \emph{Hadamard condition} on the union of the exterior and the interior region, thus ruling out infinite accumulation of energy at the event horizon. The remarkable fact is that imposing the Hadamard condition singles out a state at the \emph{Hawking temperature} at the event horizon. 

The Hadamard condition was also shown recently for the Unruh state on Reissner--Nordström--de Sitter spacetime by Hollands--Wald--Zahn \cite{hwz}, who used it   as a reference  state to demonstrate the quantum instability of the Cauchy horizon in that setting. 

The essential feature of e.g.~Schwarzschild spacetime that makes it more tractable than the case of rotating black holes is the existence of a Killing vector field which is time-like in the whole exterior region. Schwarzschild spacetime has also the special structure of a \emph{static bifurcate Killing horizon}, which makes it also possible to consider a different distinguished state, the \emph{Hartle--Hawking--Israel} state \cite{HH,israel},  conjectured in the '70s to be well-defined on the whole Kruskal--Szekeres extension. While its uniqueness is known since the work of Kay--Wald \cite{KW}, its rigorous construction and the proof of its Hadamard property was established relatively recently by Sanders \cite{sanders}, followed by a generalization to the stationary case by G\'erard \cite{HHI}.  Although believed to be too idealized to describe the final black hole collapse state accurately, the Hartle--Hawking--Israel state is an important theoretical model nevertheless in view of its high level of symmetry and its connection to black hole thermodynamics \cite{thermowald}.

In the physically more realistic rotating case, however, there is no rigorous result so far that gives the existence of a distinguished Hadamard state. The absence of a global time-like Killing vector field causes severe difficulties both on the conceptual and technical level, which are expressed in the following non-existence theorems:
\begin{enumerate}
\item\label{nonexistence1} The Kay--Wald theorem  asserts the non-existence of a state which is invariant under the flow of the Killing field $v_{\sH}$ that generates the horizon, under the assumption that a certain \emph{superradiance} property holds true \cite{KW}. The latter is conjectured to be verified in the case of bosonic fields. (Note that on the other hand, there is no superradiance for fermions in this sense.)  \smallskip
\item\label{nonexistence2} A theorem due to Pinamonti--Sanders--Verch asserts that a thermal state associated with a complete Killing vector field $v$ cannot be Hadamard if there is a point at which $v$ is space-like \cite{PSV}. This result holds true in a broad setting including bosonic and fermionic non-interacting fields.  
\end{enumerate} 

Focusing  our attention on the exterior Kerr spacetime $(\MI,g)$ for the moment, the closest analogues of the time-like Killing vector field $\p_t$ on Schwarzschild are the two Killing vector fields $v_{\sH}= \p_{t}+ \Omega_{\sH}\p_{\varphi}$ (the generator of the past horizon $\sH_-$; the constant $\Omega_{\sH}$ is the angular velocity of $\sH_-$) and $v_{\sI}=\p_{t}$ (the generator of past null infinity $\sI_-$), each of which is time-like only in a subregion of the exterior.  While the result (\ref{nonexistence2}) (and in all likelihood (\ref{nonexistence1}) as well in the bosonic case) implies bad properties of any state in the exterior region that is thermal with respect to either $v_{\sH}$ or $v_{\sI}$, we propose instead to split the solution space into two parts, and use $v_{\sH}$ for solutions coming from $\sH_-$ and $v_{\sI}$ for those coming from $\sI_-$ in the sense of scattering theory. 

 This is consistent with a formal definition in terms of mode expansions, proposed by Ottewill--Winstanley in the case of scalar fields \cite{OW}  and Casals--Dolan--Nolan--Ottewill--Winstanley for fermions \cite{kermions}, cf.~\cite{CO} for electromagnetic fields.  The approach via mode expansions is advantageous for practical computations.  However, making the definition of the state rigorous and proving the Hadamard condition requires a sufficiently precise scattering theory (which distinguishes between solutions coming from $\sH_-$ and solutions coming from $\sI_-$), combined with high frequency estimates for solutions in terms of their asymptotic data. On top of that, the fact that none of the Killing vector fields is everywhere time-like makes it impossible to use the arguments employed in \cite{DMP2} to show the Hadamard condition.

\subsection{Main result}\label{ss:maintheorem} In the present paper, we provide the first rigorous result of existence of a distinguished Hadamard state on a rotating black hole spacetime. More precisely, we give a precise definition and prove the Hadamard property of the \emph{Unruh state} (or more pedantically, of the \emph{past Unruh state}) in the case of massless Dirac fields on Kerr spacetime.

\begin{figure}[H]
  \centering
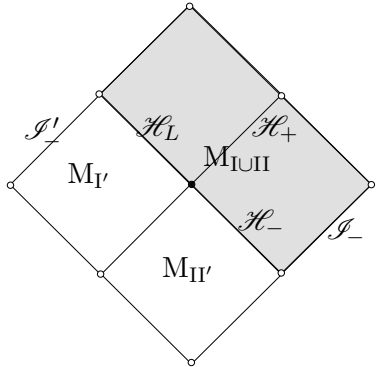
	\caption{The Carter--Penrose diagram of the spacetime $\MIUII$ (represented by the shaded region) embedded in the larger spacetime $\MK$.  Scattering in the exterior region $\MI$ refers to data at the past horizon $\sH_-\subset \sH_L$ and at past null infinity $\sI_-$. Scattering in $\MIUII$ refers to data at the long horizon $\sH_L$ and at $\sI_-$. Scattering in $\MK$ requires an extra piece of data at $\sI_-'$.}
	\label{fig8}
\end{figure}

We consider the four-dimensional spacetime $(\MIUII,g)$ which is the union of the Kerr black hole exterior and interior spacetimes with rotation parameter $a>0$  and view it as a subregion of the Kerr--Kruskal extension $(\MK,g)$ (see Figure  \ref{fig8}). 

The  Dirac operator 
$\slashed{D}$  acting on smooth sections of the canonical spinor bundle $\cS$ over $\MK$ is the differential operator defined as
\[
\slashed{D}= g^{\mu\nu}\gamma(e_{\mu})\nabla^{\cS}_{e_{\nu}},
\]
see Sections \ref{sec10} and \ref{sec9} for details. In the massless case considered here, it is well known that the whole analysis can be reduced to the \emph{Weyl equation} $\mathbb{D}\phi=0$, which accounts for half of the degrees of freedom, see Section \ref{sec10}. 

The main reason for us to consider massless fermions is that in this case the scattering theory in the exterior region $\MI$ is sufficiently well developed thanks to results by H\"afner--Nicolas \cite{HN}, and in contrast to bosons, there are no extra difficulties due to superradiance.  In particular there is a notion of scattering data at $\sH_-$ and $\sI_-$ of square-integrable solutions. Furthermore, for sufficiently regular solutions these data agree with traces at $\sH_-$ and $\sI_-$. 

 The traces at $\sH_{-}$ and $\sI_{-}$ extend to solutions in the Kerr--Kruskal spacetime $\MK$ as traces at the left long horizon $\sH_{L}$ (see Figure \ref{fig8}) and at the two conformal infinities $\sI_{-}$ and $\sI'_{-}$. Using these traces it is possible to  define a \emph{pure state} $\omega_{\MK}$ in the whole of $\MK$, which we call the {\em Unruh state}. 

As explained in Subsect.~\ref{demian}, to construct a state it suffices to specify a pair of positive semidefinite operators $C^+$ and $C^-$  acting on square-integrable solutions of $\mathbb{D}\phi=0$ and satisfying $C^+ + C^-=\one$ (these are the so-called  \emph{covariances}). We demonstrate that the \emph{Hadamard condition} can be then formulated as the requirement that for each solution $\phi$, the wavefront set of $(C^\pm)^{\12} \phi$ is confined to only one component of the characteristic set of $\mathbb{D}$, namely     
\beq\label{eq:had1}
\wf((C^\pm)^{\12} \phi)\subset \cN^\pm,
\eeq
where the two components $\cN^+$ and $\cN^-$ are defined in \eqref{sloubi1} and the definition of the wavefront set is recalled in Appendix \ref{secapp1}. The main difficulty is how to control a high-frequency condition such as \eqref{eq:had1} when $(C^\pm)^{\12}\phi$ is defined through its  asymptotic data.

\medskip

Our main result is the following theorem. Below, $\one_{\rr^\pm}$ is the characteristic function of the positive, resp.~negative half-line, $(U,V,\theta,\varphi^{\t})$ are the Kruskal--Boyer--Lindquist coordinates,  and $(t\kst,r,\theta,\varphi\kst)$ are the Kerr-star coordinates in the exterior region, see Section \ref{sec9}.  

\begin{theorem}[{cf.~Theorems \ref{thm13.2} and \ref{thm13.1}}]\label{mainmaintheorem} Let $(\MK,g)$ be the Kerr--Kruskal spacetime, and let $\mathbb{D}$ be the  Weyl operator on $(\MK,g)$.  Then there exists $0<a_{0}\leq 1$ such that if $|a|M^{-1}<a_{0}$ then the following holds:
\ben
\item there exists a unique state $\omega_\MK$ for $\mathbb{D}$, called the \emph{Unruh state}, with data $\one_{\rr^\pm}(-\i^{-1}\p_U)$ at the horizon $\sH_{L}$ and $\one_{\rr^\pm}(\i^{-1}\p_{t\kst})$, resp.~$\one_{\rr^{\mp}}(\i^{-1}\p_{t\kst})$  at null infinity $\sI_-$, resp.~$\sI'_{-}$; 
\item $\omega_{\MK}$ and its restriction $\omega_{\MIUII}$ to $\MIUII$ are pure states;
\item the restriction $\omega_{\MI}$ of $\omega_{\MK}$ to $\MI$ is thermal with respect to $v_{\sH}$ at the past horizon $\sH_{-}$ with temperature equal to the Hawking temperature $(2\pi)^{-1}\kappa_{+}$;
\item $\omega_{\MIUII}$ is a Hadamard state.
\een
\end{theorem}

The property that the state $\omega_\MK$ is  \emph{pure} is equivalent to its covariances $C_\MK^\pm$ being projections, and since $\one_{\rr^\pm}(\i^{-1} \p_U)$ and $\one_{\rr^\pm}(\i^{-1}\p_{t\kst})$ are projections, this is a direct consequence of the fact that the scattering data maps that we construct are unitary. The same arguments apply to $\omega_{\MIUII}$.

The Hawking temperature at $\sH_-$ arises exactly as anticipated. Namely, the restriction of $\one_{\rr^\mp}(\i^{-1} \p_U)$ to $\sH_-=\sH_L\cap \{ U>0\}$ is a function of $v_\sH$ (using that $v_\sH$ equals $-\kappa_+ U\p_U$ on $\sH_-$), and this function is precisely the fermionic thermal distribution $\chi^\pm_{\sH_-}(s)=(1+\e^{\mp 2\pi s/\kappa_+})^{-1}$.  This can be understood through an elementary computation presented in Appendix \ref{appC}. 

The difficult part is unquestionably the Hadamard property of $\omega_{\MIUII}$. Part of the proof relies on wavefront set estimates for solutions in terms of their characteristic data. This is a refinement of the strategy used in works including \cite{Mo2,DMP1,DMP2,radiative,characteristic}, and it applies to points lying on bicharacteristics that intersect  $\sH$ or $\sI_-$ (with some extra care if the intersection lies on the crossing sphere). In contrast to the Schwarzschild case, however, it is essential to deal with bicharacteristics that do not have this property and  that cannot be dealt with by similar arguments. 

As a way around that difficulty, we first show that in the exterior region $\MI$, the Unruh state has  covariances of the form:
\beq\label{eq:cpm}
C^\pm_{\MI}={\rm P}_{\sH_{-}}\circ \chi^{\pm}_{\sH_{-}}(\i^{-1}\cL_{\sH})+{\rm P}_{\sI_{-}} \circ \chi^{\pm}_{\sI_{-}}(\i^{-1}\cL_{\sI}),
\eeq
where ${\rm P}_{\sH_{-}}$ and ${\rm P}_{\sI_{-}}$ are projections defined using scattering theory, and $\chi^\pm_{\sI_-}(s)=\one_{\rr^\pm}(s)$.  The operator $\cL_{\sH/\sI}$  is the \emph{spinorial Lie derivative} of $v_{\sH/\sI}$, and in \eqref{eq:cpm} we interpret $\i^{-1}\cL_{\sH/\sI}$ as a self-adjoint operator acting on the Hilbert space of square integrable solutions of $\mathbb{D}\phi=0$. 

The form \eqref{eq:cpm} allows us to prove wavefront set estimates for $(C^\pm_{\MI})^{\12}\phi$ in the region where $v_{\sH/\sI}$ is time-like. More precisely, we show in Lemma \ref{key-lemma2} that in that region, modulo inessential terms, the operators $\chi^{\pm}_{\sH_{-}}(\i^{-1}\cL_{\sH})$ act    as  pseudo-differential operators that are elliptic on $\cN^\pm$ and smoothing on $\cN^\mp$.

    The use of the projections ${\rm P}_{\sH_{-}/\sI_{-}}$ is also needed to separate solutions coming from $\sH_-$ and from $\sI_-$ when using  arguments based on wavefront sets of traces. All this requires a careful examination of the relationship between scattering theory and characteristic Cauchy problems.  

Next, we show that in the slowly rotating case, i.e.~for sufficiently small $a$, all  bicharacteristics in $\MI$  that do not meet $\sH_-$ nor $\sI_-$ necessarily cross  a region where $v_{\sH}$ and $v_{\sI}$ are  time-like. Therefore, the  estimates in this good region can be propagated to the problematic points. 

Finally, in the remaining part of the larger spacetime $\MIUII$,  we propagate estimates from $\MI$ and from the analogous time-reflected block ${\rm M_{I'}}$. Then, an extra argument near the crossing sphere allows us to conclude the proof of Theorem \ref{mainmaintheorem}.

We emphasize that while the results in Theorem \ref{mainmaintheorem} are formulated for the Weyl equation, they can be  immediately  translated to  \emph{massless Dirac} fields, see for example the discussion in \cite[Sect.~17.15]{G}. 

\subsection{Physical interpretation and Hawking effect} The Unruh state is conjectured to be the final state emerging from the collapse of a star into a black hole. This is the  physical context of the celebrated \emph{Hawking effect} \cite{Haw}. Mathematically rigorous works on this process exist in  simplified models where the boundary of the star is a perfect mirror, see Bachelot \cite{Ba} for the Klein--Gordon equation in the spherically symmetric setting and Häfner \cite{Ha} for the Dirac equation on rotating black holes, including also the charged and massive case (cf.~Drouot \cite{drouot} for Klein--Gordon fields on Schwarzschild-de Sitter spacetime).  The final state arising in this process can be interpreted as a state on the  Kerr exterior spacetime, and its asymptotic data was shown in \cite[Thm.~1.1]{Ha} to be exactly  that of the Unruh state $\omega_{\MI}$ considered in the present work. Therefore, the Unruh state \emph{is} the conjectured  final state arising from the collapse in this model. Furthermore, since we now know that the final state is Hadamard, it would be interesting to see if the arguments of Fredenhagen--Haag \cite{FH} about the derivation of the Hawking radiation (formulated in \cite{FH} for scalar fields on Schwarzschild spacetime) could be generalized for fermions to the rotating case  considered in the present work.

\subsection{Bibliographical remarks and related settings} The problem of constructing Hadamard states from asymptotic data was considered in various settings by Hollands \cite{hollandsthesis}, Moretti \cite{Mo1,Mo2}, Dappiaggi--Pinamonti--Moretti \cite{DMP1,DMP2}, Dappiaggi--Siemssen \cite{DS}, Benini--Dap\-piaggi--Murro \cite{radiative}, G\'erard--Wrochna \cite{characteristic} using characteristic data methods, by G\'erard--Wrochna \cite{inout} using scattering theory and time-dependent pseudo-differential calculus, and by Vasy--Wrochna \cite{VW} using radial propagation estimates.  This includes the already mentioned result \cite{DMP1} on the Unruh state on Schwarzschild spacetime, which uses decay bounds for scalar waves due to Dafermos--Rodnianski \cite{DR} (cf.~\cite{BJ} for a variant of \cite{DMP1} for the Schwarzschild-de Sitter metric), and \cite{hwz} on Reissner--Nordström--de Sitter, which uses estimates by Hintz--Vasy \cite{HV}. 

The results of the present paper strongly rely on the scattering theory for massless Dirac fields in the Kerr exterior region $\MI$ due to H\"afner--Nicolas \cite{HN}, see also \cite{HN2} for a concise presentation. The fact that asymptotic data in the sense of scattering theory coincides with traces at $\sH_-$ and conformal traces at $\sI_-$ is what allows us to combine the formalism with wavefront set estimates from characteristic data.

For {\em massive Dirac} fields, the scattering theory 
 \cite{HN}  was extended by Daud\'e, see \cite[Ch.~IV]{Da}. In that setting, the scattering data on $\sH_{-}$ can still be interpreted as traces, but this  is no longer  true for the scattering data at spatial infinity, since the massive Dirac equation is no more conformally invariant. In this situation it is possible to define the Unruh state in $\MI$, $\MIUII$ or even $\MK$ as in the present paper, but its Hadamard property in $\MIUII$ or  $\MI$ is unknown. 

The rigorous definition and Hadamard property of the Unruh state on rotating black holes for bosonic, e.g.~{\em scalar} fields is still a major open problem. Useful insight is provided by the work of Ottewill--Winstanley \cite{OW}, which uses mode analysis. In the case of the wave equation on the Kerr metric, a scattering theory was developed by Dafermos--Rodnianski--Shlapentokh-Rothman \cite{DRS}, and similar results at fixed angular momentum were obtained for the Klein-Gordon equation on the Kerr--De Sitter metric by G\'erard--Georgescu--Häfner \cite{GGH}. What makes however the case of bosons more difficult than fermions  is that there is no obvious Hilbert space that is equally convenient from the point of view of energy estimates and quantization. It is also not fully understood whether superradiance is an obstruction for the existence of the Unruh state. 

\subsection{Structure of the paper} In Sections \ref{sec10}--\ref{sec9} we introduce the setup in detail and prove several preliminary results. This includes a Green's formula for square-integrable solutions and partially null hypersurfaces, see Proposition \ref{propcoupure.1}, and a proof of the criterion \eqref{eq:had1} for the Hadamard condition, see Theorem \ref{propcoup.2}. The purpose of Sections \ref{sec11}--\ref{sec12} is to summarize  the scattering results from \cite{HN}, translate them in a more covariant way to the setting of Weyl fields in $\MI$ and then extend them to a scattering theory on $\MK$.  An important role is played by different choice of tetrads, specially adapted either to scattering or to formulas involving traces at $\sH_-$ and $\sI_-$. The main results are proved in Section \ref{sec13}.  

Many auxiliary results are collected in the appendices. This includes the proof of global hyperbolicity and the construction of Cauchy hypersurfaces for the Kerr--Kruskal spacetime $\MK$ and the $\MI$ and $\MIUII$ regions in Appendix \ref{secG}.

\section{Dirac and Weyl equations on Lorentzian manifolds}\label{sec10}\init

\subsection{Notation} Let $(M, g)$ be an oriented and time oriented Lorentzian manifold  of dimension $4$.  We denote by $\Omega_{g}\in \largewedge^{4}M$ the volume form associated to $g$ and by $\dVol_{g}= | \Omega_{g}|$ the volume density.

If $\cS\xrightarrow{\pi}M$ is a vector bundle, we denote by $\cinf(M;\cS)$ the space of its smooth sections. In the case when $\cS$ is a complex vector bundle, we denote its anti-dual bundle by $\cS^*$. The complex conjugate bundle (obtained by considering the fibers as complex vector spaces with the opposite complex structure) is denoted by $\bar{\cS}$.

If $\cX$ is a complex vector space, we denote by $L_{\rm h}(\cX,\cX^*)$, resp.~$L_{\rm a}(\cX,\cX^*)$, the space of Hermitian, resp.~anti-Hermitian forms on $\cX$. Generally, if $\beta\in  L(\cX,\cX^*)$, i.e.~if $\beta$ is a sesquilinear form, we denote by $\overline{\psi_1}\cdot \beta \psi_2$ its evaluation on $\psi_1,\psi_2\in  \cX$. If instead  $\beta\in \cinf(M;\End(\cS,\cS^*))$, the same notation is used for the fiberwise evaluation of  $\psi_1,\psi_2\in  \cinf(M;\cS)$. 

\subsection{Spin structures}\label{sec10.1}

A Lorentzian manifold admits  spin structures if and only if  its  second Stiefel--Whitney class $w_{2}(TM)$ is trivial, see \cite{Mi, Na}.  If $n=4$ this is also equivalent to the fact that $M$ is parallelizable,  see \cite{spinors1, spinors2}. 
It admits a {\em unique} spin structure if in addition  its  first Stiefel--Whitney class $w_{1}(M)$ is trivial, which is equivalent to the fact that $M$ is orientable, see e.g.~\cite{Na}.

In our situation,   $M$ is orientable hence spin structures on $(M, g)$ are unique  if they exist. If $(M, g)$ is globally hyperbolic, then $M$ is diffeomorphic to $\rr\times \Sigma$, where $\Sigma$ is a smooth Cauchy surface. The orientation and time orientation of $M$  induce an orientation of $\Sigma$, hence $\Sigma$ is orientable.

Since any orientable $3$-manifold is parallelizable, this implies that if $(M, g)$ is globally hyperbolic and  $n=4$,  $M$ is parallelizable and by the above facts it admits a  {\em unique} spin structure.

From the spin structure one obtains in a canonical way a spinor bundle $\cS\to M$ of rank $4$, a spin connection $\nabla^{\cS}$ and a representation of the Clifford bundle ${\rm Cl}(M, g)$ in $End(\cS)$. One also obtains a Hermitian form $\beta$ and a complex conjugation $\kappa$ acting on the fibers of $\cS$. Summarizing, we obtain from the spin structure the following canonical objects:

\ben
 \item  a linear map $\gamma: \cinf(M; TM)\to \cinf(M; \End(\cS))$ such that 
 \beq\label{e10.1}
\gamma(X)\gamma(Y)+ \gamma(Y)\gamma(X)= 2X\dual gY \one, \ \ X, Y\in \cinf(M; TM),
\eeq
 and  for each $x\in M$, $\gamma_{x}$ induces a faithful irreducible representation of the Clifford algebra ${\rm  Cl}(T_{x}M, g_{x})$ in $\cS_{x}$; \medskip
 \item a section $\beta\in \cinf(M; \End(\cS, \cS^{*}))$ such that $\beta_{x}$ is Hermitian non-degenerate for each $x\in M$ and 
 \beq\label{e10.2}
 \bear
i)&\gamma(X)^{*}\beta = - \beta \gamma(X), \ \ \forall X\in \cinf(M; TM), \\[2mm]
ii)& \i \beta \gamma(e)>0, \hbox{ for }e\hbox{ a time-like, future directed vector field on }M;
\eear
\eeq
\item a section $\kappa\in \cinf(M; \End(\cS, \bar{\cS}))$ such that
\beq\label{e10.3}
\kappa \gamma(X)= \gamma(X)\kappa, \ \ \kappa^{2}=\one;
\eeq
\item a connection $\nabla^{\cS}$ on $\cS$, called a {\em spin connection}, such that:
\beq\label{e10.4}
\bear
i)&\nabla_{X}^{\cS}(\gamma(Y)\psi)= \gamma(\nabla_{X}Y)\psi+ \gamma(Y)\nabla_{X}^{\cS}\psi,\\[2mm]
ii)&X(\bar{\psi}\dual \beta \psi)= \bar{\nabla_{X}^{\cS}\psi}\dual \beta \psi+ \bar{\psi}\dual \beta \nabla_{X}^{\cS}\psi,\\[2mm]
iii)&\kappa \nabla_{X}^{\cS}\psi= \nabla_{X}^{\cS}\kappa \psi,
\eear
\eeq
for all $X, Y\in \cinf(M; TM)$ and $\psi\in \cinf(M; \cS)$, where $\nabla$ is the Levi-Civita connection on $(M, g)$.
 \een
 \medskip
 
It is well known that if \eqref{e10.2} holds for some time-like future directed vector field $e$, then it holds for all such vector fields.

 \begin{remark} A linear map $\gamma$ as in (1) is called a {\em Clifford representation}.  A section $\beta$ as in (2)  is called a {\em positive energy Hermitian form} (for the Clifford representation $\gamma$), while a section $\kappa$ as in (3) is called a {\em charge conjugation} (for the Clifford representation $\gamma$).

The properties listed in \eqref{e10.4} are usually summarized by saying that $\gamma, \beta, \kappa$ are {\em covariantly constant} w.r.t.~the connection $\nabla^{\cS}$.
\end{remark}
 
 \subsubsection{Spinor bundles from global frames}\label{globiglobo}
 Let us now recall a concrete construction of the above objects, starting from a global frame on $(M, g)$.
 
 Suppose that $(M, g)$ has a global oriented and time oriented orthonormal frame $(e_{0}, \dots, e_{3})$. This will be in particular the case for all the Kerr spacetimes considered in later sections.

Let  $(\rr^{1, 3}, \eta)$ be the Minkowski space with canonical basis $(u_{0}, \dots ,u_{3})$ and ${\rm Cliff}(1, 3)$   the associated Clifford algebra. We  fix a faithful and irreducible representation $\rho_{0}: {\rm Cliff}(1, 3)\to M_{4}(\cc)$.  We denote by $\gamma_{i}\in M_{4}(\cc)$ the gamma matrices associated to $u_{i}$ for $0\leq i\leq 3$, and fix a Hermitian  sesquilinear form $\beta_{0}$ on $\cc^{4}$, and a complex conjugation $\kappa_{0}$ such that
\[
\gamma_{i}^{*}\beta_{0}= - \beta_{0}\gamma_{i}, \ \ \i \beta \gamma_{0}>0, \ \  \gamma_{i}\kappa_{0}= \kappa_{0}\gamma_{i}.
\]
Then we can construct a spinor bundle $\cS\xrightarrow{\pi}M$ by setting:
\[
\begin{array}{rl}
i)&\cS= M\times \cc^{4}, \\[2mm]
ii)&\gamma(e_{i})= \gamma_{i}, \  \beta (x)= \beta_{0},  \ \kappa(x)= \kappa_{0}, \\[2mm]
iii)&\nabla^{\cS}_{e_{i}}\psi= e_{i}\dual \nabla \psi+ \sigma_{i}\psi\hbox{ where }\\[2mm]
&\sigma_{i}= \frac{1}{4}\Gamma_{ij}^{k}\gamma_{k}\gamma^{j}, 
 \ \ \gamma^{k}= \eta^{kl}\gamma_{l}, \  \ \nabla_{e_{i}}e_{j}\eqdef \Gamma_{ij}^{k}e_{k}.
 \end{array}
\]
 \subsection{Dirac operators}
If $\cS\xrightarrow{\pi}M$ is a spinor bundle and $m\in \cinf(M; \rr)$, the {\em Dirac operator} $\slashed{D}$  acting on $\cinf(M; \cS)$ is the differential operator defined as:
\[
\slashed{D}= g^{\mu\nu}\gamma(e_{\mu})\nabla^{\cS}_{e_{\nu}} + m.
\]
where $(e_{0}, \dots ,e_{3})$ is a local frame of $TM$. We assume throughout the paper that $m\equiv 0$, in which case $\slashed{D}$ is called the {\em massless} Dirac operator.

\subsection{Weyl spinors}\label{sec10.2}
 If $(e_{0}, \dots, e_{3})$ is an oriented orthonormal local frame of $TM$ one defines the volume element  $\eta= \gamma(e_{0})\cdots \gamma(e_{3})$, which is independent of the choice of such a local frame. The section  $\eta\in \cinf(M; \End(\cS))$ satisfies:
 \begin{equation}
\label{e10.5}
\eta^{2}= -\one, \ \ \eta \gamma(X)= - \gamma(X)\eta.
\end{equation}
 Using \eqref{e10.5} and \eqref{e10.4} together with the Clifford relations \eqref{e10.1} we get:
 \begin{equation}
\label{e10.5b}
\eta^{*}\beta= \beta \eta, \ \ \nabla_{X}^{\cS}(\eta \psi)= \eta\nabla_{X}^{\cS}\psi.
\end{equation}
It follows that $\cS_{x}= \cW_{{\rm e}, x}\oplus \cW_{{\rm o}, x}$, where $\cW_{{\rm e} ,x}= \Ker (\i\eta(x)-\one)$, $\cW_{{\rm o} ,x}= \Ker (\i\eta(x)+\one)$, and the bundle $\cS$ splits as $\cW_{\rm e}\oplus \cW_{\rm o}$, where $\cW_{\rm e/o}$ is the bundle of {\em even/odd Weyl spinors}. One can check that 
\beq\label{e10.6}
\bar{\psi}_{1}\dual \beta \kappa \psi_{2}= - \bar{\psi}_{2}\dual \beta \psi_{1}, \ \ \psi_{i}\in \cinf(M; \cS),
\eeq
see e.g.~\cite[Sect.~17.5]{G}.
If $\cW$ is a vector bundle, we denote by $\cW^{\t}$ its dual bundle, so that $\cW^*=\overline{\cW^{\t}}$ in the complex case. Note that $\kappa$ maps sections of $\cW_{\rm o}$ to sections of $\overline{\cW_{\rm e}}$.  Setting
\[
\SS\defeq \cW_{\rm e}^{*},
\]
one identifies $\cS$ with $\SS^{*}\oplus \SS^{\t}$ by the map
\[
\cinf(M; \cS)\ni \psi\mapsto \psi_{\rm e}\oplus \kappa \psi_{\rm o}\eqdef \chi\oplus \phi \in \cinf(M;\SS^{*})\oplus \cinf(M;\SS^{\t}).
\]

From \eqref{e10.6} one sees that $\SS$ is equipped with the symplectic form
\[
\epsilon\defeq \frac{1}{\sqrt{2}}(\beta\kappa)^{-1}\in \cinf(M; \End(\SS, \SS^{\t})).
\]
One can also identify $T_{x}M$ with $L_{\rm a}(\SS^{*}, \SS)$, as real vector spaces by the map:
\[
T_{x}M\ni v\mapsto \beta_{x}\circ \gamma_{x}(v)\in L_{\rm a}(\cW_{{\rm e}, x}, \cW_{{\rm e, }x}^{*}).
\]
This map is injective since the representation in \eqref{e10.1} is faithful and bijective since both spaces have dimension $4$. By complexification we obtain an isomorphism
\[
\tau_{x}: \cc T_{x}M\ni z\mapsto \beta _{x}\gamma_{x}(z)\in L_{\rm a}(\cW_{{\rm e}, x}, \cW_{{\rm e}, x}^{*})\sim \SS_{x}\otimes \bar{\SS}_{x},
\]
and hence an isomorphism
\begin{equation}
\label{e10.6b}
\tau : \cinf(M; \cc TM)\ni v\mapsto \beta \gamma(v)\in \cinf(M; \SS\otimes\bar{\SS}) 
\end{equation}
If we extend $g$ to $\cc TM$ as a {\em bilinear} (not sesquilinear) form, one can show that
\[
\tau^{\t}\circ (\epsilon\otimes \bar{\epsilon})\circ \tau= g.
\]
\subsection{Null tetrads and associated frames}\label{sec10.2b}
A {\em normalized null tetrad} is a global frame   $(l, n, m, \bar{m})$  of $\cc TM$ such that:
\begin{equation}
\label{e4.1}
\begin{array}{l}
l, n \hbox{ are real} ,\ \  l\dual g l= n\dual g n=0,\  \  l\dual g n= -1 \\
 m\dual g m=l\dual g m= n\dual g m=0, \ \  m\dual g \bar{m}=1.
\end{array}
\end{equation}
It follows that if
\[
u_{0}= \frac{1}{\sqrt{2}}(l+n), \ \  u_{1}= \frac{1}{\sqrt{2}}(l-n), \  \ u_{2}= \sqrt{2}\Re m, \ \ u_{3}= \sqrt{2}\Im m,
\]
then $(u_{0}, \dots, u_{3})$ is a global  orthonormal frame for $(M, g)$ and we  associate to a spinor bundle as in \ref{globiglobo}.

\subsubsection{Time orientation}\label{fincko.3}
A normalized null tetrad induces a time orientation of $(M, g)$ by declaring that  the time-like vector $\frac{1}{\sqrt{2}}(l+n)$ is future directed. Note that for any smooth functions $\alpha, \beta$ with $\alpha, \beta\geq 0, \alpha+ \beta>0$, $\alpha l+ \beta n$ is also future directed, in particular $l,n$ are future directed null vectors. 

\subsubsection{Spin frames}
 Using the isomorphism $\tau$ in \eqref{e10.6b}
 one associates to a normalized null tetrad a  global frame $(\mo, \mi)$ of $\SS$ such that:
 \beq\label{e4.2}
 \begin{array}{l}
 \i \tau(l)= \mo\otimes \bar{\mo}, \ \ \i \tau(n)= \mi\otimes\bar{\mi}, \\[2mm]
  \i \tau(m)= \mo\otimes\bar{\mi}, \ \ \i \tau(\bar{m})= \mi\otimes\bar{\mo},\\[2mm]
  \mo\dual \epsilon \mi= 1.
 \end{array}
 \eeq
  If $s_{1}, s_{2}\in \SS$ we denote by $|s_{1})(s_{2}|\in L(\SS^{*}, \SS)$ the map $w\mapsto (s_{2}|w)s_{1}$. Using this notation we have:
  \beq\label{e4.3}
 \i \Gamma(l)= |\mo)(\mo|, \ \ \i \Gamma(n)= |\mi)(\mi|, \ \ \i \Gamma(m)= |\mo)(\mi|, \ \ \i \Gamma(\bar{m})= |\mi)(\mo|,
 \eeq
where we have set:
\beq\label{e10.6c}
\Gamma(X)= \beta \gamma(X)\in\cinf(M, \End(\SS^{*}, \SS)), \ \ X\in \cinf(M; TM).
\eeq
\subsection{Dirac and Weyl equations}\label{sec10.3}
Let  $\slashed{D}$  be the massless Dirac operator acting on $\cinf(M; \cS)$.  From \eqref{e10.5}, \eqref{e10.5b} we obtain that $\eta\circ \slashed{D}= - \slashed{D}\circ \eta$, hence after identifying $\cS$ with $\cW_{\rm e}\oplus \cW_{\rm o}$ we can write $\slashed{D}$ as:
\[
\slashed{D}= \mat{0}{\slashed{D}_{\rm o}}{\slashed{D}_{\rm e}}{0},
\] 
for $\slashed{D}_{\rm e/o}= \slashed{D}_{| \cinf(M; \cW_{\rm e/o})}$. Using  Subsect.~\ref{sec10.2} we define the {\em Weyl operator}:
\[
\DD\defeq (\beta \slashed{D})_{|\cinf(M; \cW_{\rm e})} : \cinf(M; \SS^{*})\to \cinf(M; \SS).
\]
We can write
\beq\label{e10.8}
\DD= g^{\mu\nu}\Gamma(e_{\mu})\nabla_{e_{\nu}}^{\cS}.
\eeq
The {\em Weyl equation} is
\begin{equation}
\label{e10.9}
\DD \phi=0.
\end{equation}
\begin{remark} In the physics literature, it is customary to study the Weyl anti-neutrino equation. Here we rather stick to the Weyl neutrino equation \eqref{e10.9} and we use the notation $\phi$ (usually reserved for anti-neutrinos) for its solutions.
\end{remark}
\subsection{Conformal transformations}\label{sec10.4}
Let $\hat{g}= c^{2}g$ be a metric conformal to $g$.  Denoting with hats the objects attached to $\hat{g}$ we have in particular $\hat{\gamma}(X)= c \gamma(X)$ for $X\in \cinf(M; TM)$. To fix the spin connection $\hat{\nabla}^{\cS}$ one has to fix the Hermitian form $\hat{\beta}$ and the charge conjugation $\hat{\kappa}$. It is natural to choose $\hat{\kappa}=\kappa$, but several choices of $\hat{\beta}$ are possible. A choice which is natural when one considers Weyl spinors is to set
\[
\hat{\beta}= c^{-1}\beta,
\]
so that the isomorphism $\tau$ in \eqref{e10.6b} is unchanged, i.e.~$\hat{\tau}= \tau$. A straightforward computation shows then that
\[
\hat{\nabla}^{\cS}_{X}= \nabla_{X}^{\cS}+ \12 c^{-1}\gamma(X)\gamma(\nabla c)- c^{-1}X\dual dc \one.
\]
Consequently, if $\hat{\slashed{D}}, \hat{\DD}$ are  the associated Dirac  and Weyl operators, they are related by:
\beq\label{e20.1a} 
\hat{\slashed{D}}= c^{-2}\slashed{D}c, \ \ \hat{\DD}= c^{-3}\DD c. 
\eeq

  \subsection{Lie derivative of spinors}\label{sec10.5}
The notion of Lie derivative of spinors was introduced by  Kosman \cite{K}.  We recall its definition below.

For any $X\in\cinf(M;TM)$, one sets:
\beq\label{e10b.1}
\cL_{X}\psi= \nabla_{X}^{\cS}\psi+ \frac{1}{8}((\nabla_{a}X)_{b}- (\nabla_{b}X)_{a})\gamma^{a}\gamma^{b}\psi, \ \ \psi\in \cinf(M, S).
\eeq
Note the sign change w.r.t.~\cite[(I.16)]{K}, coming from the  convention for the Clifford algebra in \cite[(I.1)]{K}.

One also defines the Lie derivative of $\gamma$ by:
\[
\cL_{X}(\gamma(v)\psi)\eqdef (\cL_{X}\gamma)(v)\psi+ \gamma(\cL_{X}v)\psi+ 
\gamma(v)\cL_{X}\psi, \ \ \psi\in \cinf(M; \cS),
\]
for all $v\in \cinf(M; TM)$. Properties of $\cL_{X}$ which are relevant for us are discussed in Appendix \ref{appLie}.

\section{Weyl equation on globally hyperbolic spacetimes}\label{sec4}\init
If $v\in \coinf(M; \SS)$ and $\phi\in \cinf(M; \SS^{*})$ we set
\begin{equation}
\label{ecoup.1}
(\phi|v)_{M}= \overline{(v| \phi)_{M}}\defeq \int_{M}\bar{\phi(x)}\dual v(x)\dVol_{g},
\end{equation}
and extend this notation to other natural cases, like  for example $v\in \cE'(M; \SS)$ and $\phi\in \cinf(M; \SS^{*})$, etc.

\subsection{Space-compact solutions}\label{sec4.1}
From now on we assume that $(M,g)$ is a globally hyperbolic spacetime (of dimension 4), equipped with its unique spin structure and denote by $\DD$ the associated Weyl operator.

\subsubsection{Space-compact solutions} 
Let $\Sol_{\rm sc}(M)$ be the space of smooth space-compact solutions of $\DD\phi=0$, $\phi\in \cinf(M; \SS^{*})$ ($\phi$ \emph{space-compact} means that the intersection of $\supp\phi$ with a space-like Cauchy surface is compact). The   {\em current}  $J(\phi_{1}, \phi_{2})\in \cinf(M; T^{*}M)$ defined by
\beq\label{e10b.0}
J(\phi_{1}, \phi_{2})\dual X\defeq \bar{\phi}_{1}\dual \Gamma(X)\phi_{2},\ \ X\in \cinf(M; TM), \  \phi_{i}\in \Sol_{\rm sc}(M),
\eeq
satisfies
\begin{equation}
\label{e10b.00}
\nabla^{a}J_{a}(\phi_{1}, \phi_{2})=0, \ \ \phi_{i}\in \Sol_{\rm sc}(M).
\end{equation}
Since $\nabla^{a}J_{a}\Omega_{g}= d (g^{-1}J\lrcorner \Omega_{g})$, we deduce from Stokes' formula that
\beq\label{stoker}
\int_{\p {U}}i^{*}(g^{-1}J(\phi_{1}, \phi_{2})\lrcorner \Omega_{g})=0, \ \phi_{i}\in \Sol_{\rm sc}(M),
\eeq
if $U$ is any open set whose boundary $\p U$ is a union of smooth hypersurfaces, such that $\supp J(\phi_{1}, \phi_{2})\cap \p U$ is compact.

Let $S\subset M$ be a smooth hypersurface and let $i: S\to M$ be the canonical injection. To express 
\[
\int_{S}i^{*}(g^{-1}J\lrcorner \Omega_{g})
\]
for $J$ a $1$-form on $M$,  we choose a vector field $l= l^{a}$ transverse to $S$, future pointing and a $1$-form $\nu= \nu_{a}dx^{a}$ on $M$ such that $TS= \Ker \nu$, normalized so that $\nu\dual l=1$. Splitting $g^{-1}J$ as
 \[
g^{-1}J= (\nu\dual g^{-1}J) l+ R, \hbox{ where }R\hbox{ is tangent to }S,
\]
we obtain:
\beq\label{corr.e-1}
i^{*}(g^{-1}J\lrcorner\Omega_{g})= (\nu\dual g^{-1}J)i^{*}(l\lrcorner \Omega_{g}).
\eeq

\subsubsection{Characteristic manifold}\label{sss:char} The \emph{principal symbol} of  $\DD$ is the section $\sigma_\DD\in\cinf(T^*M\setminus \zero;\End(\SS^*,\SS))$ given by
\beq\label{eq:sigmadd}
\sigma_\DD(x,\xi)=\Gamma(g^{-1}(x)\xi), \ \ (x,\xi)\in T^*M\setminus \zero.
\eeq

\begin{lemma}\label{prenormal} The Weyl operator $\DD$ is pre-normally hyperbolic, i.e., there exists a differential operator ${\mathbb{D}'}$ such that $(\sigma_{\DD}\circ\sigma_{{\DD}'})(x,\xi)=(\xi\cdot g^{-1}(x)\xi) \one$. 
\end{lemma}
\proof The Dirac operator $\slashed{D}$ is pre-normally hyperbolic because 
\beq\label{eq:dsquared}
\sigma_{\!\slashed{D}}^2=(\xi\cdot g^{-1}(x)\xi) \one,
\eeq
as can be checked using $\sigma_{\!\slashed{D}}=\gamma(g^{-1}(x)\xi)$ and the Clifford relations. After identification of $\cS$ with $\cW_{\rm e}\oplus \cW_{\rm o}$, $\sigma_{\!\slashed{D}}$ is anti-diagonal, therefore we can conclude from \eqref{eq:dsquared} that $\slashed{D}_{\rm o}$ and $\slashed{D}_{\rm e}$ are pre-normally hyperbolic. It follows that $\DD$ is pre-normally hyperbolic.
\qeds

The \emph{characteristic manifold} of $\DD$ is defined as
\[
\Char(\DD)=\{ (x,\xi)\in T^*M\setminus\zero \,:\, \sigma_\DD(x,\xi) \mbox{ is not invertible} \}.
\]
By Lemma \ref{prenormal}, 
\[
\Char(\DD)=\{ (x,\xi)\in T^*M\setminus\zero \,:\,  \xi\cdot g^{-1}(x)\xi = 0 \}\eqdef \cN.
\]
Its two connected components are 
\begin{equation}
\label{sloubi1}
\cN^{\pm}\defeq\cN\cap \{(x, \xi)\in T^*M\setminus\zero \,:\, \pm v\dual \xi>0\ \forall v\in T_{x}M\hbox{ future directed time-like}\}.
\end{equation}

 \subsubsection{Propagators and Cauchy problem}\label{sec4.1.2} An adaptation of an argument due to Dimock \cite{dimock} to the case of general pre-normally hyperbolic operators (see \cite[Thm.~1]{muehlhoff}) gives the existence and uniqueness of  \emph{retarded} and \emph{advanced propagators}, $\GG_{\rm ret}$ and $\GG_{\rm adv}$. Recall that $\GG_{\rm ret/ \rm adv}$ is by definition a two-sided inverse of $\DD$ (on test sections) such that
\[
\supp \GG_{\rm ret/ \rm adv} v \subset J_\pm(\supp v), \ \ v\in\coinf(M;\SS),
\]
where $J_\pm(K)$ stands for the causal future/past of $K\subset M$. The \emph{Pauli-Jordan} or \emph{causal propagator} is the difference 
\[
\GG= \GG_{\rm ret}-\GG_{\rm adv},
\]
We have
\beq\label{etrou.3}
(v_{1}| \GG v_{2})_{M}= - (\GG v_{1}|v_{2})_{M}, \ \ v_{i}\in \coinf(M; \SS),
\eeq
i.e.~$\GG^{*}= - \GG$ for the pairing $(\cdot | \cdot)_{M}$ defined in \eqref{ecoup.1}.

This allows to extend $\GG$ as a continuous operator $\GG: \cE'(M;\SS)\to \cD'(M;\SS^*)$ acting on compactly supported distributional sections $\cE'(M;\SS)$ by setting \[
(\phi| \GG v)_{M}\defeq (-\GG \phi| v)_{M}, \ \ \phi\in \coinf(M; \SS), \ v\in \cE'(M; \SS).
\]

 If $S$ is a space-like Cauchy surface, the Cauchy problem
\begin{equation}
\label{e4.0}
\begin{cases}
\DD \phi= 0, \\
r_{S}\phi= \varphi\in \coinf(S; \SS^{*}_{S}),
\end{cases}
\end{equation}  where $\SS^{*}_{S}$ is the restriction of $\SS^{*}$ to $S$ and  $r_{S}\phi= \phi_{| S}$,  has a unique solution $\phi\eqdef \mathbb{U}_{S}\varphi\in \Sol_{\rm sc}(M)$ (see e.g.~\cite[Thm.~2]{muehlhoff}). 

Using the formal notation $\mathbb{G}(x,y)$ for the Schwartz kernel of $\mathbb{G}$, for all $\phi\in\Sol_{\rm sc}(M)$ one has:
\beq\label{corr.e9}
\phi(x)= - \int_{S}\mathbb{G}(x, y) \Gamma(g^{-1}\nu)(y)\phi(y)i^{*}_{l}(\dVol_{g})(y),
\eeq
where $l$ and $\nu$ are as in \eqref{corr.e-1}. Note that since $\phi\in \Sol_{\rm sc}(M)$ and $S$ is spacelike $\supp \phi\cap S$ is compact and we can interpret the r.h.s. of \eqref{e4.0} as $\mathbb{G}(\delta_{S}\otimes \phi)$, where $\delta_{S}$ is defined in \eqref{etrou.2}. 
 Choosing $l= n, \nu= - g n$, where $n$  the future directed vector field normal to $S$, this can be rewritten as
\beq\label{corr.e10}
\phi(x)= - \int_{S} \mathbb{G}(x, y) \Gamma(n(y)) \phi(y)\dVol_{h}(y),
\eeq
where $h$ is the induced  Riemannian metric on $S$.

If $S$ is given by $\{f=0\}$ for some function $f$ with $df\neq 0$ on $S$ and if we can complete $f$ near $S$ with coordinates $y^{1}, \dots, y^{n-1}$ such that $df\wedge dy^{1}\cdots \wedge dy^{n-1}$ is direct, $\p_{f}$ future pointing, we take $l= \p_{f}$, $\nu= df$ 
in \eqref{corr.e-1}  and  obtain:
\beq\label{corr.e11}
\phi(x)= -\int_{S}\mathbb{G}(x, y)\Gamma (\nabla f)(y) \phi(y) |g|^{\12}\diff y^{1}\dots \diff y^{n-1}.
\eeq
\subsection{$L^{2}$ solutions}\label{sec4.1b}

\subsubsection{Hilbertian scalar product}\label{sec4.1.1}
Let now $S\subset M$ be any smooth Cauchy surface (with the convention that a Cauchy surface does not need to be space-like). We set 
\beq\label{e4.02}
\bea
(\phi_{1}| \phi_{2})_{\DD}&\defeq \i\int_{S} i^{*}(g^{-1}J(\phi_{1}, \phi_{2})\lrcorner \Omega_{g})\\[2mm]
&=\i\int_{S}\bar{\phi}_{1}\dual \Gamma(g^{-1}\nu)\phi_{2} \mathop{}\,i^{*}_{l}d{\rm vol}_{g},
\eea
\eeq
where $i^{*}_{l}d{\rm vol}_{g}= |i^{*}(l\lrcorner \Omega_{g})|$.

From \eqref{stoker}  we see that the r.h.s.~in \eqref{e4.02} is independent of the choice of  the Cauchy surface $S$. 

If $S$ is space-like, we obtain as in \eqref{corr.e10}:
\beq\label{trif.1}
(\phi_{1}| \phi_{2})_{\DD}= \i \int_{S} \bar{\phi_{1}}\dual \Gamma(n)\phi_{2} \dVol_{h}.
\eeq
 By \eqref{e10.2} $\i \Gamma(n)$ is positive definite, which shows that 
 $(\cdot| \cdot)_{\DD}$ is  a Hilbertian scalar product on $\Sol_{\rm sc}(M).$

If $S$ is given by $\{f=0\}$ for some function $f$  as in \eqref{corr.e11} we obtain:
\[
(\phi_{1}| \phi_{2})_{\DD}= \i \int_{S}\bar{\phi}_{1}\dual \Gamma(\nabla f) \phi_{2} |g|^{\12}\diff y^{1}\dots \diff y^{n-1}.
\]

\begin{definition}\label{def4.1}
The Hilbert space   $\Sol_{{\rm L}^{2}}(M)$, called the {\em space of $L^{2}$ solutions}, is  the completion of $\Sol_{\rm sc}(M)$ for the scalar product $(\cdot| \cdot)_{\DD}$.
 \end{definition}
Note that $\Sol_{{\rm L}^{2}}(M)\subset L^{2}_{\rm loc}(M; \SS)$ continuously, so  elements of $\Sol_{{\rm L}^{2}}(M)$ are distributional solutions of $\DD \phi=0$.
\subsubsection{Conformal transformations}\label{conforama}
Let us denote by $\hat{M}$ the spacetime $M$ equipped with the metric $\hat{g}= c^{2}g$. From Subsect.~\ref{sec10.4} and \eqref{trif.1} we obtain that
\begin{equation}
\label{trif.2}
\Sol_{{\rm L}^{2}}(M)\ni \phi\mapsto \hat{\phi}= c^{-1}\phi\in \Sol_{{\rm L}^{2}}(\hat{M})
\end{equation}
is unitary. 
 \subsubsection{Equivalent Hilbert spaces}
 We now recall  other  Hilbert spaces unitarily equivalent to $\Sol_{{\rm L}^{2}}(M)$, see e.g. \cite[Sect.~17.14]{G}. Let $\Sigma$ be a smooth space-like Cauchy surface and $\SS_{\Sigma}$, $r_{\Sigma}$ defined in \ref{sec4.1.2}.
 
\begin{proposition}\label{prop4.1b}
 The  following maps are unitary
 \beq\label{e15c.6}
\begin{CD}
 \big(\frac{\coinf(M;\SS)}{\DD\coinf(M;\SS^{*})}, \i \GG\big)@>\GG>> \big(\Sol_{\sc}(M), (\cdot| \cdot)_{\DD}\big) @>
r_{\Sigma}>>\big(\coinf(\Sigma; \SS^{*}_{\Sigma}), \nu_{\Sigma}\big),
\end{CD}
\eeq
where
\[
\overline{\varphi}_{1}\dual\nu_{\Sigma}\varphi_{2}=\i \int_{\Sigma} \bar{\varphi}_{1}\dual \Gamma(n)\varphi_{2}\,d\Vol_{h}, \ \varphi_{i}\in \coinf(\Sigma; \SS^{*}_{\Sigma}).
\]
\end{proposition}

The proof is analogous to the Dirac case, which follows from the arguments in \cite{dimock} (see e.g.~\cite[\S II.3.2]{hack}  or \cite[Thm. 19.65]{DG} for  details). One can remark that \eqref{e15c.6} yields an exact  complex
$$
\begin{CD}
0 \to \coinf(M;\SS) @>\DD>> \coinf(M;\SS) @>\GG>> \Sol_{\sc}(M)  @>
\DD >> \Sol_{\sc}(M).
\end{CD}
$$

As a consequence of Proposition \ref{prop4.1b} we have the identity
\begin{equation}
\label{e15c.6b}
(\phi_{1}|\phi_{2})_{\DD}= (v_{1}| \i \GG v_{2})_{M}= (\i \GG v_{1}| v_{2})_{M}, \hbox{ for }\phi_{i}= \GG v_{i}, \ v_{i}\in\coinf(M; \SS), 
\end{equation}
which extends to
\begin{equation}
\label{e15c.16c}
(v| \phi)_{M}= (\i \GG v| \phi)_{\DD}, \ v\in \coinf(M; \SS), \ \phi\in \Sol_{{\rm L}^{2}}(M). 
\end{equation}
\subsection{Traces of $L^{2}$ solutions on hypersurfaces}\label{ss:green}
Let $S\subset M$ be a piecewise smooth hypersurface, equipped with a piecewise smooth density $dm$. If $u\in \cE'(S; \SS_{S})$ we denote by $\delta_{S}\otimes u\in \cD'(M; \SS)$ the distribution defined by
\beq\label{etrou.2}
(v| \delta_{S}\otimes u)_{M}\defeq \int_{S}\bar{v}_{| S}\dual u \,dm, \ v\in \coinf(M).
\eeq
If $S$ is a smooth space-like Cauchy hypersurface and $\phi\in \Sol_{\rm sc}(M)$,  we can rewrite \eqref{corr.e10} or \eqref{corr.e11} as
\beq\label{ecorr.3}
\phi= - \GG (\delta_{S}\otimes \Gamma_{S}r_{S}\phi),\ \phi\in \Sol_{\rm sc}(M), 
\eeq
where $ S$ is equipped with the appropriate density  and $\Gamma_{S}: S\to End(\SS^{*}_{S}, \SS_{S})$ denotes the factor $\Gamma(n)$ or $\Gamma(\nabla f)$ in \eqref{corr.e10} or \eqref{corr.e11}.

 Note that \eqref{ecorr.3} extends to $\phi\in \Sol_{{\rm L}^{2}}(M)$. In fact   the trace $r_{S}\phi$ of $\phi$ on $S$ is well defined using  that $\WF(\phi)\subset \cN$.  

In this subsection we will show   how to extend  \eqref{ecorr.3}   for $\phi\in \Sol_{{\rm L}^{2}}(M)$ to cases where $S$ is not necessarily smooth nor everywhere space-like. We recall that $J(K)$ for $K\subset M$ is the causal shadow of $K$.

\begin{proposition}\label{propcoupure.1}
Let $S\subset M$ be a piecewise smooth hypersurface and  $\SS_{S}$ the restriction of $\SS$ to $S$. Let $dm$ be a  piecewise smooth density on $S$ and $\i \Gamma_{S}: S\to End(\SS^{*}_{S}, \SS_{S})$ a piecewise smooth positive Hermitian structure on $\SS^{*}_{S}$. We denote by $L^{2}(S; \SS^{*}_{S})$ the  Hilbert space of $L^{2}$ sections of $\SS^{*}_{S}$ equipped with the scalar product:
\[
(u| u)_{S}\defeq \i \int_{S}\bar{u}\dual \Gamma_{S}u \,dm.
\]
We assume the following: 
\ben
\item $J(K)\cap S$ is compact for all  $K\Subset M$.
 \item there exists a map $r_{S}: \Sol_{\rm sc}(M)\to L^{2}(S; \SS^{*}_{S})$ such that
 \[
\phi= - \GG(\delta_{S}\otimes \Gamma_{S}r_{S}\phi), \ \forall \phi\in \Sol_{\rm sc}(M).
\]
\item the map $r_{S}$ extends as a bounded operator
\[
T_{S}: \Sol_{{\rm L}^{2}}(M)\to L^{2}(S; \SS^{*}_{S}).
\]
\een
Then for each $\chi\in \coinf(M)$ and  $\chi_{S}\in \coinf(S)$ such that $\chi_{S}\equiv 1$ on $J(\supp \chi)\cap S$ one has:
\[
\chi \phi= - \chi \GG(\delta_{s}\otimes \Gamma_{S}\chi_{S}T_{S}\phi), \ \forall \phi\in \Sol_{{\rm L}^{2}}(M).
\]
\end{proposition}
\proof   Let us recall  the notation $\delta_{S}\otimes \cdot$ introduced in \eqref{etrou.2}.

For $v\in \coinf(M; \SS)$,  $u\in L^{2}(S; \SS^{*}_{S})$ and $\chi_{S}\in \coinf(S)$, we have
\[
\bea
(v| \i \GG(\delta_{S}\otimes \Gamma_{S}\chi_{S}u))_{M}&= (\i \GG v| \delta_{S}\otimes \Gamma_{S}\chi_{S}u)_{M}\\
&= (\i r_{S}\GG v| \chi_{S}u)_{S}= (\i T_{S} \GG v| \chi_{S}u)_{S}\\
&= (\i \GG v| T_{S}^{*}\chi_{S}u)_{\DD}= (v| T_{S}^{*}\chi_{S}u)_{M}.
\eea
\]
In the first line we use \eqref{etrou.3}, in the second line we use that $\GG v\in \Sol_{\rm sc}(M)$ and in the third line we use that $T_{S}^{*}: L^{2}(S; \SS^{*}_{S})\to \Sol_{{\rm L}^{2}}(M)\subset \cD'(M; \SS^{*})$ and the identity \eqref{e15c.16c}. 

It follows that 
\beq\label{etrou.4}
\i \GG (\delta_{S}\otimes \Gamma_{S}\chi_{S}u)=T_{S}^{*}\chi_{S}u, \ u\in L^{2}(S, \SS^{*}_{S}).
\eeq
Let now  $\chi\in \coinf(M)$ and $\chi_{S}\in \coinf(S)$ such that $\chi_{S}\equiv 1$ on $J(\supp \chi)\cap S$. Because of the support of properties of $\GG$ we have
\begin{equation}
\label{etrou.1}
\bea
\chi \phi&= - \chi \GG(\delta_{S}\otimes \Gamma_{S}r_{S}\phi)\\
&=  -\chi \GG(\delta_{S}\otimes \Gamma_{S}\chi_{S}r_{S}\phi)= \i \chi T_{S}^{*}\chi_{S}T_{S}\phi, \ \phi\in \Sol_{\rm sc}(M).
\eea
\end{equation}
Observing that $\chi: \Sol_{{\rm L^{2}}}(M)\to L^{2}(M; \SS^{*})$ is bounded,  the identity \eqref{etrou.1}  extends to $\phi\in \Sol_{{\rm L^{2}}}(M)$ by density. \qed

 \subsection{Action of Killing vector fields}\label{sec4.2}
 Let $X$ be a complete Killing vector field on $(M, g)$.  By Lemmas \ref{lemma1.2}, \ref{lemma1.3},   $\cL^{*}_{X}$  preserves $\cinf(M; \SS)$ and $\DD\coinf(M; \SS^{*})$, and $\cL_{X}$ preserves $\Sol_{\rm sc}(M)$.

\begin{proposition}\label{prop10.1}
The operator $\i^{-1}\cL_{X}$ with domain  $\Sol_{\rm sc}(M)$ is essentially selfadjoint on the Hilbert space $\Sol_{{\rm L}^{2}}(M)$. 
\end{proposition}
 \proof 
 Let $\phi_{1}, \phi_{2}\in \Sol_{\rm sc}(M)$.
  Since $X$ is Killing, using the properties of the spinorial Lie derivative listed in Appendix \ref{appLie}, we obtain
 \[
 X(\bar{\phi}_{1}\dual \Gamma(v)\phi_{2})= \bar{\cL_{X}\phi}_{1}\dual \Gamma(v)\phi_{2}+ \bar{\phi}_{1}\dual \Gamma(v)\cL_{X}\phi_{2}+ \bar{\phi}_{1}\dual \Gamma(\cL_{X}v)\phi_{2},
 \]
 so  defining  the 1-form $K(\phi_{1}, \phi_{2})\in \cinf(M; T^{*}M)$ by 
 \[
 K(\phi_{1}, \phi_{2})\dual v\defeq\bar{\cL_{X}\phi}_{1}\dual \Gamma(v)\phi_{2}+ \bar{\phi}_{1}\dual \Gamma(v)\cL_{X}\phi_{2},
 \]
 we have $K(\phi_{1}, \phi_{2})= \cL_{X}J(\phi_{1}, \phi_{2})$, where $J(\phi_{1}, \phi_{2})$ is the conserved current defined in \eqref{e10b.0}.
 
Since $X$ is Killing we have:
 \[
 (g^{-1}\cL_{X}J)\lrcorner\: \Omega_{g}= \cL_{X}(g^{-1}J)\lrcorner\: \Omega_{g}= \cL_{X}(g^{-1}J\lrcorner\: \Omega_{g}),
 \]
 and by Cartan's formula:
 \[
  \cL_{X}(g^{-1}J\lrcorner\: \Omega_{g})= X\lrcorner\:d(g^{-1}J\lrcorner\: \Omega_{g})+ d(X\lrcorner\: g^{-1}J\lrcorner\: \Omega_{g}).
 \]
 Since $J$ is conserved we have $d(g^{-1}J\lrcorner\: \Omega_{g})=0$ hence:
 \[
 g^{-1}K\lrcorner\:\Omega_{g}= d(X\lrcorner\: g^{-1}J\lrcorner\: \Omega_{g}).
 \]
  and since $i^{*}d \omega=d i^{*}\omega$ we get:
 \[
 \int_{S}i^{*}(g^{-1}K\lrcorner\: \Omega_{g})= \int_{S}di^{*}(X\lrcorner\: g^{-1}J\lrcorner\: \Omega_{g})=0.
 \]
 If $S$ is space-like this yields:
 \[
 \int_{\Sigma}(\bar{\cL_{X}\phi}_{1}\dual \Gamma(n)\phi_{2}+ \bar{\phi}_{1}\dual \Gamma(n)\cL_{X}\phi_{2})\dVol_{h}=0,
 \]
  i.e.~$\i^{-1}\cL_{X}$ is symmetric on $\Sol_{\rm sc}(M)$. The one parameter group $\{\e^{s \cL_{X}}\}_{s\in \rr}$ preserves  $\Sol_{\rm sc}(M)$, is isometric and strongly continuous on $\Sol_{\rm sc}(M)$. Therefore,  by Nelson's invariant domain theorem, it extends to a unitary group $\{U_{s}\}_{s\in \rr}$ on $\Sol_{{\rm L}^{2}}(M)$, whose generator is the closure of $\i^{-1}\cL_{X}$ on $\Sol_{\rm sc}(M)$. \qeds

In the sequel we will also denote by $\i^{-1}\cL_{X}$ the corresponding selfadjoint extension on $\Sol_{{\rm L}^{2}}(M)$.

\subsection{Use of null tetrads}\label{sec4.3}

Let   $(l, n, m, \bar{m})$ be a normalized null tetrad (see Subsect.~\ref{sec10.2b})  and    $(\mo, \mi)$  the associated frame of $\SS$. For $\phi\in \cinf(M; \SS^{*})$ one sets then:
 \beq\label{e4.3b}
 \phi_{0}= \phi\dual \mo, \ \ \phi_{1}= \phi\dual \mi, \ \ \mathit{U} \phi= \col{\phi_{0}}{\phi_{1}}\in \cinf(M; \cc^{2}),
 \eeq
 so that  $\phi= \phi_{0}\mo^{*}+ \phi_{1}\mi^{*}$ if $(\mo^{*}, \mi^{*})$ is the dual frame of $\SS^{*}$.   If the tetrad is chosen such that $l+n$ is normal to  a space-like Cauchy surface $S$ , then from \eqref{e4.3} we obtain
 \begin{equation}
 \label{e4.4}
 (\phi| \phi)_{\DD}= \frac{1}{\sqrt{2}}\int_{S}\left(|\phi_{0}|^{2}+ |\phi_{1}|^{2}\right)\dVol_{h}.
 \end{equation}

\subsubsection{Action of Killing vector fields}\label{sec4.3.1}
Let $X$ be a complete Killing vector field, and $\i^{-1}\cL_{X}$ its selfadjoint action  on $\Sol_{{\rm L}^{2}}(M)$. We would like to compute $\mathit{U}\circ \cL_{X}\circ \mathit{U}^{-1}$, where $\mathit{U}$ is defined in \eqref{e4.3b}. Assume that the null tetrad $(l,n,m, \bar{m})$ is invariant under $X$. Then by Lemma \ref{lemma1.5} we have
 $\cL_{X}^{*}\mo= \cL_{X}^{*}\mi=0$. We have $\cL_{X}\mo^{*}= \cL_{X}\mi^{*}=0$ so 
 \[
 \cL_{X}(\phi_{0}\mo^{*}+ \phi_{1}\mi^{*}) = X(\phi_{0})\mo^{*}+ X(\phi_{1})\mi^{*}.
 \]
 Therefore
 \beq\label{e4.9a}
\mathit{U}\circ \cL_{X}\circ \mathit{U}^{-1}= X,
 \eeq
 where on the right we mean the trivial  action of the vector field $X$ on $\cinf(M; \cc^{2})$.
 
 \section{Hadamard states for the Weyl equation}\label{quanto}\init
 
\subsection{The $\CAR$ $*$-algebra and states}\label{demian}
 We denote by $\CAR(M)$ the   $\CAR$ $*$-algebra associated to any of the equivalent pre-Hilbert spaces in Proposition \ref{prop4.1b} (see Appendix \ref{algebraic} for the definition). 
  
 We can use any of the three pre-Hilbert spaces in  Proposition \ref{prop4.1b}  to  specify a quasi-free state on $\CAR(M)$.  If one considers the first one as the starting point (which is possibly the most natural choice), then one can show that  to define a quasi-free state $\omega$ it suffices to specify a pair of   {\em spacetime covariances} (or \emph{two-point functions} if one speaks of the associated Schwartz kernels), i.e.~a pair of operators  $\LL^{\pm}$ satisfying:
\begin{equation}
\label{e15c.7}
\begin{array}{rl}
 i)&\LL^{\pm}:\coinf(M; \SS)\to \cD'(M; \SS^{*}) \hbox{ is linear continuous},\\[2mm]
ii)&\LL^{\pm}\,\geq 0, \\[2mm]
iii)&\LL^{+}\!+ \LL^{-}= \i \GG,\\[2mm]
iv)&\DD\!\LL^{\pm}=\,\LL^{\pm}\!\DD=0.
\end{array}
\end{equation}
Alternatively,  one can  define the state $\omega$ by  its {\em solution space covariances},   i.e.~operators $C^{\pm}\in B(\2Sol(M))$ such that
\begin{equation}
\label{frip.1}
C^{\pm}\geq 0, \ \ C^{+}+ C^{-}= \one.
\end{equation}
By Proposition \ref{prop4.1b}, the two types of covariances are related as follows: 
\beq\label{frip.6}	
\bar{v}\,\dual \LL^{\pm}\!v= (\GG v| C^{\pm}\GG v)_{\DD}, \ v\in \coinf(M; \SS).
\eeq

\subsection{Hadamard states}\label{quanto.ss3}
The definition of Hadamard states for Weyl fields is analogous to the  well-known case of Dirac fields, see e.g.~\cite{hollands}.
\begin{definition}\label{defcoupure.1}
 A quasi-free state $\omega$ on $\CAR(M)$ is a \emph{Hadamard state} if it satisfies (the \emph{Hadamard condition}):
 \beq\label{eq:hc}
\WF(\LL^{\pm})'\subset \cN^{\pm}\times \cN^{\pm},
\eeq
where $\cN^+$ and $\cN^-$ are the two components of the characteristic set   defined in \eqref{sloubi1}.
\end{definition}

Several equivalent definitions of the wavefront set $\wf(\phi)$ of a distribution $\phi$ are recalled in Appendix \ref{secapp1}. Here we only sketch very briefly the characterization of $\wf(\phi)$ in terms of what we call \emph{generalized oscillatory functions}. 

Namely, in the simplest $\rr^n$ setting, an \emph{oscillatory function at} $q_0=(x_0,\xi_0)\in T^*\rr^n$  is a function  (strictly speaking, family of functions) on $\rr^n$ of the form   
$$
w_q^\lambda (x)= \chi(x) \e^{\i\lambda(x-y)\cdot \eta}, \  \lambda\geq 1, \ q=(y,\eta)\in T^*\rr^n,
$$
where $\chi\in C_{\c}^\infty(\rr^n)$ and $\chi(x_0)\neq 0$. This can be extended to the setting of manifolds in the obvious way. Next, we say that $v_q^\lambda$ is a \emph{generalized oscillatory function at} $q_0\in T^*M\setminus\zero$ if it is of the form $v_q^\lambda=A^*w_q^\lambda$, where $v_q^\lambda$ is an oscillatory function at $q_0$ and $A$ is a properly supported pseudo-differential operator of order $0$ and elliptic at $q_0$.

With these definitions, $q_0=(x_0,\xi_0)\notin \wf(\phi)$ iff there exists a generalized oscillatory test function $v_q^\lambda$ at $q_0$ such that for all $N\in \nn$, 
$$
|(v^\lambda_q | \phi )_M| \leq C_N \lambda^{-N}, \ \lambda\geq 1, 
$$
uniformly for $q$ in a neighborhood of $(x_0,\xi_0)$ in $T^*M\setminus \zero$. If $\LL\,: \coinf(M; \SS)\to \cD'(M; \SS^{*})$ is linear and continuous, then $\wf(\LL)$ is by definition the wavefront set of the Schwartz kernel of $\LL$, and $\wf(\LL)'$ is defined by the usual convention
$$ 
(x,\xi,y,\eta)\in\wf(\LL)' \,  \Leftrightarrow\,   (x,\xi,y,-\eta)\in\wf(\LL). 
$$

\medskip 
The fact that the phase space for  Weyl or Dirac fields is a (pre)-Hilbert space has some important consequences for  Hadamard states, which we will now explain.

Let us assume that the  quasi-free state $\omega$  is defined by its solution space covariances  $C^{\pm}$, see Subsect.~\ref{demian}.

\begin{lemma}\label{lemmacoup.1}
 Suppose that for any $q_{0}\in\cN^{\mp}$ there exists a generalized oscillatory test function $v_{q}^{\lambda}$ at $q_{0}$ such that if $\phi_{q}^{\lambda}= \GG v_{q}^{\lambda}$ one has
 \[
 \| (C^{\pm})^{\12}\phi_{q}^{\lambda}\|_{\DD}\leq C_{N}\lambda^{-N}, \ \forall N\in \nn
 \]
 uniformly for $q$ in a neighborhood of $q_{0}$ in $T^{*}M\setminus\zero$. Then  $\omega$  is a Hadamard state.
 \end{lemma}
\proof 
By the same argument as for Klein-Gordon fields, see e.g.~\cite{Ra}, it suffices to prove that $\WF(\LL^{\pm})'\cap \Delta\subset \cN^{\pm}\times \cN^{\pm}$, where $\Delta\subset T^{*}M\times T^{*}M$ is the diagonal.  If $q_{0}\in\cN^{\mp}$  and $v_{q}^{\lambda}$ are as in the lemma,  in view of   $C^{\pm}\in B(\2Sol(M))$ and  $C^{\pm}\geq 0$ we have
\[
\bar{v_{q}}^{\lambda}\,\dual \LL^{\pm} \! v_{q}^{\lambda}= (\phi_{q}^{\lambda}| C^{\pm}\phi_{q}^{\lambda})_{\DD}=  \| (C^{\pm})^{\12}\phi_{q}^{\lambda}\|^{2}_{\DD}\in O(\lambda^{-N}), \  N\in \nn,
\]
so $(q_{0}, q_{0})\not \in \WF(\LL^{\pm})'$, which by the remark above implies that $\omega$ is a Hadamard state. \qed

\begin{theorem}\label{propcoup.2}
 Suppose that
 \beq\label{ecoup.6}
 \WF((C^{\pm})^{\12}\phi)\subset \cN^{\pm} \ \ \forall \phi\in \Sol_{{\rm L}^{2}}(M).
 \eeq
 Then the state $\omega$  is a Hadamard state.
\end{theorem}
\proof Let $q_{0}\in \cN^{\mp}$ and $N\in \nn$. By \eqref{ecoup.6} and Lemma \ref{lem:osc4}, there exists a generalized oscillatory test function $v_{q}^{\lambda}$ at $q_{0}$ such that
\[
\sup_{\lambda\geq 1}\lambda^{N}|(v_{q}^{\lambda}|((C^{\pm})^{\12}\phi )_{M}|<\infty \  \ \forall \phi\in \Sol_{{\rm L}^{2}}(M).
\]
Applying the uniform boundedness principle to the family of linear forms
\[
T_{\lambda}: \Sol_{{\rm L}^{2}}(M)\ni \phi\mapsto \lambda^{N}(v_{q}^{\lambda}|((C^{\pm})^{\12}\phi )_{M}\in \cc
\]
we obtain that
\begin{equation}
\label{ecoup.6b}
\sup_{\lambda\geq 1, \| \phi\|_{\DD}=1} \lambda^{N}|(v_{q}^{\lambda}|((C^{\pm})^{\12}\phi )_{M}|<\infty.
\end{equation}
Denoting $\phi_{q}^{\lambda}= \GG v_{q}^{\lambda}$  and using also \eqref{e15c.16c} this gives
\begin{equation}
\label{ecoup.7}
\bea
\| (C^{\pm})^{\12} \phi_{q}^{\lambda}\|_{\DD}&= \sup_{\| \phi\|_{\DD}=1}|((C^{\pm})^{\12}\phi_{q}^{\lambda}| \phi)_{\DD}|\\
&= \sup_{\| \phi\|_{\DD}=1}|(\phi_{q}^{\lambda}| (C^{\pm})^{\12}\phi)_{\DD}|=  \sup_{\| \phi\|_{\DD}=1}|(v_{q}^{\lambda}| (C^{\pm})^{\12}\phi)_{M}|\in O(\lambda^{-N})
\eea
\end{equation}
which by Lemma \ref{lemmacoup.1} implies that $\omega$ is a Hadamard state. \qeds

The first, obvious advantage of Theorem \ref{propcoup.2} is that it gives a criterion in terms of solutions rather than  bi-solutions. On top of that, the  merit is that we now only need to consider square-integrable solutions rather than  arbitrary distributional ones. This will become crucial in the next chapters, where the use of scattering theory will force us to work in an $L^2$ setting.

 \section{The  Weyl equation on Kerr spacetime}\init\label{sec9}
\def\MIII{M_{\rm III}}

\subsection{Boyer--Lindquist blocks}  We now recall the relevant facts on Kerr black hole geometry.

We fix $a\in \rr, M>0$ with $0<|a|<M$, i.e.~we consider the \emph{slowly rotating} Kerr case.  One sets
\[
\begin{array}{l}
\Delta = r^{2}- 2Mr+a^2,\ \ 
\rho^{2} = r^{2}+ a^{2}\cos^{2}\theta,\\[2mm]
 \sigma^{2} = (r^{2}+a^{2})^{2}- a^{2}\Delta\sin^{2}\theta= (r^{2}+a^{2})\rho^{2}+ 2 a^{2}Mr\sin^{2}\theta,
\end{array}
\]
 and $r_{\pm}= M\pm \sqrt{M^{2}- a^{2}}$ for the two roots of $\Delta$ as a function of $r$.

The  Boyer--Lindquist blocks  are the manifolds $(\MI, g)$, $(\MII, g)$, where
\[
\begin{array}{l}
\MI = \rr_{t}\times\open{r_{+},+\infty}_r\times\mathbb{S}^{2}_{\theta, \varphi},\\[2mm]
\MII = \rr_{t}\times\open{r_{-},r_{+}}_r\times\mathbb{S}^{2}_{\theta, \varphi},\\[2mm]
\end{array}
\] 
$\theta\in \closed{0, \pi}$, $\varphi\in \rr/2\pi \zz$ are the spherical coordinates on $\mathbb{S}^{2}$, and
\[
g=-\left(1-\frac{2Mr}{\rho^2}\right)dt^2-\frac{4aMr\sin^2\theta}{\rho^2}\diff td\varphi +\frac{\rho^2}{\Delta}\diff r^2+\rho^2d\theta^2+\frac{\sigma^2}{\rho^2}\sin^2\theta \diff\varphi^2.
\]
(We will not need to consider the third Boyer--Lindquist block $\MIII$ corresponding to $r\in \open{-\infty, r_{-}}$). The global coordinates $(t, r, \theta, \varphi)$ on $\MI, \MII$  are called {\em Boyer--Lindquist coordinates}.   We have $\det g= -\rho^{4}\sin^{2} \theta$ and
the inverse metric $g^{-1}$ has components:
\beq\label{e5.1a}
\begin{array}{l}
g^{tt}= - \dfrac{\sigma^{2}}{\Delta\rho^{2}},\ \ g^{rr}= \dfrac{\Delta}{\rho^{2}},\\[2mm]
g^{t\varphi}=-\dfrac{2aMr}{\Delta\rho^{2}},\ \
g^{\varphi\varphi}= \dfrac{\Delta- a^{2}\sin^{2}\theta}{\Delta\rho^{2}\sin^{2}\theta},\ \ 
g^{\theta\theta}= \dfrac{1}{\rho^{2}},
\end{array}
\eeq
all other being equal to $0$. 
Since $g^{tt}<0$ in ${\rm M}_{\rm I}$, the vector field $\nabla t$ is time-like and one fixes the  time orientation of $(\MI, g)$   by declaring $-\nabla t$ to be future directed.   By convention the time orientation of ${\rm M}_{\rm II}$ is the one inherited from its embedding into ${\rm K}\kst$, see \ref{eclap} below. 

The following fact appears to be folklore knowledge. For lack of a reference that show the precise statement, a proof is given in Proposition \ref{prop1}.
\begin{proposition}\label{debile.1}
 $(\MI, g)$ is globally hyperbolic. 
\end{proposition}

\subsection{Time reversal} When considering the gluing of various blocks into a larger spacetime  it is useful to introduce the following notation. 

If $(M, g)$ is a spacetime we denote by $(M', g)$ the same Lorentzian manifold with the opposite time orientation. So, $\id : (M, g)\to (M', g)$ is an isometric involution reversing the time orientation. If $(M, g)$ admits a spinor bundle $\cS\xrightarrow{\pi}M$, then so does $(M', g)$. We will generally decorate with primes the objects associated to $(M', g)$. 

Consequences of the time reversal on the level of classical and quantized Weyl fields are discussed in Section \ref{quanto.ss2} in the appendix. 

\subsection{The ${\rm K}\kst$ and $\stk{\rm K}$ spacetimes }\label{turluto} We now recall the {\em Kerr-star} and {\em star-Kerr} coordinates, which allow to glue together ${\rm M}_{\rm I}$ and  ${\rm M}_{\rm II}$ along parts of $\{r= r_{+}\}$.   As pre-announced, we will remove from ${\rm K}\kst$ and $\stk{\rm K}$ the parts corresponding to the third Boyer--Lindquist block $\mathrm{M}_{\rm III}$.
\subsubsection{The ${\rm K}\kst$ spacetime} We set 
\[
{\rm K}\kst= \rr_{t\kst}\times\open{r_{-}, +\infty}_{r}\times \mathbb{S}^{2}_{\theta, \varphi\kst}, 
\]
equipped with the metric
\[
g= g_{tt}dt^{*2}+ 2 g_{t\varphi}dt\kst d\varphi\kst+ g_{\varphi\varphi}d\varphi^{*2}+ 2 dt\kst dr- 2a\sin^{2}\theta d\varphi\kst dr+ \rho^{2}d\theta^{2}.
\]
The global coordinates $(t\kst, r, \theta, \varphi\kst)$ are called {\em Kerr-star coordinates}. 

We will denote by $\p_{r\kst}$, $\p_{\theta\kst}$, the coordinate vector fields $\p_{r}$, $\p_{\theta}$ in Kerr-star coordinates. We time orient ${\rm K}\kst$ by declaring that the null vector $-\p_{r\kst}$ is future directed.
We have $\det g= - \rho^{4}\sin^{2}\theta$. The inverse metric $g^{-1}$ has components:
 \beq\label{e5.1aa}
 \begin{array}{l}
 g^{t\kst t\kst}= \dfrac{a^{2}\sin^{2}\theta}{\rho^{2}},\ \
 g^{t\kst r\kst}= \dfrac{r^{2}+ a^{2}}{\rho^{2}},\ \
 g^{r\kst r\kst}= \dfrac{\Delta}{\rho^{2}},\\[2mm]
 g^{t\kst\varphi\kst}= \dfrac{a}{\rho^{2}},\ \ 
 g^{r\kst\varphi\kst}=\dfrac{a}{\rho^{2}},\\[2mm]
 g^{\varphi\kst\varphi\kst}=\dfrac{1}{\rho^{2}\sin^{2}\theta},\ \ g^{\theta\kst\theta\kst}= \dfrac{1}{\rho^{2}},
 \end{array}
  \eeq
  all other being equal to $0$. 
\subsubsection{Embedding ${\rm M}_{\rm I}$ and ${\rm M}_{\rm II}$ into ${\rm K}\kst$}\label{eclap}

Let  $x(r)$ and $\Lambda(r)$ for $r\in \open{r_{-}, r_{+}}\cup \open{r_{+}, +\infty}$ be such that 
\beq\label{eq:TLambda}
\frac{dx}{dr}=\frac{r^2+a^2}{\Delta}, \ \ \frac{d\Lambda}{d r}=\frac{a}{\Delta}.
\eeq
 We  choose, see \cite[Lem.~3.4.2]{N}:
 \beq\label{defdeix}
 \begin{array}{l}
  x(r)= r+ \frac{1}{2\kappa_{+}}\ln(|r-r_{+}|)+\frac{1}{2\kappa_{-}}\ln(|r-r_{-}|),\\[2mm]
  \Lambda(r)= \frac{a}{r_{+}- r_{-}}\ln(\dfrac{|r-r_{+}|}{|r-r_{-}|}),
 \end{array}
 \eeq
 where \beq\label{defodekappa}
\kappa_{\pm}= \frac{r_{\pm}- r_{\mp}}{2(r_{\pm}^{2}+a^{2})},
\eeq
so that $\kappa_{-}<0<\kappa_{+}$ and $\kappa_{+}\kappa_{-}^{-1}= - r_{-}r_{+}^{-1}$.

We define the map  $j\kst: \MI\cup \MII\to {\rm K}\kst$ by:
\[
t\kst\circ j\kst= t+ x(r), \ \  r\circ j\kst= r, \ \ \theta\circ j\kst= \theta, \ \ \varphi\kst\circ j\kst= \varphi+ \Lambda(r),
\]
which allows to identify  isometrically $(\MI, g)$ resp.~ $(\MII, g)$ with   $(\MI\kst, g)$, resp.~$(\MII\kst, g)$,  where
\[
\MI\kst= {\rm K}\kst\cap\{r_{+}<r\}, \  \ \MII\kst= {\rm K}\kst\cap \{r_{-}<r<r_{+}\}.
\]
Using $j\kst$ one can glue  ${\rm M}_{\rm I}$ with ${\rm M}_{\rm II}$ inside ${\rm K}\kst$ along the {\em future horizon}:
\[
\sH_+ = \rr_{t\kst}\times\{ r= r_{+} \} \times \mathbb{S}^{2}_{\theta, \varphi}.
\]

The embedding of ${\rm M}_{\rm I}$ and ${\rm M}_{\rm II}$  into ${\rm K}\kst$ respects the time orientation. For coherence of notation we will use the following definition. 

\begin{definition}\label{defdeMIunionMII}
We set   \[
{\rm M}_{{\rm I}\cup {\rm II}}\defeq (j\kst)^{-1}({\rm K}\kst)= \MI\cup\MII\cup \sH_{+}, 
\]
with the spacetime structure inherited from ${\rm K}\kst$.
\end{definition}
The following fact will be checked in Appendix \ref{secG}, see Proposition \ref{pripoto}.
\begin{proposition}\label{pripo}
  $({\rm M}_{{\rm I}\cup {\rm II}}, g)$ is globally hyperbolic. 
\end{proposition}

\subsubsection{The $\stk{\rm K}$ spacetime}
Similarly we set
\[
\stk{\rm K}=  \rr_{\stk t}\times \open{r_{-}, +\infty}_{r}\times \mathbb{S}^{2}_{\theta, \stk\varphi},
\]
equipped with the metric
\[
g= g_{tt}d\stk t^{2}+ 2 g_{t\varphi}d\stk t d\stk\varphi+ g_{\varphi\varphi}d\stk\varphi^{2}- 2 d\stk t dr+ 2a\sin^{2}\theta d\stk\varphi dr+ \rho^{2}d\theta^{2}.
\]
The global coordinates $(\stk t, r, \theta, \stk\varphi)$ are called {\em star-Kerr coordinates} and as before $\p_{{}^{*}\!r}$, $\p_{{}^{*}\!\theta}$ denote the coordinate vector fields $\p_{r}$, $\p_{\theta}$ in star-Kerr coordinates.  We time orient $\stk{\rm K}$ by declaring that the null vector $\p_{\stk r}$ is future directed. We have $\det g= - \rho^{4}\sin^{2}\theta$.

\subsubsection{Embedding ${\rm M}_{\rm I}$ and ${\rm M}_{\rm II}$ into $\stk{\rm K}$}
Again, the map  $\stk j: \MI\cup \MII\to {\rm K}\kst$ defined by:
\[
\stk t\circ \stk j= t- x(r), \ \  r\circ \stk j= r, \ \ \theta\circ \stk j= \theta, \ \ \stk \varphi\circ \stk j= \varphi- \Lambda(r),
\]
allows to identify  isometrically $(\MI, g)$ resp.~$(\MII, g)$  with   $(\stk\MI, g)$, resp.~$(\stk\MII, g)$, where
\[
\stk\MI= \stk{\rm K}\cap \{r_{+}<r\}, \ \ \stk\MII= \stk{\rm K}\cap \{r_{-}<r<r_{+}\}.
\]
Using $\stk j$ one can glue  ${\rm M}_{\rm I}$ with ${\rm M}_{\rm II}$ inside $\stk{\rm K}$ along the \emph{past horizon}:
\beq\label{corr.e2}
\sH_- =\rr_{\stk t}\times\{ r= r_{+} \} \times \mathbb{S}^{2}_{\theta, \stk\varphi}.
\eeq
The embedding of ${\rm M}_{\rm I}$ (resp.~${\rm M}_{\rm II}$)  into $\stk{\rm K}$ respects (resp.~reverses) the time orientation, see \cite[Lem.~3.1.3]{N}.

In  the original Boyer--Lindquist coordinates, the future and past horizons $\sH_+$ and $\sH_-$ are reached at infinite values of $t$.

\subsection{Conformal extension of ${\rm M}_{\rm I}$}\label{confconf}
\begin{figure}[H]
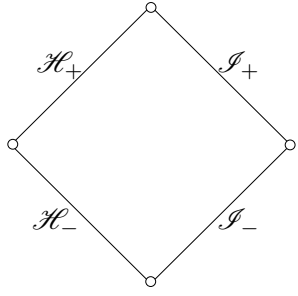\caption{The conformal extensions of $\MI$.}\label{fig10}
\end{figure}

The Penrose conformal extension of ${\rm M}_{\rm I}$ is obtained by setting $\hat g=w^{2} g$, where $w=r^{-1}\in \open{0, r_{+}^{-1}}$. The metric $\hat{g}$ expressed in coordinates $(t\kst, w, \theta, \varphi\kst)$  equals , see e.g.~\cite[(8.36)]{HN}:
\[
\bea
\hat{g}&= -\Big(w^{2}- \dfrac{2Mw^{3}}{1+ a^{2}w^{2}\cos^{2}\theta}\Big)dt^{*2}- \dfrac{4Maw^{3}\sin^{2}\theta}{1+ a^{2}w^{2}\cos^{2}\theta}dt^{*}d\varphi^{*}\\[2.5mm]
&\phantom{=}\, +\Big(1+ a^{2}w^{2}+\dfrac{2Ma^{2}w^{3}\sin^{2}\theta}{1+ a^{2}w^{2}\cos^{2}\theta}\Big)\sin^{2}\theta d\varphi^{*2}\\[2.5mm]
&\phantom{=}\, + \big(1+a^{2}w^{2}\cos^{2}\theta\big)d\theta^{2}- 2 dt^{*}dw+ 2 a\sin^{2}\theta d\varphi^{*}dw.
\eea
\]
We have $\det\hat{g}= - (1+ a^{2}w^{2}\cos^{2}\theta)^{2}\sin^{2}\theta$, so $\hat{g}$ extends as a smooth, non-degenerate metric to $w\in \open{-\infty, r_{+}^{-1}}$. Clearly we can fix $\epsilon_{0}>0$ small enough so that $\hat{g}$ is Lorentzian for $w\in \open{-\epsilon_{0}, r_{+}^{-1}}$.
We define the conformal extension of ${\rm M}_{\rm I}$ as 
\[
\hat{{\rm M}}_{\rm I}\defeq\rr_{t\kst}\times \opencl{-\epsilon_{0},r_{+}^{-1}}_{w}\times \ss^2_{\theta,\varphi\kst}.
\]
\emph{Past null infinity} is then the null hypersurface
\beq\label{corr.e3}
\sI_-=\rr_{t\kst}\times \{ w=0 \} \times \ss^2_{\theta,\varphi\kst}.
\eeq
In an analogous way, extending $\hat{g}$ expressed in coordinates $(\stk t, w, \theta,\stk \varphi)$ to
\[
\rr_{\kst t}\times\open{-\epsilon_{0}, r_{+}^{-1}}_{w}\times \bS^{2}_{\theta \stk \varphi}
\]
allows to define \emph{future null infinity} $\sI_+=\rr_{\kst t}\times \{w=0\}\times \bS^{2}_{\theta\stk\varphi}$ , see Figure \ref{fig10} for the conformal diagram.
The time orientations of the conformal extensions of ${\rm M}_{\rm I}$ are  inherited from the time orientation of ${\rm M}_{\rm I}$.
\subsection{The Kerr--Kruskal extension}\label{subsec.kk}
Recall that  $\kappa_{\pm}$ was defined in \eqref{defodekappa}.
The {\em Kerr--Kruskal extension} ${\rm M}$ is the manifold $({\rm M}, g)$ with:
\[
{\rm M}= \rr_{U}\times \rr_{V}\times \bS^{2}_{\theta,\varphi^{\t}}.
\]
The global coordinates $(U, V, \theta, \varphi^{\t})$ on ${\rm M}$ are called {\em KBL coordinates}.
One  defines  $r$ as a smooth function on ${\rm M}$ by the equation:
 \[
\frac{ r- r_{+}}{UV}= G(r), \hbox{ for }G(r)= \e^{- 2\kappa_{+}r}(r-r_{-})^{r_{-}/r_{+}},
 \]
 see \cite[Lem.~3.4.9]{N}.
${\rm M}$ is then equipped with the metric \cite[Prop.~3.5.3]{N}:
 \beq\label{trif.5}
\bea
g&=\frac{G^{2}(r)a^{2}\sin^{2}\theta}{4\kappa_{+}^{2}\rho^{2}}\frac{(r-r_{-})(r+r_{+})}{(r^{2}+ a^{2})(r_{+}^{2}+a^{2})}\Big(\frac{\rho^{2}}{r^{2}+a^{2}}+\frac{\rho_{+}^{2}}{r_{+}^{2}+a^{2}}\Big)(U^{2}dV^{2}+ V^{2}dU^{2})\\
&\phantom{=}\,+ \frac{G(r)(r-r_{-})}{2\kappa_{+}^{2}\rho^{2}}\Big(\frac{\rho^{4}}{(r^{2}+a^{2})^{2}}+\frac{\rho_{+}^{4}}{(r_{+}^{2}+a^{2})^{2}}\Big)dUdV\\
&\phantom{=}\,+\frac{G(r)a\sin^{2}\theta}{\kappa_{+}^{2}\rho^{2}(r_{+}^{2}+a^{2})}\big(\rho_{+}^{2}(r-r_{-})+ (r^{2}+a^{2})(r+r_{+})\big)(UdV- VdU)d\varphi^{\sharp}\\
&\phantom{=}\,+ \rho^{2}d\theta^{2}+ \Big(r^{2}+ a^{2}+ \frac{2Mra^{2}\sin^{2}\theta}{\rho^{2}}\Big)\sin^{2}\theta d\varphi^{\sharp 2},
\eea
 \eeq
 where $\rho_{+}= \rho(r_{+}, \theta)$.
 Again, the following result is proved in Proposition \ref{propitoli1}.
 \begin{proposition}\label{debile.2}
 $({\rm M}, g)$ is globally hyperbolic. 
\end{proposition}
 \subsubsection{Null tetrad on ${\rm M}$}
 One sets:
 \beq\label{fincko.2}
\bea
&\mathfrak{l}= \frac{\e^{- \kappa_{+}r}(r-r_{-})^{M/r_{+}}}{(r-r_{-})\sqrt{2\rho^{2}}}\left(\frac{2\kappa_{+}(r^{2}+a^{2})}{G(r)}\p_{V}- \frac{a(r+r_{+})}{r_{+}^{2}+a^{2}}U\p_{\varphi^{\t}}\right),\\
&\mathfrak{n}= \frac{\e^{- \kappa_{+}r}(r-r_{-})^{M/r_{+}}}{(r-r_{-})\sqrt{2\rho^{2}}}\left(-\frac{2\kappa_{+}(r^{2}+a^{2})}{G(r)}\p_{U}- \frac{a(r+r_{+})}{r_{+}^{2}+a^{2}}V\p_{\varphi^{\t}}\right),\\
&m= \dfrac{1}{\sqrt{2}p}\left(\i a \sin \theta \p_{t}+ \p_{\theta}+ \dfrac{\i}{\sin \theta}\p_{\varphi}\right),
\eea
\eeq
for $p= r+ \i a \cos \theta$. Then $(\mathfrak{l}, \mathfrak{n}, m, \bar{m})$ is a global null tetrad on ${\rm M}$. We time orient $({\rm M}, g)$ by saying that $\mathfrak{l}+ \mathfrak{n}$ is future directed.

The  submanifolds $\sH_{R}\defeq\{U=0\}$, $\sH_{L}\defeq\{V=0\}$ are called the {\em long horizons}, and they intersect at the {\em crossing sphere} $S(r_{+})\defeq \{U= V= 0\}$.

 The following changes of coordinates allow to embed isometrically  ${\rm M}_{\rm I}, {\rm M}_{\rm II}, {\rm M}_{\rm I}', {\rm M}_{\rm II}'$ into ${\rm M}$, respecting the time orientations:
 \beq\label{e2}
 \begin{array}{l}
 U= \e^{-\kappa_{+}\stk t}, \ V= \e^{\kappa_{+}t\kst}, \ \hbox{ on }\MI, \\[2mm]
 U= -\e^{-\kappa_{+}\stk t}, \ V= \e^{\kappa_{+}t\kst}\hbox{ on }\MII,\\[2mm]
  U= -\e^{-\kappa_{+}\stk t}, \ V= -\e^{\kappa_{+}t\kst}\hbox{ on }\MI',\\[2mm]
 U= \e^{-\kappa_{+}\stk t}, \ V= -\e^{\kappa_{+}t\kst}\hbox{ on }\MII',\\[2mm]
 \varphi^{\sharp}= \12(\varphi\kst+ \stk\varphi- \frac{a}{r_{+}^{2}+a^{2}}(t\kst+ \stk t))= \varphi- \frac{a}{r_{+}^{2}+a^{2}}t.
\end{array}
  \eeq
We recall that  if $M$ is a spacetime, $M'$ stands for the Lorentzian manifold $M$ with the reversed time orientation. By slightly altering this rule we denote by
  \[
  \bea
 &{\rm M}_{\rm I}= \{U>0, \ V>0\}, \ \ {\rm M}_{\rm II}= \{U<0, \ V>0\}, \\ &{\rm M}_{\rm I'}=\{U<0, \ V<0\}, \ \ {\rm M}_{\rm II'}= \{U>0, \ V<0\},
 \eea
 \]
 the four quadrants of    ${\rm M}\setminus(\sH_{L}\cup \sH_{R})$. 
 
 \subsubsection{The wedge reflection}
 The map
 \beq\label{defdeR}
 R: \begin{array}{l}
 \MK\to \MK\\
 (U, V, \theta, \varphi^{\t})\mapsto (-U, -V, \theta, \varphi^{\t})
 \end{array}
  \eeq
 is called the {\em wedge reflection}. It preserves the orientation, it reverses the time orientation and it gives an identification of $\MI'$ and ${\rm M}_{\rm I'}$ (resp.~of $\MII'$ and ${\rm M}_{{\rm II}'}$)   as spacetimes.

 \begin{figure}[H]
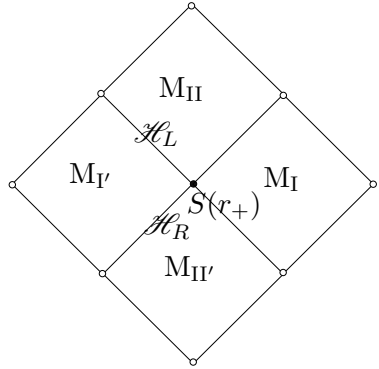\caption{The Kerr--Kruskal spacetime $\MK$.}\label{fig5}
\end{figure}

\subsection{The Weyl equation on  the Kerr--Kruskal spacetime}\label{sec9.1}
From the global null tetrad $(\mathfrak{l}, \mathfrak{n}, m, \bar{m})$ defined in \eqref{fincko.2}, we obtain a global orthonormal frame as in Subsect.~\ref{sec10.2b} and hence a spinor bundle as in \ref{globiglobo}.  We denote by $\DD$ the associated Weyl operator and by  
$\Sol_{{\rm L}^{2}}({\rm M})$  the space of $L^{2}$ solutions as in Subsect.~\ref{sec4.1b}. 

The restriction of $\DD$ on ${\rm M}_{\rm I}$, ${\rm M}_{\rm I'}$ etc will be denoted by the same letter, and the corresponding spaces of solutions by $\Sol_{{\rm L}^{2}}({\rm M}_{\rm I})$, $\Sol_{{\rm L}^{2}}({\rm M}_{\rm I'})$, etc. 

 \subsubsection{Killing vector fields}
An important role is played by the two Killing vector fields:
\[
\begin{array}{l}
v_{\sI}= \p_{t}= \kappa_{+}(- U \p_{U}+ V\p_{V})-\Omega_{\sH}\p_{\varphi^{\t}}, \\[2.5mm]
  v_{\sH}= \p_{t}+ \Omega_{\sH}\p_{\varphi}= \kappa_{+}(- U \p_{U}+ V\p_{V}),
\end{array}
\]
where $\Omega_{\sH}= \frac{a}{r_{+}^{2}+a^{2}}$  is the \emph{angular velocity} of the horizon. The vector field  $v_{\sH}$ is tangent to  $\sH_{-}$ (in  $\stk{\rm K}$), while $v_{\sI}$ is tangent to $\sI_{-}$ (in the conformal extension $\hat{{\rm M}}_{\rm I}$ of ${\rm M}_{\rm I}$).

The self-adjoint generators of their unitary actions on  $\Sol_{{\rm L}^{2}}(\MI)$ will be  denoted by $\i^{-1}\cL_{\sH}$ and $\i^{-1}\cL_{\sI}$  (so that $\i^{-1}\cL_{\sH}=\i^{-1}\cL_{v_{\sH}}$ and $\i^{-1}\cL_{\sI}=\i^{-1}\cL_{v_{\sI}}$ in the notation of Prop.~\ref{prop10.1}), and we also set $\i^{-1}\cL_{\varphi}\defeq\i^{-1}\cL_{\p_{\varphi}}$.

\section{Evolutionary form of the Weyl equation in $\MI$}\label{sec11}\init

\subsection{The HN reduction}\label{sec11.2} Following \cite{HN}, we now explain how  the Weyl equation $\DD\phi=0$ can be reduced to an equation of the form $\p_{t}\Psi- \i H \Psi=0$ with $H$ a $t$-independent differential operator. 

 In \cite{HN} two related null tetrads are used over $\MI$. The basic  null tetrad is given by:
  \begin{equation}
\label{e4.2b}
\bea
l&=\dfrac{1}{\sqrt{2\Delta\rho^{2}}}\left((r^{2}+a^{2})\p_{t}+ \Delta\p_{r}+ a\p_{\varphi}\right),\\[2mm]
n&=\dfrac{1}{\sqrt{2\Delta\rho^{2}}}\left((r^{2}+a^{2})\p_{t}- \Delta\p_{r}+ a\p_{\varphi}\right),\\[2mm]
m&= \dfrac{1}{\sqrt{2}p}\left(\i a \sin \theta \p_{t}+ \p_{\theta}+ \dfrac{\i}{\sin \theta}\p_{\varphi}\right),
\eea
\end{equation}
for $p= r+ \i a \cos \theta$. As in Subsect.~\ref{sec4.3} we set 
\[
\phi_{0}= \phi\dual \mo, \ \ \phi_{1}= \phi\dual \mi .
\]
 Another null 
tetrad $({\bf l}, {\bf n}, {\bf m}, \bar{\bf m})$ adapted to the foliation of ${\rm M}_{\rm I}$ by the hypersurfaces $\Sigma_{s}=\{t= s\}$  is used for the scattering theory arguments, with the property that  
\[
{\bf l}+ {\bf n}= 2 {\bf t},
\]
 where ${\bf t}$ is the future directed  unit normal vector field to this foliation. Concretely we have, see \cite[2.5.1]{HN}:
 \begin{equation}
\label{e4.2c}
\bea
{\bf l}&= \dfrac{\sigma}{\sqrt{2 \Delta \rho^{2}}}\left(\p_{t}+ \dfrac{ 2aMr}{\sigma^{2}}\p_{\varphi}\right)+ \sqrt{\dfrac{\Delta}{2\rho^{2}}}\p_{r},\\[2mm]
{\bf n}&= \dfrac{\sigma}{\sqrt{2 \Delta \rho^{2}}}\left(\p_{t}+ \dfrac{ 2aMr}{\sigma^{2}}\p_{\varphi}\right)- \sqrt{\dfrac{\Delta}{2\rho^{2}}}\p_{r},\\[2mm]
{\bf m}&= \dfrac{1}{\sqrt{2\rho^{2}}}\left(\p_{\theta}+ \dfrac{\i \rho^{2}}{ \sigma \sin \theta}\p_{\varphi}\right).
\eea
\end{equation} 

 If $({\bf o}, {\bf i })$ is the spinor basis associated to $({\bf l}, {\bf n}, {\bf m}, \bar{\bf m})$, one sets then:
 \begin{equation}
\label{e4.2d}
\Psi\defeq  \left(\dfrac{\Delta\rho^{2}\sigma^{2}}{(r^{2}+a^{2})^{2}}\right)^{\frac{1}{4}}\col{\phi\dual {\bf o}}{\phi\dual {\bf i}}= 
\col{\Psi_{0}}{\Psi_{1}}\eqdef \cV \phi,
\end{equation}
and uses the coordinates $(t, x, \theta, \varphi)$ where
$x(r)$ is defined in \eqref{eq:TLambda}.  Let us define   the matrix $\bf{U}\in M_{2}(\cc)$ corresponding to the above change of spinor basis by 
\begin{equation}
\label{corr.e6}
\col{\phi\dual {\bf o}}{\phi\dual {\bf i}}= {\bf U}\col{\phi\dual \mo}{\phi\dual \mi}.
\end{equation}
We refer the reader to  \cite[(2.50)]{HN} for the  concrete expression of ${\bf U}$. We have:
\beq\label{e4.2dd}
\col{\Psi_{0}}{\Psi_{1}}= \left(\dfrac{\Delta\rho^{2}\sigma^{2}}{(r^{2}+a^{2})^{2}}\right)^{\frac{1}{4}}{\bf U}\col{\phi_{0}}{\phi_{1}}.
\eeq
Finally, one sets 
\[
D= \cV\DD\cV^{-1}, 
\]
and
\beq\label{corr.e1}
(\Psi|\Psi)_{D}= \frac{1}{\sqrt{2}}\int_{\Sigma}\left(|\Psi_{0}|^{2}+ |\Psi_{1}|^{2}\right) dx d^{2}\omega
\eeq
 where $\Sigma=\Sigma_s$ for some $s\in \rr$.   From \eqref{e4.4} and the fact that the scalar product $(\cdot| \cdot)_{\DD}$ is independent on the choice of a Cauchy surface, we obtain that the r.h.s~in \eqref{corr.e1} is independent of $s$.

The equation $D\Psi=0$ can be rewritten as
\[
\p_{t}\Psi- \i H \Psi=0,
\]
which we will call the {\em reduced Weyl equation}.
A concrete expression for the $t$-independent differential operator $H$ can be found in \cite[(2.56)]{HN} (the operator $H$ is denoted by $\slashed{D}_{K}$ therein).

We  denote by  $\Solit_{L^{2}}(\MI)$ the closure of the space $\Solit_{\rm sc}(\MI)$ of space-compact solutions of $D\Psi=0$ for the scalar product $(\cdot| \cdot)_{D}$ defined in \eqref{corr.e1}. Using   \eqref{e4.4}  and the identity
\[
\dVol_{\Sigma}= \sqrt{\frac{\sigma^{2}\rho^{2}\Delta}{r^{2}+a^{2}}} \diff xd^{2}\omega,
\]
we obtain that:
\beq\label{e4.10}
\cV: \Sol_{{\rm L}^{2}}(\MI)\tosim \Solit_{L^{2}}(\MI)\hbox{ is unitary}.
\eeq
\subsubsection{Cauchy evolution}\label{sec11.2.1}  Let $\cH$ be the Hilbert space associated to $(\Psi|\Psi)_{D}$. We set \[
 \rho_{\Sigma}: \Solit_{L^{2}}(\MI)\ni \psi\mapsto \psi(0)\in \cH
 \]
  and denote by $\Psi= U_{\Sigma}f$ the  unique solution of the Cauchy problem
  \[
  \begin{cases}
  D\Psi=0, \\
  \rho_{\Sigma}\Psi= f\in \cH.
  \end{cases}
  \]
\subsubsection{Killing vector fields}\label{sec11.2.2}
The  Killing vector fields $v_{\sH_{-}/ \sI_{-}}$  preserve $({\bf l}, {\bf n}, {\bf m}, \bar{\bf m})$ and 
$v_{\sH/\sI}\left(\frac{\Delta\rho^{2}\sigma^{2}}{(r^{2}+a^{2})^{2}}\right)^{\frac{1}{4}}=0$, hence by \eqref{e4.9a} we have, in the sense of unitary equivalence of selfadjoint operators:
\begin{equation}
\label{e4.9b}
\cV\circ \i^{-1}\cL_{\sH_{-}/ \sI_{-}}\circ \cV^{-1}= \i^{-1}v_{\sH/\sI},
\end{equation}
where we also denote by $v_{\sH/\sI}$ the first order differential operators associated to $v_{\sH_{-}/ \sI_{-}}$, acting on $\Solit_{L^{2}}(\MI)$. 

Furthermore, we have:
\beq\label{e9.1}
\i^{-1} v_{\sH}\circ U_{\Sigma}=U_{\Sigma}\circ (H+ \Omega_{\sH}\i^{-1}\p_{\varphi}), \ \ \i^{-1}v_{\sI}\circ U_{\Sigma}= U_{\Sigma}\circ H.
\eeq

 \subsection{Decomposition of $\Solit_{L^{2}}(\MI)$}\label{sec11.3}
 \subsubsection{Asymptotic velocity}\label{sec1.3.1}
The first important result of \cite{HN} is the existence of the {\em asymptotic velocity} for solutions of the reduced Weyl equation. For the sake of brevity we will consider only the past asymptotics $t\to-\infty$. 
The following result is proved in \cite[Thm.~2.1]{HN} (see also \cite{HN2} for an introductory account).
\begin{theorem}\label{thm1.1} There exists  a selfadjoint operator $v\in B(\cH)$, called the {\em past asymptotic velocity} such that:
\[
\chi(v)=\slim_{t\to-\infty}\e^{-\i t H} \chi\left(\frac{x}{t}\right)\e^{\i t H}, \ \ \forall \chi\in \coinf(\rr).
\]
The spectrum of $v$ is ${\rm sp}(v)= \{-1, 1\}$, and
\begin{equation}
\label{corr.e4}
 [v, H]= [v, \p_{\varphi}]=0.
\end{equation}
\end{theorem}
\begin{definition}\label{def1.1}
 We set 
 \[
\pi_{\sH_{-}}\defeq \one_{\{1\}}(v), \ \ \pi_{\sI_{-}}\defeq \one_{\{-1\}}(v).
\]
\end{definition}
 The orthogonal decomposition $\one= \pi_{\sH_{-}}+ \pi_{\sI_{-}}$ allows to split an initial data $f\in \cH$  as $f_{\sH_{-}}+ f_{\sI_{-}}$ with $\e^{\i tH}f_{\sH_{-}}$ resp.~$\e^{\i tH}f_{\sI_{-}}$ moving towards $\sH_{-}$, resp.~$\sI_{-}$ when $t\to -\infty$.

 It follows that we can decompose $\Solit_{L^{2}}(\MI)$ as an orthogonal direct sum:
\[
\Solit_{L^{2}}(\MI)= \Solit_{L^{2}, \sH_{-}}(\MI)\oplus \Solit_{L^{2}, \sI_{-}}(\MI),
\]
where
\[
\Solit_{L^{2}, \sH_{-}/\sI_{-}}(\MI)\defeq U_{\Sigma}\circ \pi_{\sH_{-}/ \sI_{-}}\cH
\]
correspond to solutions going entirely through $\sH_{-}$, resp.~$\sI_{-}$ in the past. We will denote by
\beq\label{trouloulou}
\Pi_{\sH_{-}/ \sI_{-}}= U_{\Sigma}\circ \pi_{\sH_{-}/ \sI_{-}}\circ  \rho_{\Sigma}
\eeq
the orthogonal projections on $\Solit_{L^{2}, \sH_{-}/\sI_{-}}(\MI)$.
\subsubsection{Killing vector fields}\label{sec11.3.1}
By \eqref{corr.e4}, we know that $\pi_{\sH_{-}/\sI_{-}}$ commute with $H$ and $\p_{\varphi}$ hence by  \eqref{e9.1}  $\i^{-1}v_{\sH}$, resp.~$\i^{-1}v_{\sI}$ preserves $\Solit_{L^{2}, \sH_{-}}(\MI)$ resp.~$\Solit_{L^{2}, \sI_{-}}(\MI)$.
\subsection{Traces at infinities}\label{sec11.4}
Another important fact proved in \cite{HN} is the existence of traces of solutions on  $\sH_{-}$ and $\sI_{-}$ and their relationship with the   spaces $\Solit_{L^{2}, \sH_{-}/\sI_{-}}(\MI)$.
\subsubsection{Traces on $\sH_{-}$}\label{sec11.4.0}
For $\Psi\in \Solit_{\rm sc}(\MI)$, the trace
\[
T_{\sH_{-}}\Psi\defeq \Psi_{1|\sH_{-}}\in \cinf(\sH_{-}; \cc)
\]
is well defined, see \cite[Subsect.~8.2]{HN}. The following result is shown in \cite[Thm.~8.2]{HN}.
\begin{proposition}\label{prop9.0}
 $T_{\sH_{-}}$ uniquely extends as a bounded operator
\[
T_{\sH_{-}}: \Solit_{L^{2}}(\MI)\to L^{2}(\sH_{-}, \dVol_{\sH_{-}}),
\]
where   $\sH_{-}$ is identified with $\rr_{\stk t}\times \SS^{2}_{\theta, \stk \varphi}$, see \eqref{corr.e2} and  $\dVol_{\sH_{-}}=  \sin \theta d\stk t d\theta d\stk \varphi$.
One has:
\beq\label{e9.0}\begin{array}{l}
\Ker T_{\sH_{-}}= \Solit_{L^{2}, \sI_{-}}(\MI), \ \  \Ran T_{\sH_{-}}= L^{2}(\sH_{-}, \dVol_{\sH_{-}}),\\[2mm]
(\Psi| \Psi)_{D}= \frac{1}{\sqrt{2}}\int_{\sH_{-}}|T_{\sH_{-}}\Psi|^{2}\dVol_{\sH_{-}}, \ \ \Psi\in \Solit_{L^{2}, \sH_{-}}(\MI).
\end{array}
\eeq
\end{proposition}

Note that at $\sH_-$, only the $\Psi_1$ component is relevant for the trace, whereas at $\sI_-$ only the $\Psi_0$ component is relevant.

\subsubsection{Killing vector field on $\sH_{-}$}\label{sec11.4.1}
The Killing vector field $v_{\sH}$  is  null and tangent to $\sH$. On $\sH_{-}$  it equals $\p_{\stk t}+ \Omega_{\sH}\p_{\stk \varphi}$, resp.~$-\kappa_{+}U\p_{U}$ in star-Kerr, resp.~KBL coordinates. 

If we also denote by  $\i^{-1}v_{\sH}$ its  selfadjoint realization on  $L^{2}(\sH_{-},\dVol_{\sH_{-}})$ we have:
\[
T_{\sH_{-}} \circ (\i^{-1}v_{\sH})= (\i^{-1}v_{\sH})\circ T_{\sH_{-}}\hbox{ on }\Solit_{L^{2}, \sH_{-}}(\MI).
\]
\subsubsection{Traces on $\sI_{-}$}
For $\Psi\in \Solit_{\rm sc}(\MI)$, the trace
\[
T_{\sI_{-}}\Psi\defeq \Psi_{0|\sI_{-}}\in \cinf(\sI_{-}; \cc)
\]
is well defined, see \cite[Subsect.~8.3]{HN}. The following result is shown  in \cite[Thm.~8.3]{HN}. 
\begin{proposition}\label{prop9.0b}
 $T_{\sI_{-}}$ uniquely extends as a bounded operator
\[
T_{\sI_{-}}: \Solit_{L^{2}}(\MI)\to L^{2}(\sI_{-}, \dVol_{\sI_{-}}),
\]
where $\sI_{-}$ is identified with $\rr_{t\kst}\times \mathbb{S}^{2}_{\theta, \varphi\kst}$ and $\dVol_{\sI_{-}}= \sin \theta \diff t\kst d\theta d\varphi\kst$. One has
\begin{equation}
\label{e9.2}
\begin{array}{l}
\Ker T_{\sI_{-}}= \Solit_{L^{2}, \sH_{-}}(\MI), \ \ \Ran T_{\sI_{-}}= L^{2}(\sI_{-}, \dVol_{\sI_{-}})\\[2mm]
(\Psi| \Psi)_{D}= \frac{1}{\sqrt{2}}\int_{\sI_{-}} |T_{\sI_{-}}\Psi|^{2}\dVol_{\sI_{-}}, \ \ \Psi\in \Solit_{L^{2}, \sI_{-}}(\MI).
\end{array}
\end{equation}

\end{proposition}
  \subsubsection{Killing vector field on $\sI_{-}$}\label{sec11.5.1}
The Killing vector field $v_{\sI}$  is  null and tangent to $\sI$. On $\sI_{-}$  it equals $\p_{t\kst }$ in Kerr-star coordinates. 
If we also denote by  $\i^{-1}v_{\sI}$  its selfadjoint realization on  $L^{2}(\sI_{-}, \dVol_{\sI_{-}})$, we have:
\[
T_{\sI_{-}} \circ (\i^{-1}v_{\sI})= (\i^{-1}v_{\sI})\circ T_{\sI_{-}}\hbox{ on }\Solit_{L^{2}, \sI_{-}}(\MI).
\]
We summarize this subsection in the following theorem.
\begin{theorem}[\cite{HN}]\label{thm9.1}
  $T= T_{\sH_{-}}\oplus T_{\sI_{-}}$ from $\Solit_{L^{2}}(\MI)=  \Solit_{L^{2}, \sH_{-}}(\MI)\oplus  \Solit_{L^{2}, \sI_{-}}(\MI)$
 to $L^{2}(\sH_{-}, \dVol_{\sH_{-}})\oplus L^{2}(\sI_{-}, \dVol_{\sI_{-}})$ is unitary with
 \[
T \,\i^{-1}v_{\sH}\Pi_{\sH_{-}}= (\i^{-1}v_{\sH}\oplus 0)T, \ \  T\, \i^{-1}v_{\sI}\Pi_{\sI_{-}}= (0\oplus \i^{-1}v_{\sI})T.
\]
\end{theorem}

\section{Traces on horizons and at infinity}\init\label{sec12}
In Section \ref{sec11}, scattering theory was formulated for \emph{vectors} $\Psi\in\Solit_{L^{2}}(\MI)$. In this section we first re-express these results as the existence of traces on the horizon and infinity for   \emph{spinors}  $\phi\in\Sol_{{\rm L}^{2}}(\MI)$. We then extend the traces at the horizon and at infinity to $\2Sol(\MK)$.
\subsection{Decomposition of $\Sol_{{\rm L}^{2}}(\MI)$}\label{sec11.6}
The map $\cV:\phi\mapsto\Psi$ in \eqref{e4.2d} allows to construct the corresponding orthogonal decomposition of $\Sol_{{\rm L}^{2}}(\MI)$:
\beq\label{e4.10ccc}
\Sol_{{\rm L}^{2}}(\MI)= \Sol_{{\rm L}^{2}, \sH_{-}}(\MI)\oplus \Sol_{{\rm L}^{2}, \sI_{-}}(\MI),
\eeq
where 
\[ \Sol_{{\rm L}^{2}, \sH_{-}/\sI_{-}}(\MI)\defeq \cV \Solit_{L^{2}, \sH_{-}/\sI_{-}}(\MI).
\]
We denote  by 
\begin{equation}
\label{e4.10cc}
{\rm P}_{\sH_{-}}= \cV \Pi_{\sH_{-}}\cV^{-1},\ \ {\rm P}_{\sI_{-}}= \cV \Pi_{\sI_{-}}\cV^{-1},
\end{equation}
 the orthogonal projections on $\Sol_{{\rm L}^{2}, \sH_{-}/\sI_{-}}(\MI)$. 
By \ref{sec11.3.1}, the subspaces $\Sol_{{\rm L}^{2}, \sH_{-}/ \sI_{-}}(\MI)$ are  invariant under $\i^{-1}\cL_{\sH}$, $\i^{-1}\cL_{\sI}$. 

\subsection{Traces on $\sH_{-}$}\label{sec11.7}
For $\phi\in \Sol_{\rm sc}(\MI)$ we set
\beq\label{corr.e7}
{\rm T}_{\sH_{-}}\phi= \phi_{| \sH_{-}}\in \cinf(\sH_{-};\cc^2),
\eeq
which is clearly well defined. We denote by ${\rm L}^{2}(\sH_{-})$ the completion of $\coinf(\sH_{-}; \cc^{2})$ for the (degenerate) scalar product
\beq\label{e12.7}
(\phi|\phi)_{\sH_{-}}= -\i\int_{\sH_{-}}\bar{\phi}\dual\Gamma(\nabla V)\phi |g|^{\12} \diff U d\theta d\varphi^{\t},
\eeq
where we recall that $(U, V, \theta, \varphi^{\t})$ are the KBL coordinates,  see Subsect.~\ref{subsec.kk}, in terms of which  
$\sH_{-}= \{0\}_{V}\times\open{0, +\infty}_{U}\times \SS^{2}_{\theta, \varphi^{\t}}$.

We will see in the proof of Prop.~\ref{prop4.1} below that $(\cdot| \cdot)_{\sH_{-}}$ is positive semidefinite.

\begin{proposition}\label{prop4.1}
${\rm T}_{\sH_{-}}$ uniquely extends to a bounded operator ${\rm T}_{\sH_{-}}: \Sol_{{\rm L}^{2}}(\MI)\to {\rm L}^{2}(\sH_{-})$ with:
\begin{equation}
\label{fincko.1}
\Ker {\rm T}_{\sH_{-}}= \Sol_{{\rm L}^{2}, \sI_{-}}(\MI),\ \ \Ran {\rm T}_{\sH_{-}}= {\rm L}^{2}(\sH_{-}),
\end{equation}
\beq\label{fincko}
\begin{array}{rl}
(\phi| \phi)_{\DD}&= -\i \int_{\sH_{-}}{\rm T}_{\sH_{-}}\bar{\phi}\dual \Gamma(\nabla V){\rm T}_{\sH_{-}}\phi|g|^{\12} \diff Ud\theta d\varphi^{\t} \\[2mm]
 & =\i\int_{\sH_{-}}i^{*}(g^{-1}J(\phi, \phi)\lrcorner \Omega_{g}), \ \ \phi\in \Sol_{{\rm L}^{2}, \sH_{-}}(\MI).
 \end{array}
\eeq
\end{proposition}

  \proof
If $\Psi= \cV \phi\in \Solit_{L^{2},\sH_{-}}(\MI)$ for $\phi\in \Sol_{{\rm L}^{2}, \sH_{-}}(\MI)$, we have by \eqref{e4.10}, \eqref{e9.0} after passing to KBL coordinates:
\beq\label{e12.5}
\|\phi\|^{2}_{\DD}=\| \Psi\|^{2}_{D}= \kappa_{+}^{-1}\dfrac{1}{\sqrt{2}}\int_{\rr^{+}_{U}\times \SS^{2}_{\theta, \varphi^{\t}}}|T_{\sH_{-}}\Psi|^{2}U^{-1}\sin \theta \diff U d\theta d\varphi^{\t},
\eeq
and hence by \eqref{e4.10ccc} we have:
\beq\label{e12.6}
\kappa_{+}^{-1}\dfrac{1}{\sqrt{2}}\int_{\rr^{+}_{U}\times \SS^{2}_{\theta, \varphi^{\t}}}|T_{\sH_{-}}\Psi|^{2}U^{-1}\sin \theta \diff U d\theta d\varphi^{\t}\leq \| \phi\|^{2}_{\DD}, \ \ \phi\in \Sol_{{\rm L}^{2}}(\MI).
\eeq
We would like to reexpress the l.h.s.~above  in terms of $\phi$, for $\phi\in \Sol_{\rm sc}(\MI)$ i.e.~$\Psi\in \Solit_{\rm sc}(\MI)$.
One associates  to the KBL coordinates the new  null tetrad $(\mathfrak{l}, \mathfrak{n}, m, \bar{m})$
given by:
\beq\label{e4.10a}
\mathfrak{l}= \dfrac{U}{\sqrt{\Delta}}\e^{-\kappa_{+}r}(r-r_{-})^{M/r_{+}}l, \ \ \mathfrak{n}=\dfrac{V}{\sqrt{\Delta}}\e^{-\kappa_{+}r}(r-r_{-})^{M/r_{+}}n,
\eeq 
which has the advantage of extending smoothly to  $\sH_{+}\cup \sH_{-}$. The associated spinor basis $(\mathfrak{o}, \mathfrak {i})$  is:
\beq\label{e4.10b}
\mathfrak{o}= \left(\dfrac{U}{\sqrt{\Delta}}\e^{-\kappa_{+}r}(r-r_{-})^{M/r_{+}}\right)^{\12}\!\!\!\mo, \ \
\mathfrak{i}=\left(\dfrac{V}{\sqrt{\Delta}}\e^{-\kappa_{+}r}(r-r_{-})^{M/r_{+}}\right)^{\12}\!\!\!\mi.
\eeq
If we set 
\beq\label{e9.2b}
\mathfrak{f}_{0}= \phi\dual \mathfrak{o}, \ \ \mathfrak{f}_{1}= \phi\dual \mathfrak{i},
\eeq
it is shown in \cite[Corr. 8.1]{HN} that for $\Psi\in \Solit_{\rm sc}(\MI)$ one has:
\beq\label{e4.10c}(T_{\sH_{-}}\Psi)(U, \theta, \varphi^{\t})= \Psi_{1}(U, 0, \theta, \varphi^{\t})= C_{1}\sqrt{\bar{p}_{+}(\theta)U}\,\mathfrak{f}_{1}(U, 0, \theta, \varphi^{\t}),
\eeq
where:
\beq\label{e4.10d}
C_{1}= \e^{-\kappa_{+}r_{+}/2}(r_{+}- r_{-})^{M/2r_{+}}, \ \ p_{+}(\theta)= r_{+}+ \i a \cos \theta.
\eeq
Therefore we have for $\Psi\in \Solit_{\rm sc}(\MI)$:
\beq\label{e9.3}
\bea
&\kappa_{+}^{-1}\dfrac{1}{\sqrt{2}}\int_{\rr^{+}_{U}\times \SS^{2}_{\theta, \varphi^{\t}}}|T_{\sH_{-}}\Psi|^{2}U^{-1}\sin \theta \diff U \theta d\varphi^{\t} \\ &=\kappa_{+}^{-1}\dfrac{1}{\sqrt{2}}C_{1}^{2}\int_{\rr^{+}_{U}\times \SS^{2}_{\theta, \varphi^{\t}}}|p_{+}|(\theta)|\mathfrak{f}_{1}|^{2}\sin \theta \diff U d\theta d\varphi^{\t}
\eea
\eeq
On the other hand over $\sH_{-}$ we have $\mathfrak{n}= \frac{C}{|p_{+}(\theta)|}\p_{U}$ (see \cite[Subsect.~8.2.2]{HN}), for
\[
C= -\kappa_{+}(r_{+}^{2}+a^{2})\e^{\kappa_{+}r_{+}}(r_{+}- r_{-})^{-M/r_{+}}.
\]
Since the matrix of $\i \Gamma(\mathfrak{n})$ in the spinor basis $(\mathfrak{o}, \mathfrak{i})$ equals $\mat{0}{0}{0}{1}$, this implies that
\[
|\mathfrak{f}_{1}|^{2}= C|p_{+}(\theta)|^{-1}\bar{\phi}\dual \i \Gamma(\p_{U})\phi,
\]
and hence:
\beq\label{e4.9c}
\bea
&\kappa_{+}^{-1}\dfrac{1}{\sqrt{2}}C_{1}^{2}\int_{\rr^{+}_{U}\times \SS^{2}_{\theta, \varphi^{\t}}}|p_{+}|(\theta)|\mathfrak{f}_{1}|^{2}\sin \theta \diff U d\theta d\varphi^{\t} \\
&= - \i\sqrt{2}(r_{+}^{2}+a^{2})\int_{\rr^{+}_{U}\times \SS^{2}_{\theta, \varphi^{\t}}}\bar{\phi}\cdot \Gamma(\p_{U})\phi \sin \theta \diff U d\theta d\varphi^{\t}.
\eea
\eeq
Finally over $\sH_{-}$ we have, see \cite[(8.21)]{HN},
 \[
g= 2g_{UV}dUdV+ g_{VV}dV^{2}+ 2 g_{V\varphi^{\t}}dVd\varphi^{\t}+ g_{\theta\theta}d\theta^{2}+ g_{\varphi^{\t}\varphi^{\t}}(d\varphi^{\t})^{2},
\]
where $g_{\theta\theta}= \rho^{2}$, $g_{\varphi^{\t}\varphi^{\t}}= \rho^{-2}\sigma^{2}\sin^{2}\theta$.
It follows that:
\beq\label{e4.9d}
|g|^{\12}= |g_{UV}|(r_{+}^{2}+a^{2})\sin \theta, \ \ \nabla V= (g_{UV})^{-1}\p_{U}.
\eeq
Therefore we have:
\beq\label{e9.4b}
\bea
&\kappa_{+}^{-1}\dfrac{1}{\sqrt{2}}C_{1}^{2}\int_{\rr^{+}_{U}\times \SS^{2}_{\theta, \varphi^{\t}}}|p_{+}|(\theta)|\mathfrak{f}_{1}|^{2}\sin \theta \diff U d\theta d\varphi^{\t} \\
&=  -\i\int_{\rr^{+}_{U}\times \SS^{2}_{\theta, \varphi^{\t}}}\bar{\phi}\cdot\Gamma(\nabla V)\phi|g|^{\12}  \diff U d\theta d\varphi^{\t},
\eea
\eeq
for $\phi\in \Sol_{\rm sc}(\MI)$.   
From \eqref{e9.3} and  \eqref{e12.6}  we obtain:
\begin{equation}
\label{e12.7b}
 -\i \int_{\rr^{+}_{U}\times \SS^{2}_{\theta, \varphi^{\t}}}\bar{\phi}\cdot \Gamma(\nabla V)\phi|g|^{\12}  \diff U d\theta d\varphi^{\t}\leq \| \phi\|^{2}_{\DD}, \ \ \phi\in \Sol_{\rm sc}(\MI),
\end{equation}
Thus, ${\rm T}_{\sH_{-}}$ uniquely extends as a bounded operator on $\Sol_{{\rm L}^{2}}(\MI)$. From \eqref{e12.5} we obtain that $\Ker {\rm T}_{\sH_{-}}= \Sol_{{\rm L}^{2}, \sI_{-}}(\MI)$ and 
\[
(\phi| \phi)_{\DD}= -\i \int_{\sH_{-}}{\rm T}_{\sH_{-}}\bar{\phi}\dual \Gamma(\nabla V){\rm T}_{\sH_{-}}\phi|g|^{\12} \diff Ud\theta d\varphi^{\t}, \ \ \phi\in \Sol_{{\rm L}^{2}, \sH_{-}}(\MI).
\]
It remains to check  that $\Ran {\rm T}_{\sH_{-}}$ is  the closure of $\coinf(\sH_{-}; \cc^{2})$ for the scalar product $(\cdot| \cdot)_{\sH_{-}}$ defined in \eqref{e12.7}.    This follows from the fact  that  $\Ran T_{\sH_{-}}$, where $T_{\sH_{-}}$ was introduced in  \ref{sec11.4.0}, is equal to $L^{2}(\sH_{-}, \dVol_{\sH_{-}})$, which has $\coinf(\sH_{-})$ as a dense subspace. This completes the proof of \eqref{fincko.1} and of the first identity of \eqref{fincko}.

 On $\sH_{-}$ we have $\nabla V= g_{UV}^{-1}\p_{U}= - \lambda \mathfrak{n}$ for $\lambda>0$.
By the time orientation of ${\rm M}$, (see Subsect.~\ref{subsec.kk})  $\mathfrak{l}+\mathfrak{n}$ is future directed, hence  by \ref{fincko.3} $\mathfrak{n}$ is  also future directed. It follows that  $-\nabla V$ is future directed on $\sH_{-}$. We then apply  \eqref{e4.02} to obtain the second identity of \eqref{fincko}. \qed

\subsubsection{Killing vector field on $\sH_{-}$}\label{sec11.7.1}
 As in \ref{sec11.4.1}, we have since $v_{\sH}$ is tangent to $\sH_{-}$:
 \beq\label{e9.5}
{\rm T}_{\sH_{-}}\circ \i^{-1}\cL_{\sH}\phi= \i^{-1}\cL_{\sH}\circ {\rm T}_{\sH_{-}}\phi,\ \phi\in \Sol_{\rm sc}(\MI).
\eeq
We would like to express $\i^{-1}\cL_{\sH}\circ {\rm T}_{\sH_{-}}\phi$  using the decomposition \eqref{e9.2b}. 
Let us set:
\beq\label{e12.1}
\cS_{\sH_{-}}: \Sol_{\rm sc}(\MI)\ni \phi\mapsto \phi\dual \mathfrak{i}_{| \sH_{-}}\in \cinf(\sH_{-}; \cc),
\eeq
so that
\[
\cS_{\sH_{-}}\phi= {\rm T}_{\sH_{-}}\phi\dual \mathfrak{i}.
\]

This new trace operator is particularly useful because thanks to the choice of tetrad \eqref{e4.10a}, \eqref{e4.10b}, it leads to simpler expressions.
\begin{definition}\label{deffrip.1}
  We denote by $L^{2}(\sH_{-}; \cc)$ the closure of $\coinf(\sH_{-}; \cc)$ for the scalar product
  \[
  (\mathfrak{f}_{1}| \mathfrak{f}_{1})= \kappa_{+}^{-1}\dfrac{1}{\sqrt{2}}C_{1}^{2}\int_{\rr^{+}_{U}\times \SS^{2}_{\theta, \varphi^{\t}}}|p_{+}|(\theta)|\mathfrak{f}_{1}|^{2}\sin \theta \diff U d\theta d\varphi^{\t}.
  \]
  
\end{definition}
\begin{proposition} \label{prop12.1}
The map
 \[
 \cS_{\sH_{-}}: \Sol_{{\rm L}^{2}, \sH_{-}}(\MI)\to L^{2}(\sH_{-}; \cc )\hbox{ is unitary},
 \] 
 and one has:
 \beq\label{e9.6}
\cS_{\sH_{-}}\circ \i^{-1}\cL_{\sH}= -\i^{-1}\kappa_{+}(U\p_{U}+ \12)\circ \cS_{\sH_{-}},
\eeq
in the sense of unitary equivalence of selfadjoint operators. 
\end{proposition}
\proof  Since $ \phi\cdot \mathfrak{i}=\mathfrak{f}_{1}$, the identity \eqref{e9.4b} means that  $\cS_{\sH_{-}}$ is unitary from $\Sol_{{\rm L}^{2}, \sH_{-}}(\MI)$ to $L^{2}(\sH_{-}; \cc)$. Next  we use that  $\cV\cL_{\sH}\phi= v_{\sH}\cV \phi$ by \eqref{e4.9b}.  From \eqref{e4.10c} we obtain then that
\[
U^{\12}((\cL_{\sH}\phi)\dual \mathfrak{i})_{| \sH_{-}}= - \kappa_{+}U\p_{U}(U^{\12} \phi\dual \mathfrak{i}_{| \sH_{-}}),
\]
which implies \eqref{e9.6} since $U^{-\12}\circ (U \p_{U})\circ U^{\12}= U\p_{U}+ \12$. \qed

\subsection{Traces on $\sI_{-}$}\label{sec11.8}
\subsubsection{Conformal rescaling}\label{sec11.8.1}
Following \cite[Subsect.~8.3]{HN} we consider the conformally rescaled metric 
\[
\hat{g}= c^{2}g, \ \ c= r^{-1}\eqdef w,
\]
which can be smoothly extended to the domain $\rr_{t\kst}\times [0, r_{+}^{-1}]_{w}\times \SS^{2}_{\theta, \varphi\stk}$, with $\sI_{-}= \rr_{t\kst}\times\{0\}_{w}\times \SS^{2}_{\theta, \varphi\kst}$. Denoting with hats the objects canonically attached to $(\MI,\hat{g})$ we have by Subsect.~\ref{sec10.4}:
\begin{equation}
\label{e4.2e}
\hat{\Gamma}(v)= \Gamma(v), \ \ \hat{\tau}=\tau, \ \ \hat{\epsilon}= c\epsilon, \ \ \hat{\DD}= c^{-3}\DD c.
\end{equation}
In the coordinates $(t\kst, w, \theta, \varphi\kst)$ we have:
\[
\bea
l&= \dfrac{1}{\sqrt{2\Delta\rho^{2}}}\left(2 (r^{2}a^{2})\p_{t\kst}- \Delta w^{2}\p_{w}+ 2 a \p_{\varphi\kst}\right),\\[2mm]
n&=\dfrac{\Delta w^{2}}{\sqrt{2\Delta\rho^{2}}}\p_{w},\\[2mm]
m&= \dfrac{1}{\sqrt{2}p}\left(\i a \sin \theta \p_{t}+ \p_{\theta}+ \dfrac{\i}{\sin \theta}\p_{\varphi}\right).
\eea
\]
This tetrad extends smoothly up to $\sI_{-}=\{w=0\}$ with
\[
l_{|\sI_{-}}= \sqrt{2}\p_{t\kst}, \ \ \pm n_{|\sI_{-}}=m_{|\sI_{-}}=0,
\]
so the tetrad degenerates on $\sI_{-}$. A  normalized null tetrad for $(\MI, \hat{g})$  which is non degenerate near $\sI_{-}$ is given by (see \cite[8.3.1]{HN}):
\begin{equation}
\label{e9.9b}
(\hat{l}, \hat{n}, \hat{m}, \bar{\hat{m}})= (l, c^{-2}n, c^{-1}m, c^{-1}\bar{m}).
\end{equation}
 Using \eqref{e4.2e} we see that the associated spinor basis is 
 \beq\label{e9.9bb}
 (\hat{\mo}, \hat{\mi})= (\mo, c^{-1}\imath).
 \eeq

\subsubsection{Traces on $\sI_{-}$}
For $\phi\in \Sol_{\rm sc}(\MI)$ we  set:
\beq\label{corr.e8}
{\rm T}_{\sI_{-}}\phi\defeq \hat{\phi}_{| \sI_{-}}, \ \ \hat{\phi}= c^{-1}\phi\in \Sol_{\rm sc}(\hat{\DD}).
\eeq
We denote by ${\rm L}^{2}(\sI_{-})$ the completion of $\coinf(\sI_{-}; \cc^{2})$ for the scalar product:
\[
(\hat{\phi}| \hat{\phi})_{\sI_{-}}=  -\i\int_{\sI_{-}}\bar{\hat{\phi}}\dual\hat{\Gamma}(\hat{\nabla} w)\hat{\phi}|\hat{g}|^{\12} \diff t\stk d\theta d\varphi\stk.
\]
As before we will see in the proof of Prop.~\ref{prop4.2} that $(\cdot| \cdot)_{\sI_{-}}$ is positive semidefinite.
\begin{proposition}\label{prop4.2}
${\rm T}_{\sI_{-}}$ uniquely extends to a bounded operator ${\rm T}_{\sI_{-}}: \Sol_{{\rm L}^{2}}(\MI)\to {\rm L}^{2}(\sI_{-})$ with:
\begin{equation}
\label{fincki.1}
\Ker {\rm T}_{\sI_{-}}= \Sol_{{\rm L}^{2}, \sH_{-}}(\MI), \ \ \Ran {\rm T}_{\sI_{-}}= {\rm L}^{2}(\sI_{-}),\\
\end{equation}
\beq\label{fincki}
\begin{array}{rl}
(\phi| \phi)_{\DD}\!\!\!&= -\i \int_{\sI_{-}}{\rm T}_{\sI_{-}}\bar{\phi}\dual \hat{\Gamma}(\hat{\nabla} w){\rm T}_{\sI_{-}}\phi \sin \theta \diff t\kst  d \theta d\varphi\kst\\[2mm]
  &=\i\int_{\sI^{-}}i^{*}(\hat{g}^{-1}\hat{J}(\hat{\phi}, \hat{\phi})\lrcorner \Omega_{\hat{g}}), \ \phi\in \Sol_{{\rm L}^{2}, \sI_{-}}(\MI).
\end{array}
\eeq
\end{proposition}

\proof
Let $\Psi= \cV \phi\in \Solit_{L^{2}, \sI_{-}}(\MI)$ for $\phi\in \Sol_{{\rm L}^{2}, \sI_{-}}(\MI)$.   From \eqref{e4.10} and Prop.~\ref{prop9.0b} we have
\beq\label{e12.10}
(\phi| \phi)_{\DD}=(\Psi| \Psi)_{D}=\dfrac{1}{\sqrt{2}}\int_{\rr_{t\stk}\times \SS^{2}_{\theta, \varphi\stk}}|\Psi_{0}|^{2} \sin \theta \diff t\stk d \theta d \varphi \kst,
\eeq
hence by \eqref{e4.10ccc} we have:
\begin{equation}
\label{e12.8}
\dfrac{1}{\sqrt{2}}\int_{\rr_{t\stk}\times \SS^{2}_{\theta, \varphi\stk}}|\Psi_{0}|^{2} \sin \theta \diff t\stk d \theta d \varphi \kst\leq\| \phi\|^{2}_{\DD}, \ \ \phi\in \Sol_{{\rm L}^{2}}(\MI).
\end{equation}
As in the proof of Prop.~\ref{prop4.1} we re-express the l.h.s.~above in terms of $\phi$ for $\phi\in \Sol_{\rm sc}(\MI)$.
From  the definition \eqref{e4.2dd} of $\Psi$, we obtain:
\[
\Psi= {\bf U}c\left(\dfrac{\Delta\rho^{2}\sigma^{2}}{(r^{2}+a^{2})^{2}}\right)^{\frac{1}{4}}\col{\hat{\phi}\dual \mo}{\hat{\phi}\dual \mi},
\]
where  we recall that $\hat{\phi}= c^{-1}\phi$. On $\sI_{-}$ the matrix ${\bf U}$ introduced in \eqref{corr.e6} is equal to $\one_{2}$ and $c\left(\dfrac{\Delta\rho^{2}\sigma^{2}}{(r^{2}+a^{2})^{2}}\right)^{\frac{1}{4}}=1$, hence
\beq\label{e9.8}
\dfrac{1}{\sqrt{2}}\int_{\rr_{t\stk}\times \SS^{2}_{\theta, \varphi\stk}}|\Psi_{0}|^{2} \sin \theta \diff t\stk d \theta d \varphi \kst=\dfrac{1}{\sqrt{2}}\int_{\rr_{t\stk}\times \SS^{2}_{\theta, \varphi\stk}}|\hat{\phi}\dual \mo|^{2} \sin \theta \diff t\stk d \theta d \varphi \kst.
\eeq
Since $\mo= \hat{\mo}$   and the matrix of $\i \hat{\Gamma}(l)$ in the basis $(\hat{\mo}, \hat{\mi})$ equals $\mat{1}{0}{0}{0}$ we have  
\[
|\hat{\phi}\dual \mo|^{2}= |\hat{\phi}\dual \hat{\mo}|^{2}=\i \bar{\hat{\phi}}\dual \hat{\Gamma}(l)\hat{\phi}= \i\sqrt{2}\bar{\hat{\phi}}\dual \hat{\Gamma}(\p_{t^{\stk}})\hat{\phi},
\]
hence:
\[
\dfrac{1}{\sqrt{2}}\int_{\rr_{t\stk}\times \SS^{2}_{\theta, \varphi\stk}}|\Psi_{0}|^{2} \sin \theta \diff t\stk d \theta d \varphi \kst=\i\int_{\rr_{t\stk}\times \SS^{2}_{\theta, \varphi\stk}}\bar{\hat{\phi}}\dual \hat{\Gamma}(\p_{t^{\stk}})\hat{\phi}\sin \theta \diff t\stk d \theta d \varphi \kst.
\]
 Denoting by $\hat{\nabla}$ the gradient w.r.t.~$\hat{g}$, we obtain that  on $\sI_{-}$, $|\hat{g}|= \sin^{2}\theta$ and $\hat{\nabla}w= - \p_{t\stk}$. Therefore if we use the coordinates $(t\stk, w, \theta, \varphi\stk)$  near $\sI_{-}$ we obtain:
 \beq\label{e9.9}
\dfrac{1}{\sqrt{2}}\int_{\rr_{t\stk}\times \SS^{2}_{\theta, \varphi\stk}}|\Psi_{0}|^{2} \sin \theta \diff t\stk d \theta d \varphi \kst= -\i \int_{\sI_{-}}\bar{\hat{\phi}}\dual \hat{\Gamma}(\hat{\nabla}w)\hat{\phi}\sin \theta \diff t\stk d\theta d\varphi\stk.
\eeq
From \eqref{e12.8} and \eqref{e9.9} we obtain:
\begin{equation}
\label{e12.9}
 -\i \int_{\sI_{-}}\bar{\hat{\phi}}\dual \hat{\Gamma}(\hat{\nabla}w)\hat{\phi}\sin \theta \diff t\stk d\theta d\varphi\stk\leq \| \phi\|^{2}_{\DD}, \ \ \phi\in \Sol_{\rm sc}(\MI),
\end{equation}
so ${\rm T}_{\sI_{-}}$ uniquely extends as a bounded operator on $\Sol_{{\rm L}^{2}}(\MI)$. From \eqref{e12.10} we obtain that $\Ker {\rm T}_{\sI_{-}}= \Sol_{{\rm L}^{2}, \sH_{-}}(\MI)$ and
\[
(\phi| \phi)_{\DD}=  -\i \int_{\sI_{-}}\bar{{\rm T}_{\sI_{-}}\phi}\dual \hat{\Gamma}(\hat{\nabla}w){\rm T}_{\sI_{-}}\phi\sin \theta \diff t\stk d\theta d\varphi\stk, \ \ \phi\in \Sol_{{\rm L}^{2}, \sI_{-}}(\MI).
\]
As in  Prop.~\ref{prop4.1} we obtain \eqref{fincki.1} and the first identity in \eqref{fincki}. On $\sI_{-}$ we have $\hat{\nabla}w= - \p_{t\kst}$ and $\p_{t\kst}$ is future directed. Applying \eqref{e4.02} we  obtain the second identity of \eqref{fincki}. \qed

\subsubsection{Killing vector field on $\sI_{-}$}
As in \ref{sec11.5.1} we have 
\[
{\rm T}_{\sI_{-}}\circ\i^{-1}\cL_{\sI}\phi= \i^{-1}\cL_{\sI}\circ {\rm T}_{\sI_{-}}\phi, \ \ \phi\in \Sol_{\rm sc}(\MI),
\]
since $v_{\sI}= \p_{t\kst}$ is tangent to $\sI_{-}$ and the conformal factor $c= r^{-1}$ is invariant under $v_{\sI}$. 
Let us set
\beq\label{e12.2}
\cS_{\sI_{-}}: \Sol_{\sc}(\MI)\ni \phi\mapsto \hat{f}_{0}= \hat{\phi}\dual \hat{\mo}_{| \sI_{-}}\in \cinf(\sI_{-}; \cc),
\eeq
so that
\[
\cS_{\sI_{-}}\phi= {\rm T}_{\sI_{-}}\phi\dual \hat{\mo}.
\]
\begin{definition}\label{deffrip.2}
 We denote by $L^{2}(\sI_{-}; \cc)$ the closure of $\coinf(\sI_{-}; \cc)$ for the scalar product
 \[
 (\hat{f}_{0}| \hat{f}_{0})=\dfrac{1}{\sqrt{2}}\int_{\rr_{t\stk}\times \SS^{2}_{\theta, \varphi\stk}}|\hat{f}_{0}|^{2} \sin \theta \diff t\stk d \theta d \varphi \kst.
 \]
 \end{definition}
\begin{proposition} \label{prop12.2}
The map
 \[
 \cS_{\sI_{-}}: \Sol_{{\rm L}^{2}, \sI_{-}}(\MI)\to L^{2}(\sI_{-}, \sin \theta \diff t\kst d\theta d \varphi\kst)\hbox{ is unitary},
 \] 
 and one has:
 \beq\label{e9.6b}
\cS_{\sI_{-}}\circ (\i^{-1}\cL_{\sI})= \i^{-1}\p_{t\kst}\circ \cS_{\sI_{-}},
\eeq
in the sense of unitary equivalence of selfadjoint operators. 
\end{proposition}
\proof
The identities \eqref{e9.8}, \eqref{e9.9} mean that $\cS_{\sI_{-}}$ is unitary.
 \eqref{e9.6b}  follows from the fact that the null tetrad $(\hat{l}, \hat{n}, \hat{m}, \bar{\hat{m}})$ in \eqref{e9.9b} is invariant under $\cL_{\sI}$, and from the identities in  \ref{sec4.3.1}.\qed

We can summarize the results of Subsect.~\ref{sec11.7} and \ref{sec11.8} as follows.  
\begin{proposition}\label{thm12.1}
\ben
\item The map
  ${\rm T}_{\rm M_I}= {\rm T}_{\sH_{-}}\oplus {\rm T}_{\sI_{-}}$ from $\Sol_{{\rm L}^{2}}(\MI)$
 to ${\rm L}^{2}(\sH_{-})\oplus {\rm L}^{2}(\sI_{-})$ is unitary.

 \item The map $\cS_{\MI}= \cS_{\sH_{-}}\oplus \cS_{\sI_{-}}$ from $\2Sol(\MI)$
 to $L^{2}(\sH_{-}; \cc)\oplus L^{2}(\sI_{-}; \cc)$ is unitary. 
 \item One has
 \[
 \begin{array}{l}
\cS_{\sH_{-}}\circ \i^{-1}\cL_{\sH}= -\i^{-1}\kappa_{+}(U\p_{U}+ \12)\circ \cS_{\sH_{-}}, \\[2mm]
 \cS_{\sI_{-}}\circ \i^{-1}\cL_{\sI}= \i^{-1}\p_{t\kst}\circ \cS_{\sI_{-}},
\end{array}
\]
 in the sense of unitary equivalence of selfadjoint operators.
 \een
 \end{proposition}
 
 \subsection{Decomposition of $\2Sol(\MK)$}
By Proposition \ref{propitoli1}, $\Sigma_{\MK}= \{U= V\}$ is a space-like Cauchy surface in $\MK$.  Let us also set
\[
\Sigma_{\rm I}\defeq \Sigma_{\MK}\cap \MI= \{t=0\}\cap \MI, \ \ \Sigma_{\rm I'}\defeq \Sigma_{\MK}\cap \MIp= \{t=0\}\cap \MIp,
\]
so that by \eqref{etoo.1}
\[
\Sigma_{\MK}= \Sigma_{\rm I}\cup \Sigma_{\rm I'}\cup S(r_{+}).
\]
Since $S(r_{+})$ is of measure $0$ in $\Sigma_{\MK}$ for the induced Riemannian metric, this implies for the  corresponding $L^{2}$ spaces of Cauchy data that:
\[
L^{2}(\Sigma_{\MK})\ni f\mapsto f_{| \Sigma_{\rm I}}\oplus f_{| \Sigma_{\rm I'}} \in L^{2}(\Sigma_{\rm I})\oplus L^{2}(\Sigma_{\rm I'}) 
\]
is unitary, where we denote by $L^{2}(\Sigma)$ the completion of $\coinf(\Sigma; \SS^{*}_{\Sigma})$ for the scalar product $\nu_{\Sigma}$, see \eqref{e15c.6}. By Proposition \ref{prop4.1b} this yields a unitary map:
\beq\label{frip.7}
\2Sol(\MK)\ni \phi\mapsto \phi_{\rm I}\oplus \phi_{\rm I'}\in \2Sol(\MI)\oplus \2Sol(\MIp)
\eeq
where $\phi_{\mathrm{I}}$, resp.~$\phi_{\mathrm{I'}}$ are the restrictions of $\phi$ to $\MI$, resp.~$\MIp$. 

Next, we observe  that while $\MI'$ and $\MIp$ are strictly speaking different, they can be identified as spacetimes through the isometry
\[
R: \MIp\ni (U, V, \theta, \varphi^{\t})\mapsto (-U, -V, \theta, \varphi^{\t})\in \MI'
\]
which preserves the orientation and time-orientation.  We also denote  by $R: \2Sol(\MIp)\to \2Sol(\MI')$ the unitary map $$R\phi= \phi \circ R.$$ 

\subsection{Traces on the long horizon and at infinity}
\subsubsection{Traces on the long horizon}
Recall that $\sH_{L}=\{V=0\}$ is the left long horizon in $\MK$, which decomposes as
\[
\sH_{L}= \sH_{-}\cup \sH_{-}'\cup S(r_{+}), \  \ \sH_{-}= \{V=0, \ U>0\}, \ \ \sH_{-}'=\{V= 0, \ U<0\}= R(\sH_{-}).
\]
To ease notation $\sH_{L}$ will be simply denoted by $\sH$ in the sequel.

 Let us denote by ${\bf L}^{2}(\sH)$ the completion of $\coinf(\sH\setminus S(r_{+}); \cc^{2})$ for the scalar product
\[
(\phi| \phi)_{\sH}= - \i \int_{\sH}\bar{\phi}\dual \Gamma(\nabla V) \phi | g|^{\12}dU d\theta d\varphi^{\t}.
\]
Clearly we have
\[
{\bf L}^{2}(\sH)\sim {\bf L}^{2}(\sH_{-})\oplus {\bf L}^{2}(\sH_{-}')
\]
with obvious notations.   We denote
\[
{\rm T}_{\sH_{-}}\phi\defeq {\rm T}_{\sH_{-}}\phi_{\rm I}, \ \ {\rm T}_{\sH_{-}'}\phi\defeq R {\rm T}_{\sH_{-}}R \phi_{\rm I'}, 
\]
and
\[
{\rm T}_{\sH}\defeq  {\rm T}_{\sH_{-}}\oplus {\rm T}_{\sH_{-}'}: \2Sol(\MK)\to {\bf L}^{2}(\sH)
\]

Similarly we denote by $L^{2}(\sH; \cc)$ the closure of $\coinf(\sH; \cc)$ for the scalar product
 \[
  (\mathfrak{f}_{1}| \mathfrak{f}_{1})= \kappa_{+}^{-1}\dfrac{1}{\sqrt{2}}C_{1}^{2}\int_{\rr_{U}\times \SS^{2}_{\theta, \varphi^{\t}}}|p_{+}|(\theta)|\mathfrak{f}_{1}|^{2}\sin \theta \diff U d\theta d\varphi^{\t},
  \]
so that
\[
L^{2}(\sH; \cc)\sim L^{2}(\sH_{-}; \cc)\oplus L^{2}(\sH_{-}'; \cc).
\]
 We set
 \[
 \cS_{\sH_{-}}\phi\defeq \cS_{\sH_{-}}\phi_{\mathrm{I}}, \ \ \cS_{\sH_{-}'}\phi\defeq R \cS_{\sH_{-}} R\phi_{\rm I'}, 
 \]
 and
 \[
 \cS_{\sH}\defeq \cS_{\sH_{-}}\oplus \cS_{\sH_{-}'}: \2Sol(\MK)\to L^{2}(\sH; \cc).
 \]
 \subsubsection{Traces at infinity}
 We set
 \[
 {\rm T}_{\sI_{-}}\phi\defeq {\rm T}_{\sI_{-}}\phi_{\rm I}, \ \ {\rm T}_{\sI_{-}'}\phi\defeq {\rm T}_{\sI_{-}}R\phi_{\rm I'},
 \]
 \[
\cS_{\sI_{-}}\phi\defeq \cS_{\sI_{-}}\phi_{\rm I}, \ \ \cS_{\sI_{-}'}\phi\defeq \cS_{\sI_{-}}R\phi_{\rm I'},
 \]

 \subsubsection{Summary}  The results of this section are summarized in the following theorem.
\begin{theorem}
The maps
  \beq\label{frip.-3}
 \begin{array}{l}
 {\rm T}_{\MK}\defeq {\rm T}_{\sH}\oplus \mathrm{T}_{\sI_{-}}\oplus \mathrm{T}_{\sI_{-}'}: 
 \2Sol(\MK)\to {\rm L}^{2}(\sH)\oplus {\rm L}^{2}(\sI_{-})\oplus {\rm L}^{2}(\sI_{-}) \\[2mm]
 \cS_{\MK}\defeq \cS_{\sH}\oplus \cS_{\sI_{-}}\oplus\cS_{\sI_{-}'}: \2Sol(\MK)\to L^{2}(\sH; \cc)\oplus L^{2}(\sI_{-}; \cc)\oplus L^{2}(\sI_{-}; \cc)
 \end{array}
 \eeq
are unitary. 
\end{theorem}

\begin{figure}[H]
 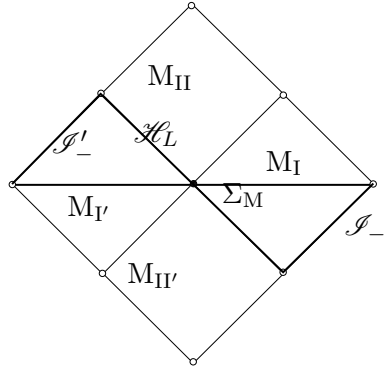\caption{The Kerr--Kruskal spacetime with  Cauchy surface $\Sigma_{\MK}$.}\label{fig9}
 \end{figure}

The geometric situation is illustrated in Figure \ref{fig9}.
 
\section{The Unruh state in  the Kerr--Kruskal spacetime}\label{sec13}\init
We recall that the solution space covariances of a quasi-free state were defined in  Subsect.~\ref{demian}. They will frequently be simply called covariances in the sequel. 
\subsection{The Unruh state in $\MK$}
\begin{definition}\label{def13.1}
 The {\em Unruh state} $\omega_{\MK}$ is the quasi-free state on $\CAR(\MK)$  with solution space covariances:
 \beq\label{sloubi2}
 C^{\pm}_{\MK}=  \cS_{\MK}^{-1}\left( \one_{\rr^{\pm}}(-\i^{-1}\p_{U})\oplus \one_{\rr^{\pm}}(\i^{-1}\p_{t\kst})\oplus \one_{\rr^{\mp}}(\i^{-1}\p_{t\kst})\right)\cS_{\MK}.
 \eeq
\end{definition}
\begin{theorem}\label{thm13.2}
\ben
\item   The Unruh state $\omega_{\MK}$ is a pure state.
\item The restriction $\omega_{\MIUII}$ of $\omega_{\MK}$ to $\MIUII$ is a pure state.
\item The restriction $\omega_{\rm M_{I}}$ of $\omega_{\MK}$  to $\MI$ has   covariances
 \[
 C^{\pm}_{\MI}= \cS_{\MI}^{-1}\left(\chi^{\pm}_{\sH_{-}}(-\i^{-1}\kappa_{+}(U\p_{U}+ \12))\oplus \chi^{\pm}_{\sI_{-}}(\i^{-1}\p_{t\kst})\right)\cS_{\MI}
 \]
 for
 \beq\label{ezlib.2}
 \chi^{\pm}_{\sI_{-}}(\lambda)= \one_{\rr^{\pm}}(\lambda), \ \ \chi^{\pm}_{\sH_{-}}(\lambda)= \big(1+ \e^{\mp T_{\rm H}^{-1}\lambda}\big)^{-1},
 \eeq
 where $T_{\rm H}= (2\pi)^{-1}\kappa_{+}$ is the  \emph{Hawking temperature}. 
 \item The restriction $\omega_{\rm M_{I'}}$ of $\omega_{\MK}$ to $\MIp$ is  the image of $\omega_{\rm M_{I}}$ under $R$, and has   covariances
 \[
C_{\MIp}^{\pm}=  R\circ C_{\MI}^{\mp}\circ R. 
 \]
  \een
\end{theorem}
\begin{theorem}\label{thm13.1}
There exists $0<a_{0}\leq 1$ such that  if $|a|M^{-1}<a_{0}$ then 
 the restriction $\omega_{\MIUII}$ of the Unruh state $\omega_{\MK}$ to $\MIUII$ is a Hadamard state.
\end{theorem}

\begin{remark} There are good reasons to expect that $\omega_{\MK}$ is not   a Hadamard state in the whole of $\MK$, and that the Hadamard condition fails precisely on the long horizon $\sH_{L}$, and in fact computations of the renormalized stress-energy tensor show that it diverges at $\sH_L$ \cite{kermions}. In our formalism, the problem comes from the wavefront set estimate in Proposition \ref{prop-key}, more precisely from \eqref{e11.2}. In general one has $N^{*}\supp u\subset \WF(\delta_{S}\otimes u)$, where $N^{*}S$ is the conormal bundle to $S$ and $N^{*}\supp u= N^{*}S\cap T^{*}_{\supp u}M$.
   Since $S$ is null, $N^{*}S$ is invariant under the bicharacteristic flow, so  $N^{*}S\subset \WF(G(\delta_{S}\otimes u))$, and therefore $\WF(G(\delta_{S})\otimes u)$ intersects both $\cN^{+}$ and $\cN^{-}$ over $S$. One applies this observation to the  long horizon $\sH_{L}$, which is included in $\MK$ (but not in $\MI$ or $\MIUII$), see for example the key identity \eqref{sloubi3} below, and one sees that the  form of the `boundary data' of $\omega_{\MK}$ on $\sH_{L}$ given in \eqref{sloubi2}  cannot  imply  the Hadamard condition over $\sH_{L}$.

  \end{remark}
  
  \begin{remark} The slow rotation assumption $|a|M^{-1}<a_{0}$ allows us to use  Lemma  \ref{lemalemo} which   asserts that null geodesics in $\MI$  that do not meet $\sH_-$ nor $\sI_-$ do necessarily cross  a region where $v_{\sH}$ and $v_{\sI}$ are  time-like. It is presently not known whether a generalisation of Theorem \ref{thm13.1} for larger $|a|$ is to be expected.
  \end{remark}

\subsection{Proof of Theorem \ref{thm13.2}}
 (1) follows from Lemma \ref{fermio-pure} (which says that the state is pure if the covariances are projections).  To prove (3), we  note that the image of $\2Sol(\MI)$ under $\cS_{\MK}$ equals $L^{2}(\sH_{-}; \cc)\oplus L^{2}(\sI_{-}; \cc)\oplus \{0\}$ and hence the restriction of $C_{\MK}^{\pm}$ to $\2Sol(\MI)$ (viewed as a sesquilinear form using the scalar product $(\cdot| \cdot)_{\DD}$) equals
 \[
 \cS_{\sH_{-}}^{-1}\left(\imath^{*}\one_{\rr^{\pm}}(-\i^{-1}\p_{U})\imath\right)\cS_{\sH_{-}}\oplus \cS_{\sI_{-}}^{-1}\one_{\rr^{\pm}}(\i^{-1}\p_{t\kst})\cS_{\sI_{-}},
 \]
 where $\imath: L^{2}(\sH_{-}; \cc)\to L^{2}(\sH; \cc)$ is the canonical injection. By 
 Lemma \ref{mirlemma} we have
 \[
 \imath^{*}\one_{\rr^{\pm}}(-\i^{-1}\p_{U})\imath= (1+ \e^{\pm 2\pi A}), 
 \]
 where $A= \i^{-1}(U\p_{U}+ \12)$ is the generator of dilations in $U$ acting on $L^{2}(\sH_{-};\cc)$, see Appendix \ref{appC}. This  completes the proof of (3).

 Let us now prove (2). We claim that the image of $\2Sol(\MIUII)$ under $\cS_{M}$ is included in  $L^{2}(\sH; \cc)\oplus L^{2}(\sI_{-}; \cc)\oplus \{0\}$, which will imply (2) by Lemma \ref{fermio-pure} and the fact that  $\one_{\rr^{\pm}}(-\i^{-1}\p_{U})\oplus \one_{\rr^{\pm}}(\i^{-1}\p_{t\kst})\oplus \one_{\rr^{\mp}}(\i^{-1}\p_{t\kst})$ restricted to $L^{2}(\sH; \cc)\oplus L^{2}(\sI_{-}; \cc)\oplus \{0\}$ is a projection.  To prove our claim we take $\phi\in \Sol_{\rm sc}(\MIUII)$, which has hence compactly supported Cauchy data on one of the space-like Cauchy surfaces $Z_{T}$ defined in \ref{slip}.  The conformal trace of $\phi$ on $\sI_{-}'$ vanishes, which by density implies our claim.

It remains to prove (4). We note that
 \[
 R C_{\MI}^{\pm}R= \cS_{\MI}^{-1}\left( \one_{\rr^{\mp}}(-\i^{-1}\p_{U})\oplus \one_{\rr^{\pm}}(\i^{-1}\p_{t\kst})\oplus \one_{\rr^{\mp}}(\i^{-1}\p_{t\kst})\right)\cS_{\MI}.
 \]
 Then (4) follows from (3) and the fact that the image of $\2Sol(\MIp)$ under $\cS_{\MK}$ equals $L^{2}(\sH_{-}'; \cc)\oplus \{0\}\oplus L^{2}(\sI_{-}; \cc)$. 
  \qed

\subsection{Preparations}
We now collect some preparations for the proof of Theorem \ref{thm13.1}.
\subsubsection{Cauchy surfaces in $\MI$}\label{pain-in-the-ass} In the proof of the Hadamard condition we will need two convenient families of Cauchy surfaces $\bar{\Sigma}_{T}$ and $\widetilde{\Sigma}_{T}$ in $\MI$. 

\begin{figure}[H]
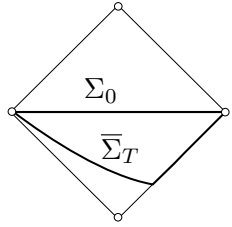
\caption{The Cauchy surfaces $\bar{\Sigma}_{T}$.}\label{fig3-bis}
\end{figure}

Since $\frac{r^{2}+a^{2}}{\Delta}= 1 - 2 r^{-1}+ O(r^{-2})$ when $r\to +\infty$  we can choose $x(r)$ in 
 \eqref{eq:TLambda} such that 
 \[
 x(r)= r- 2\ln r+ O(r^{-1}), \hbox{ when } r\to +\infty.
 \]
 It follows that the hypersurface $\{t+ r- 2 \ln r= T\}$ intersects $\sI_{-}$ (inside $\hat{{\rm M}}_{\rm I}$, the conformal extension of ${\rm M}_{\mathrm{I}}$, see \ref{confconf}),  transversally along the submanifold $\{w=0, t\kst= T\}$. 
 We  set then
\[
\bar{\Sigma}_{T}\defeq \{t+r- 2\ln r=T\}\cup \sI_{-}\cap \{t\kst\geq T\},
\]
which we smooth out near the intersection, to make it smooth and space-like inside ${\rm M}_{\rm I}$, null on $\sI_{-}$ (see Figure \ref{fig3-bis}).

To define the hypersurfaces $\tilde{\Sigma}_{T}$ we set 
 $f(r)= \dfrac{(r^{2}+a^{2})^{2}}{\Delta^{2}}- \dfrac{a^{2}+ M^2}{\Delta}$. We have
\beq\label{limi}
f>0\hbox{ in }]r_{+}, +\infty[ \hbox{ and } f^{\12}(r)- \dfrac{r^{2}+a^{2}}{\Delta}\in O(1)\hbox{ when }r\to r_{+}.
\eeq
We define the function $y(r)$ by
\beq\label{matznoni}
y'(r)= f^{\12}(r), \ \lim_{r\to r_{+}}y(r)- x(r)=0,
\eeq
noting that the second condition can be imposed due to \eqref{limi}.  Finally we define
\beq\label{matzno}
\tilde{x}(r)= \one_{]1, 3]}(r)y(r)+ \one_{]3, +\infty[}(r)y(3), 
\eeq
making $y(r)$ constant in $\{r\geq 3\}$. 

\begin{figure}[H]
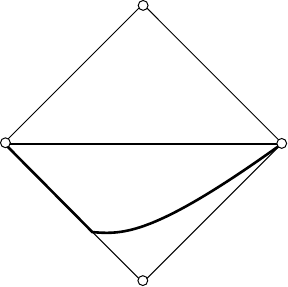\caption{The Cauchy surfaces $\widetilde{\Sigma}_{T}$.}\label{fig4}
\end{figure}

The hypersurface  $\{t-\tilde{x}(r)=T\}$ intersects $\sH_{-}$ (inside $\stk{\rm K}$, see \ref{turluto}) transversally along the submanifold $\{r=r_{+}, \stk t=T\}$. We set
\[
\widetilde{\Sigma}_{T}\defeq \{t-\tilde{x}(r)=T\}\cup \sH_{-}\cap\{\stk t\geq T\}
\]
again smoothed out near the intersection, to make it smooth and space-like inside ${\rm M}_{\rm I}$, null on $\sH_{-}$ (see Figure \ref{fig4}).

The fact that $\bar{\Sigma}_T$ and $\widetilde{\Sigma}_T$  are Cauchy surfaces follows from the constructions in Subsect.~\ref{ss:cauchy}, which provide a way of approximating $\bar{\Sigma}_T$ and $ \widetilde{\Sigma}_T$ by families of space-like Cauchy surface $\bar{\Sigma}_{T,n}$ and $ \widetilde{\Sigma}_{T,n}$. The space-like property of  $\bar{\Sigma}_T$ and $\widetilde{\Sigma}_T$  inside of $\MI$ can be also deduced from the computations in Subsect.~\ref{ss:cauchy}.

 \subsubsection{ Cauchy surfaces in $\MK$}
\begin{figure}[H]
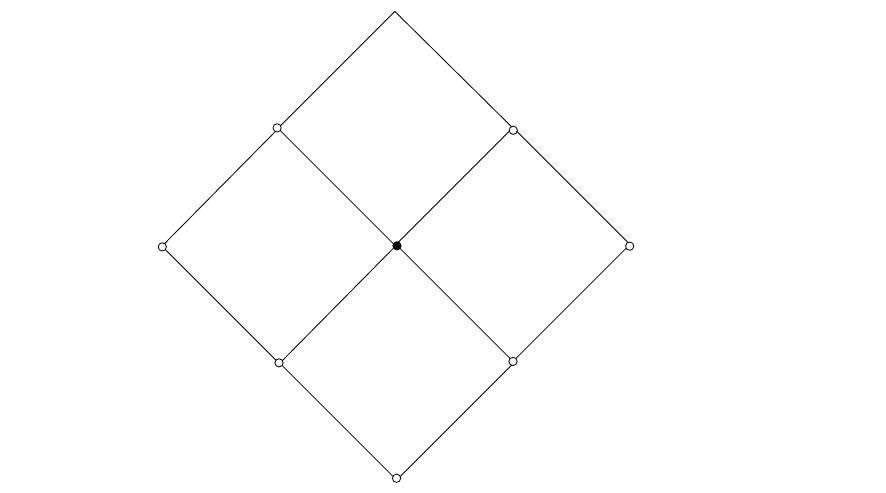\caption{The Cauchy surfaces $\widehat{\Sigma}_{T}$.}\label{fig11}
\end{figure}

 We will also need Cauchy surfaces $\widehat{\Sigma}_{T}$ in $\MK$, obtained by gluing together along $S(r_{+})$ the surface $\widetilde{\Sigma}_{T}$ with its image under the wedge reflection $R$, see Figure \ref{fig11}.
 
 \subsubsection{Wavefront set estimates}
The next lemma provides a way of producing solutions with wavefront set in $\cN^+$ or $\cN^-$ using a Killing vector field $X$, in a region where it is assumed to be time-like. For the sake of simplicity we formulate it here in the specific case of the Killing vector fields $v_{\sH}$ and $v_{\sI}$ on $\MI$ which will be needed later on. Below, $\bra \lambda\ket=(1+\lambda^2)^{\12}$.

\begin{lemma}\label{key-lemma2}
 Let  $X=v_{\sH}$ or $v_{\sI}$ and recall that $\i^{-1}\cL_{X}$ is the selfadjoint generator of a unitary group on $\Sol_{{\rm L}^{2}}(\MI)$. Let  $\chi^{\pm}\in L^\infty(\rr)$ be such that $\chi^\pm- \one_{\rr^{\pm}}\in O(\langle \lambda\rangle^{-\infty})$ and $\sing\supp \chi^\pm$ is compact. Then, if $X$ is future directed time-like near $x_{0}\in \MI$, one has:
\[
\WF(\chi^{\pm}(\i^{-1}\cL_{X})\phi)\subset \cN^{\pm}\hbox{ over a neighborhood of }x_{0},
\]
for all $\phi\in \Sol_{{\rm L}^{2}}(\MI)$.
  \end{lemma}	
\proof We consider the case of $v_{\sI}=\p_t$, the other one being analogous. Let us denote $\phi^+=\chi^+(\i^{-1}\cL_{X})\phi$. Let $q_0=(x_0,\xi_0)\in\cN^-$. 

Let $\vartheta\in \cinf(\rr)$ be such that $\vartheta(\lambda)=1$ for $\lambda \ll 0$ and $\vartheta(\lambda)=0$ for $\lambda> 0$ and  $\lambda \in \sing\supp\chi^+$. Setting  $\chi_\infty= \vartheta \chi^+ \in O(\langle \lambda\rangle^{-\infty})$, by functional calculus we have
\beq\label{eq:chizero}
\vartheta(\i^{-1}\cL_{X})\phi^+ = \chi_{\infty} (\i^{-1}\cL_{X})\phi.
\eeq
In coordinates, $\cL_{X} \phi= M(t)\p_t M(t)^{-1} \phi$ for some smooth family of invertible fiber endomorphisms $M(t)$. It follows that 
\[
\vartheta(\i^{-1}\cL_{X})\phi^+ =  M(t) \big(\vartheta(\i^{-1}\p_t)\otimes\one_y\big) M(t)^{-1}  \phi^+,
\]
where $\vartheta(\i^{-1}\p_t)\otimes\one_y$ acts as a product-type pseudo-differential operator. Let $A\in\Psi^0$ be a properly supported pseudo-differential operator (in the sense of the usual calculus on manifolds), elliptic at $q_0$ and micro\-supported in a small conic neighborhood $\Gamma_{0}$ of $q_0$. Since $q_0\in\cN^-$ and $\p_t$ is future directed time-like in the relevant region, the symbol of $\vartheta(\i^{-1}\p_t)\otimes\one_y$ equals $1$ on $\Gamma_{0}$ away from $\zero$. Therefore, the operator
\[
B\defeq A \circ M(t) \big(\vartheta(\i^{-1}\p_t)\otimes\one_y\big) M(t)^{-1}
\]
belongs to $\Psi^0$ and is elliptic at $q_0$, and similarly one can show that $A \circ \big(\chi_{\infty}(\i^{-1}\p_t)\otimes\one_y\big)$ belongs to $\Psi^{-\infty}$. Thus, by acting on both sides of \eqref{eq:chizero} with $A$ we find $B \phi^+ \in \cinf$. Consequently, $q_0\notin \wf(\phi^+)$. 

The proof of the minus sign version of the statement is analogous. \qeds

The other essential ingredient are the Green's formulas from Subsect.~\ref{ss:green} (which roughly speaking, allow to  represent a solutions locally as $\phi=-\GG (\delta_{S}\otimes v)$, with $S$ a Cauchy surface and $v$ a trace on $S$)  combined with Proposition \ref{prop-key} in the appendix (which allows to estimate $\wf\big(\GG (\delta_{S}\otimes v)\big)$ in terms of $\wf(v)$ if $v$ is compactly supported).

\subsection{Hadamard property in $\MI$} Our objective is to prove the Hadamard property of $\omega_{\MIUII}$.
We start by considering the restriction $\omega_{\MI}$  to $\MI$. From Proposition \ref{thm12.1} we obtain the key formula
\beq\label{ezlib.1}
C^{\pm}_{\MI}= {\rm P}_{\sH_{-}}\circ \chi^{\pm}_{\sH_{-}}(\i^{-1}\cL_{\sH})+ {\rm P}_{\sI_{-}} \circ \chi^{\pm}_{\sI_{-}}(\i^{-1}\cL_{\sI}),
\eeq
where the projections ${\rm P}_{\sH_{-}/\sI_{-}}$ are defined in \eqref{e4.10cc}. 
\begin{proposition}\label{propzlib.1}
There exists $0<a_{0}\leq 1$ such that  if $|a|M^{-1}<a_{0}$ then 
 $\omega_{\MI}$ is a Hadamard state.
\end{proposition}
\begin{remark}
 The Hadamard property extends to any state in $\MI$ of the form \eqref{ezlib.1} where $\chi^{\pm}_{\sH_{-}},\chi^{\pm}_{\sI_{-}}\in L^\infty(\rr)$ are such that $\sing\supp \chi^\pm$ is compact, and:
\beq\label{e5.0}
\begin{array}{l}
\chi^{+}_{\sH_{-}/ \sI_{-}}+ \chi^{-}_{\sH_{-}/ \sI_{-}}=1\\[2mm]
\chi^{\pm}_{\sH_{-}/ \sI_{-}}(\lambda)\in O(\langle \lambda\rangle^{-\infty})\hbox{ in }\rr^{\mp}.
\end{array}
\eeq
Such states can be called {\em asymptotically passive}. The class of \emph{passive states} was introduced by Pusz--Woronowicz \cite{PW}, and was investigated by Sahlmann--Verch  \cite{SV1} in settings with a global time-like Killing vector field $v$ (cf.~the work of Hack--Verch \cite{hackverch} for the  related notion of non-equilibrium steady states in a relativistic setting). Roughly speaking, a passive state is a mixture of ground and thermal states with respect to $v$. The name \emph{asymptotically passive}  indicates that at $\sH_-$, the state is passive with respect to $v_{\sH}$, and asymptotically at $\sI_-$ it is passive with respect to $v_{\sI}$. 
\end{remark}

\noindent {\bf Proof of Proposition \ref{propzlib.1}.} 
By Theorem \ref{propcoup.2} it suffices to prove that 
\begin{equation}
\label{ecoup.66}
\WF((C_{\MI}^{\pm})^{\12}\phi)\subset \cN^{\pm}, \ \forall \phi\in \Sol_{{\rm L}^{2}}(\MI).
\end{equation}
 We will prove only the $+$ case, the $-$ case being analogous. Let  $q_{0}=(x_{0}, \xi_{0})\in T^{*}\MI\cap \cN^{-}$ and let $\gamma$ be the null bicharacteristic in $T^{*}\MK$ from $q_{0}$. 
 
 \medskip
{\it Case 1:} suppose that $\pi(\gamma)$ (where $\pi:T^*\MK\to\MK$ is the base projection) intersects $\sI_{-}$  at some point $y_{0}$.  We will need  a family $\bar{\Sigma}_{T, n}$ of smooth, space-like Cauchy surfaces in ${\rm M}_{\mathrm{I}}$ that approximates $\bar{\Sigma}_{T}$. It is constructed in Proposition \ref{prop2} in the appendix. 

We will use  the conformal transformation $\hat{g}= c^{2}g$, $c= r^{-1}$, see Subsects. \ref{sec10.4} and \ref{confconf}.  We recall from Subsect.~\ref{sec10.4} that $\hat{\Gamma}(X)= \hat{\beta}\hat{\gamma}(X)= \Gamma(X)$, $\hat{\DD}= c^{-3}\DD c$ and $\hat{\mathbb{G}}$ is the Pauli-Jordan operator for $\hat{\DD}$, equal to $c^{-1}\mathbb{G}c^{3}$, see \cite[(17.51)]{G}. 

Applying \eqref{corr.e9} to   the conformally rescaled Weyl operator $\hat{\DD}$  and recalling that $\hat{\phi}= c^{-1}\phi$, this yields:
\beq\label{e.matz1}
\hat{\phi}(x)= - \int_{\bar{\Sigma}_{T, n}}\hat{\mathbb{G}}(x, y) \hat{\Gamma}(\hat{g}^{-1}\nu)(y)\hat{\phi}^{\pm}(y)i^{*}_{l}(\dVol_{\hat{g}})(y),\ \phi\in \Sol_{\rm sc}(\MI).
\eeq
  The solution $\hat{\phi}$ extends across $\sI_{-}$ to the conformal extension $\hat{\rm M}_{\rm I}$ of $\MI$  defined in Subsect.~\ref{confconf}, as a smooth solution of $\hat{\DD}\hat{\phi}=0$.  We can then take  the limit $n\to +\infty$ in \eqref{e.matz1} and obtain that:
 \beq\label{e.matz2}
\hat{\phi}(x)= - \int_{\bar{\Sigma}_{T}}\hat{\mathbb{G}}(x, y) \hat{\Gamma}(\hat{g}^{-1}\nu)(y)\hat{\phi}(y)i^{*}_{l}(\dVol_{\hat{g}})(y),\  \phi\in \Sol_{\rm sc}(\MI).
\eeq
From Subsect.~\ref{sec11.8} we recall that  the conformal trace on $\sI_{-}$, ${\rm T}_{\sI_{-}}\phi= \hat{\phi}_{|\sI_{-}}$ is well defined on $\Sol_{\rm sc}(\MI)$ and extends as a bounded operator 
${\rm T}_{\sI_{-}}:\Sol_{{\rm L}^{2}}(\MI)\to {\rm L}^{2}(\sI_{-})$.   On the other hand the trace of  $\hat{\phi}$ on $\bar{\Sigma}_{T}\setminus\sI_{-}$ is also well defined and obviously bounded from $\Sol_{{\rm L}^{2}}(\MI)$ to the appropriate $L^{2}$ space, because $\bar{\Sigma}_{T}\setminus\sI_{-}$ is a part of a smooth, space-like Cauchy surface and the conformal factor $c=r^{-1}$ is bounded with bounded inverse on $\bar{\Sigma}_{T}\setminus\sI_{-}$.

 Summarizing we obtain a bounded operator ${\rm T}_{\bar{\Sigma}_{T}}: \Sol_{{\rm L}^{2}}(\MI)\to {\rm L}^{2}(\bar{\Sigma}_{T})$, extending the map $\phi\mapsto \hat{\phi}_{| \bar{\Sigma}_{T}}$, where we denote by ${\rm L}^{2}(\bar{\Sigma}_{T})$ the appropriate $L^{2}$ space on $\bar{\Sigma}_{T}$.  Let us also recall from \eqref{trif.2} that $ \Sol_{{\rm L}^{2}}(\MI)\ni \phi\mapsto \hat{\phi}= c^{-1}\phi\in  \Sol_{{\rm L}^{2}}({\rm \hat M_I})$ is unitary. 

It follows then from Proposition \ref{propcoupure.1} that  \eqref{e.matz2} extends to $\phi\in \Sol_{{\rm L}^{2}}(\MI)$. More precisely if  $\chi\in \coinf(\MI)$ is a cutoff function supported near $x_{0}$, there exists $\chi_{1}\in\coinf(\bar{\Sigma}_{T})$ such that
\[
\chi \hat{\phi}(x)= -\int_{\bar{\Sigma}_{T}}\hat{\mathbb{G}}(x, y) \hat{\Gamma}(\hat{g}^{-1}\nu)(y)\chi_{1}(y){\rm T}_{\bar{\Sigma}_{T}}\phi(y )i^{*}_{l}(\dVol_{\hat{g}})(y),\ \phi\in \Sol_{{\rm L}^{2}}(\MI). 
\]
We then choose $T\ll -1$ such that $t^{*}(y_{0})>T$. Since one has
\[
\WF(\hat{\mathbb{G}})'\subset \{(q, q') \,:\, q\sim q'\},
\]
we can find   $\chi_{2}\in \coinf(\sI_{-})$,   equal to $1$ near $y_{0}$  such that:  
 \beq\label{etordu.1}
\chi\hat{\phi}(x)= - \int_{\bar{\Sigma}_{T}}\hat{\mathbb{G}}(x, y) \hat{\Gamma}(\hat{g}^{-1}\nu)(y)\chi_{2}(y){\rm T}_{\sI_{-}}\phi(y )i^{*}_{l}(\dVol_{\hat{g}})(y)+ u, \ \phi\in \Sol_{{\rm L}^{2}}(\MI), 
\eeq
with $q_{0}\not\in \WF(u)$.   We will drop the error term $u$ in the sequel.  Using  \eqref{corr.e11} we can rewrite 
this identity as:
\[
\chi\hat{\phi}(x)= -\int_{\sI_{-}}\hat{\mathbb{G}}(x, y)\hat{\Gamma}(\hat{\nabla}w)\chi_{2} {\rm T}_{\sI_{-}}\phi|\hat{g}|^{\12}dt\kst d \theta d\varphi\kst.
\]
We can assume that $\chi_{2}$ depends only on $t\kst$. 
Since $\hat{\nabla}w= - \p_{t\kst}$ and $|\hat{g}|^{\12}= \sin \theta $ on $\sI_{-}$, we get: 
\beq\label{corr.e12}
\chi\hat{\phi}(x)= \int_{\sI_{-}}\hat{\mathbb{G}}(x, t\kst, \theta, \varphi\kst)\hat{\Gamma}(\p_{t\kst})\chi_{2}(t\kst) {\rm T}_{\sI_{-}}\phi\sin \theta \diff t\kst d \theta d\varphi\kst.
\eeq
We use the spinor basis $(\hat{\mo}, \hat{\mi})$ associated to the null tetrad $(\hat{l}, \hat{n}, \hat{m}, \bar{\hat{m}})$ introduced in Subsect.~\ref{sec11.8}, see \eqref{e9.9b}, \eqref{e9.9bb} to  express the r.h.s.~of \eqref{corr.e12} in terms of the components
\[
\hat{f}_{0}={\rm T}_{\sI_{-}}\phi\cdot \hat{\mo}= \cS_{\sI_{-}}\phi, \ \ \hat{f}_{1}={\rm T}_{\sI_{-}}\phi\cdot \hat{\mi}.
\]
We have $\p_{t\kst}=\sqrt{2}^{-1}l$ hence the matrix of $\hat{\Gamma}(\p_{t\kst})$ equals $\i^{-1}\sqrt{2}^{-1}\mat{1}{0}{0}{0}$, see the proof of Prop.~\ref{prop4.2} . We set
\[
\hat{\mathbb{G}}_{0}(x, y)= \hat{\mathbb{G}}(x, y)\hat{\mo}^{*}(y), \ \ \hat{\mathbb{G}}_{1}(x, y)= \hat{\mathbb{G}}(x, y)\hat{\mi}^{*}(y),
\]
and obtain finally:

\beq\label{trif.e-10}
\chi\phi(x)=Cr^{-1}\int_{\sI_{-}}\hat{\mathbb{G}}_{0}(x, t\kst, \theta, \varphi\kst)\chi_{2}(t\kst) \cS_{\sI_{-}}\phi\sin \theta \diff t\kst d \theta d\varphi\kst.
\eeq
We apply \eqref{trif.e-10} to $ (C^{+}_{\MI})^{\12}\phi$, using that $\cS_{\sI_{-}}{\rm P}_{\sH_{-}}=0$ and Proposition \ref{prop12.2} to  obtain:
\[
\chi(C^{+}_{\MI})^{\12}\phi=Cr^{-1}\int_{\sI_{-}}\hat{\mathbb{G}}_{0}(x, t\kst, \theta, \varphi\kst)\chi_{2}(t\kst)(\chi^{+}_{\sI_{-}})^{\12}(\i^{-1}\p_{t\kst})\cS_{\sI_{-}}\phi\sin \theta \diff t\kst d \theta d\varphi\kst.
\]

We apply  Prop.~\ref{prop-key} with  $S=\sI_{-}$ and $X= v_{\sI}= \p_{t\kst}$. We claim that for
\[
g^{+}= \chi_{2}(t\kst)(\chi^{+}_{\sI_{-}})^{\12}(\i^{-1}\p_{t\kst})\cS_{\sI_{-}}\phi\in \cE'(\sI_{-}),
\]
we have
\[
\WF(g^{+})\subset \{(y, \eta)\in T^{*}\sI_{-}\setminus \zero \,:\, + \eta\dual v_{\sI}(y)\geq 0 \},
\]
which is immediate, since by \eqref{e5.0} $\chi^{+}_{\sI_{-}}$ decays rapidly in $\rr^{-}$. 

By Prop.~\ref{prop-key} this implies that  $q_{0}\not\in\WF(\phi^{+})$.

\medskip

{\it Case 2:} suppose that $\pi(\gamma)$ intersects $\sH_{-}$  at some $y_{0}$.  We use now the family $\widetilde{\Sigma}_{T, n}$ of Cauchy surfaces in $\MI$ and write
\beq\label{e.matz4}
\phi(x)= -\int_{\widetilde{\Sigma}_{T,n}}\mathbb{G}(x, y)\Gamma(g^{-1}\nu)(y) \phi(y)i^{*}_{l}(\dVol_{g})(y), \ \phi\in \Sol_{\rm sc}(\MI). 
\eeq
The smooth solutions $\phi$ extends across $\sH_{-}$ to $\kst {\rm K}$ as smooth solutions of $\DD \phi=0$. We can again pass to the limit $n\to +\infty$ in \eqref{e.matz4} and obtain:
\beq\label{e.matz5}
\phi(x)= -\int_{\widetilde{\Sigma}_{T}}\mathbb{G}(x, y)\Gamma(g^{-1}\nu)(y) \phi(y)i^{*}_{l}(\dVol_{g})(y),  \ \phi\in \Sol_{\rm sc}(\MI). 
\eeq
As before the traces on $\sH_{-}$ and  on $\widetilde{\Sigma}_{T}\setminus \sH_{-}$ extend as a bounded operator ${\rm T}_{\widetilde{\Sigma}_{T}}: \Sol_{{\rm L}^{2}}(\MI)\to {\rm L}^{2}(\widetilde{\Sigma}_{T})$, by  Proposition \ref{prop4.1}  for  the trace on $\sH_{-}$ and using that $\widetilde{\Sigma}_{T}\setminus \sH_{-}$ is a part of a smooth, space-like Cauchy surface for the trace on $\widetilde{\Sigma}_{T}\setminus \sH_{-}$.  

Again  we denote by ${\rm L}^{2}(\widetilde{\Sigma}_{T})$ the appropriate $L^{2}$ space on $\widetilde{\Sigma}_{T}$. 

By Proposition \ref{propcoupure.1} we can extend \eqref{e.matz5} to $\phi\in \Sol_{{\rm L}^{2}}(\MI)$ and obtain:
\[
\chi\phi(x)= -\int_{\widetilde{\Sigma}_{T}}\mathbb{G}(x, y)\Gamma(g^{-1}\nu)(y)\chi_{1}{\rm T}_{\widetilde{\Sigma}_{T}} \phi(y)i^{*}_{l}(\dVol_{g})(y),  \ \phi\in \Sol_{{\rm L}^{2}}(\MI). 
\]

We choose $T\ll -1$ such that $\stk t(y_{0})>T$.
Again we can find $\chi_{2}\in \coinf(\sH_{-})$ equal to $1$ near $y_{0}$ such that
\beq\label{trif.4}
\chi\phi(x)= -\int_{\sH_{-}}\mathbb{G}(x, y)\Gamma(g^{-1}\nu)(y) \chi_{2}(y){\rm T}_{\sH_{-}}\phi(y)i^{*}_{l}(\dVol_{g})(y)+ u, \ \phi\in \Sol_{{\rm L}^{2}}(\MI), 
\eeq
with $q_{0}\not\in \WF(u)$, and we  will forget the error term $u$ in the sequel. Using \eqref{corr.e11} we can rewrite  this identity as
\beq\label{trif.3}
\chi\phi(x)= - \int_{\sH_{-}}\mathbb{G}(x, y)\Gamma(\nabla V)\chi _{2}{\rm T}_{\sH_{-}}\phi\,|g|^{\12} \diff U d\theta d\varphi^{\t},
\eeq
where $\sH_{-}\sim \rr^{+}_{U}\times \SS^{2}_{\theta, \varphi^{\t}}$.

 We can assume that $\chi_{2}$ depends only on  the variable $U$. Since $\nabla V= (g_{UV})^{-1}\p_{U}$ and $|g|^{\12}= g_{UV}(r_{+}^{2}+a^{2})\sin \theta$, see \eqref{e4.9d}, we obtain:
\beq\label{e5.2}
\chi \phi(x)= - (r_{+}^{2}+a^{2})\int_{\sH_{-}} \GG(x, U, \theta, \varphi^{\t})\Gamma(\p_{U}) \chi_{2}(U){\rm T}_{\sH_{-}}\phi \,\sin \theta \diff U d\theta d \varphi^{\t}.
\eeq
We now use the spinor basis $(\mathfrak{o}, \mathfrak{i})$ associated to the null tetrad   $(\mathfrak{l}, \mathfrak{n}, m, \bar{m})$ introduced in  Subsect.~\ref{sec11.7}, see \eqref{e4.10a}, \eqref{e4.10b}, and 
 express \eqref{e5.2} in terms of the components
\[
\mathfrak{f}_{0}= {\rm T}_{\sH_{-}}\phi\dual \mathfrak{o}, \ \ \mathfrak{f}_{1}= {\rm T}_{\sH_{-}}\phi\dual \mathfrak{i}= \cS_{\sH_{-}}\phi.
\]
We have $\p_{U}= C| p_{+}(\theta)|\mathfrak{n}$, hence the matrix of $\Gamma(\p_{U})$ in the basis $(\mathfrak{o}, \mathfrak{i})$ is equal to $\i^{-1}C|p_{+}(\theta)|\mat{0}{0}{0}{1}$, see the proof of Prop.~\ref{prop4.1}.  We also set
\[
\GG_{0}(x, y)= \GG(x, y)\mathfrak{o}^{*}(y), \ \ \GG_{1}(x, y)= \GG(x, y)\mathfrak{i}^{*}(y), 
\]
where we recall that $(\mathfrak{o}^{*}, \mathfrak{i}^{*})$ is the dual basis of $(\mathfrak{o}, \mathfrak{i})$. This yields
\beq\label{etrif.-4}
\chi\phi(x)= C\int_{\sH_{-}}  \GG_{1}(x, U, \theta, \varphi^{\t})\chi_{2}(U) \cS_{\sH_{-}}\phi\sin \theta \diff U d\theta d\varphi^{\t}.
\eeq
Finally we apply \eqref{etrif.-4} to $(C_{\MI}^{+})^{\12}\phi$, using that $\cS_{\sH_{-}}{\rm P}_{\sI_{-}}=0$ and Proposition \ref{prop12.1}  to obtain:
\[
\chi(C_{\MI}^{+})^{\12}\phi(x)= C\int_{\sH_{-}}  \GG_{1}(x, U, \theta, \varphi^{\t})g^{+}(U, \theta, \varphi^{\t})  \sin \theta \diff U d\theta d\varphi^{\t},
\]
where 
\[
g^{+}= \chi_{2}(U)  (\chi^{+}_{\sH_{-}})^{\12}\big(-\i^{-1}\kappa_{+}(U\p_{U}+ \textstyle\12)\big)\cS_{\sH_{-}}\phi\in \cE'(\sH_{-}).
\] 
We now apply Proposition \ref{prop-key}, with $S= \sH_{-}$ and $X=v_{\sH}=  - \kappa_{+}U\p_{U}$. This yields
$ \WF(\chi(C_{\MI}^{+})^{\12}\phi)\subset \cN^{+}$ and hence $q_{0}\not\in\WF((C_{\MI}^{+})^{\12}\phi)$ provided that 
\beq\label{e5.3}
\WF(g^{+})\subset \{(y, \eta)\in T^{*}\sH_{-}\setminus \zero \,:\, +\eta\dual v_{\sH}(y)\geq 0\}.
\eeq
To prove \eqref{e5.3} we use the unitary map:
  \[
B: L^{2}(]0, +\infty[_{U}\times \SS^{2})\ni f\mapsto Bf(s, \omega)=\kappa_{+}^{\12}\e^{-\kappa_{+}s/2}f(\e^{-\kappa_{+}s}, \omega),
\]
implementing the change of coordinates $U= \e^{-\kappa_{+}s}$, which satisfies:
\[
B\circ(-\i^{-1}\kappa_{+}(U\p_{U}+ \textstyle\12))= (\i^{-1}\p_{s})\circ B.
\]
Since by \eqref{e5.0} $\chi^{+}_{\sH_{-}}(\lambda)$ decays rapidly in $\rr^{-}$, we have 
\[
\WF\big(B (\chi^{+}_{\sH_{-}})^{\12}(-\i^{-1}\kappa_{+}(U\p_{U}+ \textstyle\12))\cS_{\sH_{-}}\phi\big)\subset \{(y, \eta)\in T^{*}\sH_{-}\setminus\zero \,:\,  + \eta\dual \p_{s}\geq 0\},
\] 
which implies \eqref{e5.3} by the covariance of the wavefront set under change of coordinates.  \medskip

{\it Case 3:} suppose that $\pi(\gamma)$ does not end up at $\sH_{-}$ nor at $\sI_{-}$. 

By Lemma  \ref{lemalemo}, there exists $x_{1}\in \pi(\gamma)$ such that $v_{\sH}$ and $v_{\sI}$ are  future directed time-like near $x_{1}$.  We have
\[
C^{+}_{\MI}\phi= (\chi_{\sH_{-}}^{+})^{\12}(\i^{-1}\cL_{\sH}){\rm P}_{\sH_{-}}\phi+ (\chi_{\sI_{-}}^{+})^{\12}(\i^{-1}\cL_{\sI}){\rm P}_{\sI_{-}}\phi\eqdef \phi_{\sH}+ \phi_{\sI}
\]
By Lemma \ref{key-lemma2}  applied first to $X=v_\sH$, $\chi^\pm=(\chi_{\sH_{-}}^{+})^{\12}$ (with  ${\rm P}_{\sH_{-}}\phi$ in place of $\phi$), and then   to $X=v_\sI$, $\chi^\pm=(\chi_{\sI_{-}}^{+})^{\12}$    (with  ${\rm P}_{\sI_{-}}\phi$ in place of $\phi$), we get that 
\[
\WF(\phi_{\sH/ \sI})\subset \cN^{+} \hbox{ above a neighborhood of }x_{1}.
\]
By propagation of singularities this implies that $q_{0}\not\in \WF(C^{+}_{\MI}\phi)$. \qeds

\subsection{Hadamard property in $\MIUII$} Finally, we prove the Hadamard property in $\MIUII$. \medskip

\noindent {\bf Proof of Theorem \ref{thm13.1}.}   We only consider  the $+$ case. By Theorem \ref{propcoup.2} it suffices to prove that 
  \begin{equation}
  \label{e13.-1}
  \WF(C^{+}_{\MK}\phi)\cap T^{*}\MIUII\subset \cN^{+}, \ \phi\in \2Sol(\MK),
  \end{equation}
  where we recall that
  \[
  C^{+}_{\MK}\phi= C_{\MI}^{+}\phi_{\rm I}\oplus RC^{-}_{\MI}R\phi_{\rm I'}.
  \]
  
  Let  $q_{0}=(x_{0}, \xi_{0})\in T^{*}\MIUII\cap \cN^{-}$ and let $\gamma$ the null bicharacteristic through $q_{0}$ {\em in} $T^{*}\MK$, and let $\pi$ be the base projection.
  
  \medskip
  
{\it Case 1:}  suppose that $\pi(\gamma)$ intersects $\MI$. We have   $C^{+}_{\MK}\phi= C_{\MI}^{+}\phi_{\rm I}$ over $\MI$,  with $\phi_{\rm I}\in \2Sol(\MI)$ and $C_{\MI}^{\pm}$ the covariances of  $\omega_{\MI}$ in Theorem \ref{thm13.2}.  We already know that $\omega_{\MI}$ is  Hadamard in $\MI$ by Proposition \ref{propzlib.1}, which implies that $\WF(C^{+}_{\MK}\phi)\subset \cN^{+}$ over $\MI$, hence 
  $q_{0}\not\in \WF(C^{+}_{\MK}\phi)$ by propagation of singularities.
  
    \medskip
    
{\it Case 2:} suppose that $\pi(\gamma)$ intersects $\MIp$.  We have   $C^{+}_{\MK}\phi= R C^{-}_{\MI}R\phi_{\rm I'}$ over $\MIp$, with $R\phi_{\rm I'}\in \2Sol(\MI)$. Since $\omega_{\MI}$ is Hadamard,  we have $\WF(C^{-}_{\MI}R\phi_{\rm I'})\subset \cN^{-}$, hence $\WF(C^{+}_{\MK}\phi)\subset \cN^{+}$ over $\MIp$, and $q_{0}\not\in \WF(C^{+}_{\MK}\phi)$ by propagation of singularities.
 
 \medskip
 
{\it Case 3:}  suppose that  $\pi(\gamma)$ does not intersect $\MI$ nor $\MIp$, which means that $\pi(\gamma)$ leaves $\MIUII$ through $S(r_{+})$, see the proof of Proposition \ref{propitoli1}. By propagation of singularities, we can assume that $x_{0}$ belongs to an arbitrarily small neighborhood of $S(r_{+})$ in $\MK$.  If $\chi\in \coinf(\MK)$ then as in the proof of Proposition \ref{propzlib.1}, we can find $\chi_{1}\in \coinf(\widehat{\Sigma}_{T})$ such that
\[
\chi\phi(x)= -\int_{\widehat{\Sigma}_{T}}\mathbb{G}(x, y)\Gamma(g^{-1}\nu)(y)\chi_{1}r_{\widetilde{\Sigma}_{T}} \phi(y)i^{*}_{l}(\dVol_{g})(y),  \ \phi\in \Sol_{{\rm L}^{2}}(\MK),
\]
where $r_{\widehat{\Sigma}_{T}}: \2Sol(\MK)\to L^{2}(\widehat{\Sigma}_{T})$ is the bounded extension of the trace on $\widehat{\Sigma}_{T}$.

We can then find $\chi_{2}\in \coinf(\sH_{ L})$, equal to $1$ near $S(r_{+})$ such that
\[
\chi\phi(x)= -\int_{\widehat{\Sigma}_{T}}\mathbb{G}(x, y)\Gamma(g^{-1}\nu)(y)\chi_{2}{\rm T}_{\sH} \phi(y)i^{*}_{l}(\dVol_{g})(y)+ u,  \ \phi\in \Sol_{{\rm L}^{2}}(\MK),
\]
where $q_{0}\not\in \WF (u)$ and we will drop the error term $u$. We rewrite this identity as
\beq\label{trif.3b}
\chi\phi(x)= - \int_{\sH}\mathbb{G}(x, y)\Gamma(\nabla V)\chi _{2}{\rm T}_{\sH}\phi\,|g|^{\12} \diff U d\theta d\varphi^{\t},
\eeq
where $\sH\sim\rr_{U}\times \SS^{2}_{\theta, \varphi^{\t}}$, 
and  assume that $\chi_{2}$ depends only on  the variable $U$.   The same arguments as in the proof of Proposition \ref{propzlib.1}  yield:
\beq\label{etrif.-8}
\chi\phi(x)= C\int_{\sH}\GG_{1}(x, U, \theta, \varphi^{\t}) \chi_{2}(U)\cS_{\sH}\phi \, \sin \theta \, dU d \theta d\varphi^{\t}.
\eeq
We apply \eqref{etrif.-8} to $C^{+}\phi$ using that
\[
\cS_{\sH}C^{+}\phi= \one_{\rr^{+}}(-\i ^{-1}\p_{U})\cS_{\sH}\phi, 
\]
and obtain:
\beq\label{sloubi3}
\chi\phi(x)= C\int_{\sH}\GG_{1}(x, U, \theta, \varphi^{\t}) \chi_{2}(U) \one_{\rr^{+}}(-\i ^{-1}\p_{U})\cS_{\sH}\phi\,\sin \theta \, dU d \theta d\varphi^{\t}.
\eeq
We apply Proposition \ref{prop-key} with $S= \sH_{L}$ and $X= - \p_{U}$. Clearly for
\[
g^{+}=\chi_{2}(U) \one_{\rr^{+}}(-\i ^{-1}\p_{U})\cS_{\sH}\phi
\]
we have
\[
\WF(g^{+})\subset \{(y, \eta)\in T^{*}\sH\setminus\zero \, : \, \eta\dual X(y)\geq 0\}.
\]
Thus, Proposition \ref{prop-key} combined with \eqref{sloubi3} imply that $q_{0}\not\in \WF(\phi)$. \qed

\appendix

\section{Properties of the spinorial Lie derivative}\label{appLie}\init

\subsection{Lie derivative of spinors} We will use the notation introduced in Section \ref{sec10} and assume that we are given a spinor bundle $\cS\xrightarrow{\pi}M$ on an  oriented and time oriented Lorentzian manifold  of dimension $4$. Recall the definition:
\beq\label{e10b.1a}
\cL_{X}\psi= \nabla_{X}^{\cS}\psi+ \frac{1}{8}((\nabla_{a}X)_{b}- (\nabla_{b}X)_{a})\gamma^{a}\gamma^{b}\psi, \ \ \psi\in \cinf(M, S),
\eeq
for $X\in\cinf(M;TM)$, and:
\[
\cL_{X}(\gamma(v)\psi)\eqdef (\cL_{X}\gamma)(v)\psi+ \gamma(\cL_{X}v)\psi+ 
\gamma(v)\cL_{X}\psi, \ \ \psi\in \cinf(M; \cS),
\]
for all $v\in \cinf(M; TM)$. One has the identity (\cite[(III.10)]{K}):
\beq\label{e10b.2}
(\cL_{X}\gamma)^{b}= - \frac{1}{2}((\nabla_{a}X)^{b}+ (\nabla^{b}X)_{a})\gamma^{a}.
\eeq
If $X$ is Killing this yields:
\begin{equation}
\label{e10.2b}
\cL_{X}(\gamma(v)\psi)=  \gamma(\cL_{X}v)\psi+ 
\gamma(v)\cL_{X}\psi.
\end{equation}

We denote by $X(f)$ the action of the vector field $X$ on the scalar function $f$. Recall that the divergence $\delta X$ of $X$ is defined by $\cL_{X}\Omega_{g}=: \delta X \Omega_{g}$, where $\Omega_{g}$ is the volume form associated to $g$.

\begin{lemma}\label{lemma1.1z}
 One has 
 \[
X( \bar{\psi}_{1}\dual \beta \psi_{2})= \bar{\cL_{X}\psi_{1}}\dual \beta\psi_{2}+ \bar{\psi}_{1}\dual \beta \cL_{X}\psi_{2},
\]
for all $X\in \cinf(M; TM)$, $\psi_{i}\in \cinf(M; \cS)$. It follows that if $\delta X= 0$, then $\cL_{X}$ is formally self-adjoint for the (non-positive) Hermitian scalar product 
 \[
(\psi_{1}| \psi_{2})_{M,\beta}= \int_{M}\bar{\psi}_{1}\dual \beta \psi_{2}\dVol_{g}, \ \ \psi_{1}, \psi_{2}\in \coinf(M; \cS).
\]
 Furthermore, $\nabla_{X}^{\cS}$ is formally self-adjoint when $\delta X=0$.
\end{lemma} 
\proof For $\psi_{1}, \psi_{2}\in \cinf(M; S)$ we have
\[
\bea
X(\bar{\psi}_{1}\dual \beta \psi_{2})&= \bar{\nabla_{X}\psi_{1}}\dual \beta\psi_{2}+ \bar{\psi}_{1}\dual \beta \nabla_{X}\psi_{2}\\[2mm]
&= \bar{\cL_{X}\psi_{1}}\dual \beta\psi_{2}+ \bar{\psi}_{1}\dual \beta \cL_{X}\psi_{2}+ \bar{R(X)\psi}_{1}\dual \beta \psi_{2}+ \bar{\psi}_{1}\dual \beta R(X)\psi_{2},
\eea
\]
for $R(X)= \frac{1}{8}((\nabla_{a}X)_{b}- (\nabla_{b}X)_{a})\gamma^{b}\gamma^{a}$. Since $(\gamma^{a})^{*}\beta= - \beta \gamma^{a}$, we obtain
\[
\bea
R(X)^{*}\beta + \beta R(X)&= \frac{1}{8}(\nabla_{a}X)_{b}- (\nabla_{b}X)_{a}\beta (\gamma^{a}\gamma^{b}+ \gamma^{b}\gamma^{a})\\[2mm]
&=  \frac{1}{8}((\nabla_{a}X)_{b}- (\nabla_{b}X)_{a})g^{ab}\beta=0,
\eea
\]
since $g^{ab}= g^{ba}$. \qed
\begin{lemma}\label{lemma1.1b}
 One has  $[\cL_{X}, \kappa]=0$.
 \end{lemma}
\proof
This follows from $[\nabla^{\cS}_{X}, \kappa]= 0$, $[\gamma(v), \kappa]=0$ and \eqref{e10b.1}. \qed 
\begin{lemma}\label{lemma1.2}
 One has $[\cL_{X}, H]=0$ and thus $\cL_{X}$ preserves $\cinf(M; \cW_{{\rm e}/{\rm o}})$.
\end{lemma}
\proof We know that $[\nabla_{X}^{\cS}, H]=0$ and $\gamma(v)H= - \gamma(v)H$, so the result follows from \eqref{e10b.1}. \qed

\begin{lemma}\label{lemma1.3}
If $X$ is a Killing vector field then
 \[
[\cL_{X}, \slashed{D}]= 0.
\]
\end{lemma}
\proof  By \cite[eqs.~(V.1), (III.50)]{K}  one has
\[
[\cL_{X}, \slashed{D}]=  (\cL_{X}\gamma)^{a}\nabla^{\cS}_{a}-\12(\p_{a}(\delta X)+ \12 \nabla_{b}(\cL_{X}g)^{b}_{a})\gamma^{a}.
\]
 If $X$ is Killing then $\cL_{X}\gamma=0$ by 
\eqref{e10b.2} and $\delta X= 0$. \qed

\subsection{Lie derivative of Weyl spinors}
By Lemma \ref{lemma1.2}, $\cL_{X}$ preserves $\cinf(M; \cW_{\rm e})$. 
Since $\SS= \cW_{\rm e}^{*}$ we can define $\cL_{X}^{*}$ acting on $\cinf(M; \SS)$ by:
\[
X( (w| s))\eqdef (\cL_{X}W| s)+ (w| \cL_{X}^{*}s), \ \ w\in \cinf(M; \cW_{\rm e}), \ s\in \cinf(M; \SS).
\]
We recall that $L(\cW_{\rm e}, \cW_{\rm e}^{*})$ is identified with $\SS\otimes \bar{\SS}$ and we 
define $\tilde{\cL}_{X}$ acting on $\cinf(M; \SS\otimes\SS^{*})$ by:
\[
\tilde{\cL}_{X} |s_{1})(s_{2}|\defeq |\cL_{X}^{*}s_{1})(s_{2}| + |s_{1}) (\cL_{X}^{*}s_{2}|, \ \ s_{i}\in \cinf(M; \SS).
\]
 We also recall that
 \[
\tau: \cinf(M; \cc TM)\ni v\mapsto \beta \gamma(v)\in \cinf(M; \SS\otimes \bar{\SS})
\]
is the canonical isomorphism.

The bundle $\SS$ is equipped with the symplectic form $\epsilon= \frac{1}{\sqrt{2}}(\beta\kappa)^{-1}$. From Lemmas \ref{lemma1.1z}, \ref{lemma1.1b} we obtain that:
\begin{equation}
\label{e10b.6}
X( s_{1}\dual \epsilon s_{2})= \cL_{X}^{*}s_{1}\dual \epsilon s_{2}+ s_{1}\dual \epsilon \cL_{X}^{*}s_{2}, \ \ s_{i}\in \cinf(M; \SS).
\end{equation}
\begin{lemma}\label{lemma1.4}
 If $X$ is Killing then we have:
 \[
\tau\circ \cL_{X}= \tilde{\cL}_{X}\circ \tau.
\]
\end{lemma}
\proof
For  $w_{i}\in \cinf(M; \cW_{\rm e})$, $s_{i}\in \cinf(M; \SS)$ and  $A= |s_{1})(s_{2}|$  we have:
\beq\label{e10b.4}
\bea
X( (w_{1}| Aw_{2}))&= X( (w_{1}| s_{1})\times (s_{2}| w_{2}))\\[1mm]
&= ((\cL_{X}w_{1}| s_{1})+ (w_{1}| \cL_{X}^{*}s_{1}))(w_{2}| s_{2}) \\[1mm] & \phantom{=],} + (w_{1}|s_{1})((\cL_{X}^{*}s_{2}|w_{2})+ (s_{2}| \cL_{X}w_{2}))\\[1mm]
&= (\cL_{X}w_{1}| s_{1})(w_{2}| s_{2})+ (w_{1}|s_{1})(s_{2}| \cL_{X}w_{2})+ (w_{1}| \tilde{\cL}_{X}A w_{2}).
\eea
\eeq
If $A= \tau(v)$  for $ v\in \cinf(M; \cc TM)$, we have $(w_{1}| Aw_{2})= \bar{w}_{1}\dual \beta \gamma(v)w_{2}$ so by Lemma \ref{lemma1.1z}:
\[
X( (w_{1}| Aw_{2}))= \bar{\cL_{X}w}_{1}\dual \beta\gamma(v) w_{2}+ \bar{w}_{1}\dual \beta \cL_{X}(\gamma(v)w_{2}).
\]
If $X$ is Killing this yields using \eqref{e10.2b}:
\beq\label{e10b.5}
\bea
X( (w_{1}| Aw_{2}))&=\bar{\cL_{X}w}_{1}\dual \beta\gamma(v) w_{2}+ \bar{w}_{1}\dual \beta\gamma(\cL_{X}v)w_{2}+ \bar{w}_{1}\dual \beta \gamma(v)\cL_{X}w_{2}\\[2mm]
&=(\cL_{X}w_{1}| s_{1})(w_{2}| s_{2})+ (w_{1}|s_{1})(s_{2}| \cL_{X}w_{2})+ \bar{w}_{1}\dual \beta\gamma(\cL_{X}v)w_{2}.
\eea
\eeq
By comparing \eqref{e10b.4} with \eqref{e10b.5} we obtain that 
$\tau(\cL_{X}v)= \tilde{\cL}_{X}\tau v$. \qed
 
\begin{lemma}\label{lemma1.5}
 Let $l,n, m, \bar{m}$ be a normalized null tetrad, and let $(\mo, \mi)$ be the associated frame of $\SS$, such that 
 \[
\i \tau l= \mo\otimes \bar{\mo}, \ \ \i \tau n= \mi\otimes\bar{\mi}, \ \ \i \tau m= \mo\otimes\bar{\mi}, \ \ \i \tau \bar{m}= \mi\otimes\bar{\mo}.
\]
Then if $X$ is a Killing vector field such that $\cL_{X}l= \cL_{X}n= \cL_{X}m=0$, one has:
\[
\cL_{X}^{*}\mo= \cL_{X}^{*}\mi=0.
\]
\end{lemma}
\proof Let us denote  $\cL_{X}^{*}\mathit{o}, \cL_{X}^{*}\mathit{i}$ simply by $\mathit{o}', \mathit{i}'$. Since $\cL_{X}l= \cL_{X}n= \cL_{X}m=0$, we obtain from Lemma \ref{lemma1.4}:
\[
\bea
1) \ & \mo'\otimes \bar{\mo}+ \mo\otimes\bar{\mo}'=0,\\[2mm]
2) \ & \mi'\otimes \bar{\mi}+ \mi\otimes\bar{\mi}'=0,\\[2mm]
3) \ & \mo'\otimes \bar{\mi}+ \mo\otimes\bar{\mi}'=0,\\[2mm]
4) \ & \mi'\otimes \bar{\mo}+ \mi\otimes\bar{\mo}'=0.
\eea
\]
Using \eqref{e10b.6} and the fact that $\mo\dual \epsilon \mi= 1$ we obtain $\mo'\dual \epsilon \mi + \mo\dual \epsilon \mi'=0$ hence  $\mo'\dual \epsilon \mi= \mi'\dual \epsilon \mo$. Composing then $1)$  and $3)$ to the right with $\one \otimes \bar{\epsilon \mi}$  and $\one\otimes \bar{\epsilon \mo}$ respectively  gives:
\[
\bea
\mo'+  (\bar{\mo'\dual \epsilon \mi})\mo&=0, \\[2mm]
- \mo'+ (\bar{\mi'\dual \epsilon \mo})\mo&=0,
\eea
\]
hence $\mo'=0$ since $\mo'\dual \epsilon \mi= \mi'\dual \epsilon \mo$. Composing then 4) to the right with $\one\otimes \bar{\epsilon \mi}$ gives finally $\mi'=0$. \qed

\section{Algebraic quantization of the Weyl equation}\label{algebraic}\init
\subsection{Quasi-free states on $\CAR $ $*$-algebras}\label{quanto.ss1}
In this appendix  of algebraic quantization of fermionic fields. Below, we follow \cite[Sect.~17.14]{G}.
 Let $(\cY, \nu)$ be a pre-Hilbert space. We denote by ${\rm CAR} (\cY, \nu)$ the unital complex $*$-algebra generated by elements $\psi(y)$, $\psi^{*}(y)$, $y\in \cY$, with the relations
\begin{equation}
\label{e15.2}
\begin{array}{l}
\psi(y_{1}+ \lambda y_{2})= \psi(y_{1})+ \overline{\lambda}\psi(y_{2}),\\[2mm]
\psi^{*}(y_{1}+ \lambda y_{2})= \psi(y_{1})+ \lambda\psi^{*}(y_{2}), \ y_{1}, y_{2}\in \cY, \lambda \in \cc, \\[2mm]
[\psi(y_{1}), \psi(y_{2})]_{+}= [\psi^{*}(y_{1}), \psi^{*}(y_{2})]_{+}=0, \\[2mm]
 [\psi(y_{1}), \psi^{*}(y_{2})]_{+}= \overline{y}_{1}\cdot \nu y_{2}\one, \ y_{1}, y_{2}\in \cY,\\[2mm]
 \psi(y)^{*}= \psi^{*}(y),\ y\in\cY, 
\end{array}
\end{equation}
where $[A, B]_{+}= AB+ BA$ is the anti-commutator.

  A quasi-free state $\omega$ on $\CAR(\cY, \nu)$ is determined by its covariances $\lambda^{\pm}\in L_{\rm h}(\cY, \cY^{*})$, defined by 
\[
\omega(\psi(y_{1})\psi^{*}(y_{2}))\eqdef \overline{y}_{1}\dual \lambda^{+}y_{2},\quad \omega(\psi^{*}(y_{2})\psi(y_{1}))\eqdef \overline{y}_{1}\dual \lambda^{-}y_{2}, \quad y_{1}, y_{2}\in \cY.
\]
A pair of Hermitian sesquilinear forms $\lambda^{\pm}$ on $\cY$ are the covariances of a quasi-free state on $\CAR(\cY, \nu)$ iff
\beq\label{pisito}
\lambda^{\pm}\geq 0, \ \ \lambda^{+}+ \lambda^{-}= \nu.
\eeq
It follows that $\lambda^{\pm}$ uniquely extend to the completion $\cY^{\rm cpl}$ of $\cY$ for $\nu$. The following characterization of pure quasi-free states is well-known, see e.g.~\cite[Thm.~17.31, Subsect.~17.2.3]{DG}.

\begin{lemma}\label{fermio-pure}
The quasi-free state $\omega$ on $\CAR(\cY, \nu)$ is a pure state iff there exist projections $\pi^{\pm}$ on $\cY^{\rm cpl}$ such that
\[
\lambda^{\pm}= \nu\circ \pi^{\pm}.
\]
\end{lemma}
Note that $\pi^{\pm}$ are selfadjoint for $\nu$ and $\pi^{+}+ \pi^{-}= \one$.

\subsection{Time reversal}\label{quanto.ss2} In the remaining part of this section we discuss the notion of time reversal in the context of quantization of Weyl fields.

\subsubsection{Time reversal on $\CAR(\cY, \nu)$}\label{quanto.ss2.1}
Let $(\cY, \nu)$ a pre-Hilbert space as in Subsect.~\ref{quanto.ss1}. A   {\em  time reversal} is a unitary involution  $\tau\in U(\cY, \nu)$ i.e.~
\[
\tau^{2}= \one, \ \  \overline{\tau y_{1}}\dual \nu \tau y_{2}=\bar{y}_{1}\dual \nu y_{2},\ \ y_{1}, y_{2}\in \cY. 
\]
We associate to $\tau$ the anti-linear $*$-involution $\hat{\tau}$ on $\CAR(\cY, \nu)$ defined by
\[
\hat{\tau}(\psi(y))\defeq \psi^{*}(\tau y), \ \hat{\tau}(\psi^{*}(y))\defeq \psi(\tau y), \ y\in \cY.
\]
Note that this map is well defined in $\CAR(\cY, \nu)$ since
\[
\bea
\hat{\tau}([\psi(y_{1}), \psi^{*}(y_{2})]_{+})&= \hat{\tau}(\bar{y}_{1}\dual \nu y_{2}\one)= \bar{y}_{2}\dual \nu y_{1}\one,\\
&=[\hat{\tau}(\psi(y_{1})), \hat{\tau}(\psi^{*}(y_{2}))]_{+}\\&= [\psi^{*}(\tau y_{1}), \psi(\tau y_{2})]_{+}=  \bar{y}_{2}\dual \nu y_{1}\one.
\eea
\]
If $\omega$ is a quasi-free state on $\CAR(\cY, \nu)$ with covariances $\lambda^{\pm}$ we define $\hat{\tau}^{*}\omega= \hat{\omega}$ by
\[
\hat{\omega}(A)\defeq \omega (\hat{\tau}(A^{*})), \ A\in \CAR(\cY, \nu),
\]
which is also a quasi-free state on $\CAR(\cY, \nu)$. We have then
\[
\begin{array}{l}
\hat{\omega}(\psi(y_{1}) \psi^{*}(y_{2}))= \omega(\psi^{*}(\tau y_{2})\psi(\tau y_{1})), \\[2mm]
\hat{\omega}(\psi^{*}(y_{2})\psi(y_{1}))= \omega(\psi(\tau y_{1})\psi^{*}(\tau y_{2})),
\end{array}
\]
so the  covariances $\hat{\lambda}^{\pm}$ of $\hat{\omega}$ are:
\[
\hat{\lambda}^{\pm}= \tau^{*}\lambda^{\mp}\tau.
\]
\subsubsection{Time reversal on a spacetime}\label{quanto.ss2.2}
Recall that if $(M, g)$ is a spacetime,  $(M', g)$ is the same Lorentzian manifold with the reversed time orientation.  Denoting with primes the objects associated to $(M', g)$ we have:
\[
\begin{array}{l}
\gamma'= \gamma, \ \ \kappa'= \kappa, \ \ \beta'= - \beta, \  \ \cN^{\pm'}= \cN^{\mp},\\[2mm]
\nabla^{S'}= \nabla^{S}, \ \  \slashed{D}'= \slashed{D}, \ \ \DD'= - \DD, \ \ \GG'= \GG,\\[2mm]
(\cdot| \cdot)_{M'}= (\cdot| \cdot)_{M},\ \  (\cdot | \cdot)_{\DD'}= (\cdot | \cdot)_{\DD},\\[2mm]
\end{array}
\]
so $\id:(\Sol_{\rm sc}(M), (\cdot | \cdot)_{\DD})\to (\Sol_{\rm sc}(M'), (\cdot | \cdot )_{\DD'})$ is  a  time reversal in the sense of \ref{quanto.ss2.1}.

\subsubsection{Time reversal on quasi-free states}
If $\LL^{\pm}$ resp.~$C^{\pm}$ are  the spacetime  resp.~solution space covariances of a quasi-free state $\omega$ on $\CAR(M)$, then
 \[
 \LL^{\pm\prime}\,= \,\,  \LL^{\mp}, \ \ C^{\pm\prime}= C^{\mp}
 \]
 are the spacetime resp.~solution space covariances of the corresponding state $\omega'$ on $\CAR(M')$ (denoted by $\hat{\omega}$ in \ref{quanto.ss2.1}).
Since $\cN^{\pm'}= \cN^{\mp}$, we see that if  $\omega$ is a  Hadamard state on $\CAR(M)$, then $\omega'$ is a  Hadamard state on $\CAR(M')$ (see Subsect.~\ref{quanto.ss3} for the definition of Hadamard states).  

\section{Null geodesics  and Cauchy surfaces in Kerr spacetime}\label{secG}\init

\subsection{Summary} The  first purpose of this appendix is to show that   various hypersurfaces used throughout the text are Cauchy, and that the corresponding spacetime regions are therefore globally hyperbolic. We remark that the causal structure of the Kerr spacetime could be deduced from detailed knowledge of lower-dimensional diagrams, see \cite{chrusc} for the relevant framework. Here however we need to analyse the geodesics  directly  to treat the families of hypersurfaces that are of interest to us.

We also prove an important result used in the main text, namely  \eqref{secite} of Lemma \ref{lemalemo}, which states that for sufficiently small $a$,  null  geodesics in $\MI$  that do not meet $\sH_-$ nor $\sI_-$ always cross a region where the Killing vector fields $v_{\sH}$ and $v_{\sI}$ are  time-like

\subsection{Null geodesics and first integrals} 
We first recall some well-known facts about null geodesics in the Kerr spacetime, see \cite[Ch.~4]{N}, \cite[Ch.~7]{Ch}.  
 
 Denoting by $g_{a, M}$ the Kerr metric in the Boyer--Lindquist coordinates $(r, t, \theta, \varphi)$, the change of coordinates $r= M \tilde{r}$, $t= M \tilde{t}$ gives $g_{a, M}= M^{2}g_{aM^{-1}, 1}$. After a trivial conformal transformation, we can  assume   that $M=1$ to make the computations easier.  
 However, we will state the main results about null geodesics for arbitrary  values of $M$.
 
With the convention that $M=1$, recall that we have the sub-extremality assumption $0<a<1$, and: 
\[
\Delta= r^{2}- 2r+ a^{2}, \ \ \rho^{2}= r^{2}+ a^{2}\cos^{2}\theta,
\]
and $r_{\pm}= 1\pm\sqrt{1- a^{2}}$ are the horizon radii.

Let $\gamma: I\ni s\mapsto x(s)$ be a null geodesic, where $s$ is the affine parameter. We have the following  three independent first integrals:
\[
E= - \p_{t}\cdot g \dot{x}, \ \ L= \p_{\varphi} \cdot g \dot{x}, \ \ K=\hbox{ Carter constant},
\]
which are constant along $\gamma$. One sets $Q= K- (L -aE)^{2}$, and one uses $(E,L, Q)$ as independent first integrals.   
Any null geodesic starting away from the axis $\{\sin \theta=0\}$ is entirely determined by its initial coordinates $(r, t, \theta, \varphi)(0)$, the signs of $r'(0)$ and $\theta'(0)$ and the constants $(E, L, Q)$, see \cite[Lem.~4.2.5]{N}.

\subsection{Equations of motion}
Let:
\[
P(r)= (r^{2}+a^{2})E- aL, \ \  D(\theta)= L- aE \sin^{2}\theta.
\]
If $\gamma: I\ni s\mapsto x(s)$ is an affinely parametrized null geodesic, we  have, see \cite[Thm.~4.2.2]{N}:
\beq\label{e0.0}
\begin{cases}
\rho^{4}\dot{r}^{2}=R(r),\\[2mm]
\rho^{4}\dot{\theta}^{2}= \Theta(\theta),
\end{cases}
\eeq
for
\[
\bea
R(r)&= P^{2}(r) -\Delta(r) K, \\
&=E^{2}r^{4}+ (a^{2}E^{2}- L^{2}- Q)r^{2}+ ((aE- L)^{2}+ Q)2r- a^{2}Q,\\
\Theta(\theta)&=K- \sin^{-2}(\theta)D^{2}(\theta).
\eea
\]
Knowing the functions $r(s), \theta(s)$, the functions $t(s), \varphi(s)$ are then determined by the equations:
\beq\label{e0.0b}
\begin{cases}
\rho^{2}\dot{\varphi}= \sin^{-2}\theta D(\theta)+ a\dfrac{P(r)}{\Delta(r)},\\[2mm]
\rho^{2}\dot{t}= aD(\theta)+\dfrac{(r^{2}+a^{2})P(r)}{\Delta(r)}\eqdef T(r, \theta)
\end{cases}
\eeq
see \cite[Prop.~4.1.5]{N}. 

\subsubsection{Some more facts}
The \emph{polar planes} $P=P(t_{0}, \varphi_{0})$ are the submanifolds $\{t=t_{0}, \varphi= \varphi_{0}\}$. The \emph{axis}  is the submanifold $A= \{\sin \theta=0\}$.
We know,  see \cite[Cor.~4.2.8, Lem.~4.2.9]{N}, that  if $\gamma$ is a null geodesic  then:
\ben
\item $K\geq 0$,
\item $K=0$ iff $\gamma$ is a principal null geodesic,
\item $L=E=0$ and $K>0$ iff $\gamma$ is in a time-like polar plane (hence in ${\rm M}_{\rm II}$ (in order for the polar plane to be time-like),
\item $K= L= 0$ and $E\neq 0$ iff $\gamma$ is in $A\setminus (\sH_{+}\cup \sH_{-})$,
\item $K=L=E=0$ iff $\gamma$ is in $\sH_{+}\cup \sH_{-}$.
\een
Moreover, see e.g.~\cite[Prop.~2.5.5]{N}, the horizons $\sH_{\pm}$ are closed and totally geodesic, hence any geodesic tangent to $\sH_{\pm}$ is entirely included in $\sH_{\pm}$.

\subsubsection{Double zeros of $R$}
We now look at possible double zeros (roots) of $R$. 

If $E= 0$, then $R(r)$ is a polynomial of degree   $0$ if  $K=0$,  and $2$ if $K\neq 0$.  If $K=0$, then  $R(r)= -a^{2}Q$ so $R$ has double zeros only if $Q=0$. If $K\neq 0$, then $K>0$  and $R$ has a double zero if $L^{2}= (a^{2}-1)Q$,  which implies that $Q<0$ and $K= a^{2}Q<0$ which is a contradiction. Therefore if $E=0$, $R$ has a double zero only if $K= L= E=0$, i.e.~if $\gamma$ is included in $\sH_{+}\cup \sH_{-}$. 

If $E\neq 0$  it is convenient to set \beq\label{ez.0}
\xi= LE^{-1}, \ \ \eta= QE^{-2}, \ \ \kappa= KE^{-2}=\eta+ (\xi-a)^{2}.
\eeq
 One has the following constraints on the integrals of motions for null geodesics:
\beq\label{G.2}
\eta+(\xi-a)^2\geq 0,
\eeq
\beq\label{G.3}
\eta<0 \ \Rightarrow \   \eta^2\leq a^2, \ 0\leq \vert\xi\vert\leq \vert a\vert-\sqrt{\vert \eta\vert}, 
\eeq
see  equations  (194), (206) in \cite[Ch.~7]{Ch}. 
The following lemmas  will be needed later on.

\begin{lemma}\label{lemilemo}
\ben 
\item There are no double zeros of $R$ in $]r_{-}, r_{+}[$.
\item
There exists $0<a_{0}<1$  and $C>0$ such that for $|a|\leq a_{0}$ and $E\neq 0$ the double zeros of $R$   in $\clopen{r_{-}, +\infty}$ belong to $\closed{3-Ca, 3+Ca}$ and one has $|\xi|\leq C$, $|\eta|\leq C$. 
\een
\end{lemma}
\proof
By the computations in \cite[p.~351]{Ch} the conditions $R(r)=0,\,\partial_rR(r)=0$ imply that: 
\begin{eqnarray}
\label{G.4}
\xi=\frac{1}{a(r-1)}((r^2-a^2)-r\Delta),\\
\label{G.5}
\eta=\frac{r^3}{a^2(r-1)^2}(4a^2-r(r-3)^2),
\end{eqnarray}
see equations (224), (225) in \cite[Ch.~7]{Ch}. The conditions $\eta<0$ and \eqref{G.3} are incompatible with \eqref{G.4}, \eqref{G.5} (see \cite[p.~352]{Ch}). It  follows hence from \eqref{G.5} that if $r$ is a double zero of $R$ then
$r(r- 3)^{2}\leq 4a^{2}$.  If $r\in \open{r_{-}, r_{+}}$ we have $\Delta(r)<0$, i.e.~$a^{2}< 2r - r^{2}$ hence 
\[
r(r-3)^{2}< 4(2r-r^{2}) 
\]
and therefore $r(r-1)^{2}<0$, which is a contradiction. This proves (1).

Let us now prove (2).  By studying  the graph of $f(r)= r(r-3)^{2}$ we  obtain that the double zeros of $R$  belong to 
\[
 \Big[0, \frac{4}{9}a^{2}+ O(a^{3})\Big]\cup \Big[3- \frac{2}{\sqrt{3}}a+ O(a^{2}), 3+ \frac{2}{\sqrt{3}}a+ O(a^{2})\Big].
\]
Since $r_{-}= \12 a^{2}+ O(a^{3})$,  the first interval above is disjoint from $\clopen{r_{-}, +\infty}$ for $a$ small enough. If $r$ is a double zero in $[r_{-}, +\infty[$ then from $r= 3+ O(a)$ and \eqref{G.4} we obtain that $\xi\in O(1)$, and since $0\leq 4a^{2}- r(r-3)^{2}\leq 4a^{2}$ we obtain that $\eta\in O(1)$. \qed

\begin{lemma}\label{zbrodj}
 Let $r_{0}>r_{+}$ be a double zero of $R$. Then 
 \[
 T(r_{0}, \theta)\neq 0, \ \forall\,  \theta\in[0, \pi].
 \]
 \end{lemma}
\proof We note that 
\[
 T(r,\theta)=\frac{\sigma^2(r, \theta)E-2arL}{\Delta(r)},
 \]
 where $\sigma^2(r,\theta)=(r^2+a^2)^2-a^2(\sin^2\theta)\Delta(r)$. 
   Assume that $T(r_{0}, \theta)=0$. Setting $\sigma_{0}= \sigma(r_{0}, \theta), \Delta_{0}= \Delta(r_{0})$, we obtain $\xi= \frac{\sigma_0^2}{2ar_0}$.
  By equation (224) in \cite[Ch.~7]{Ch} we have 
 \begin{equation}
 \label{C.4.3}
 \xi=\frac{1}{a(r_0-1)}(r_0^2-a^2-r_0\Delta_0),
 \end{equation}
 which yields:
 \begin{equation*}
 \sigma_0^2(r_0-1)-2r_0(r_0^2-a^2-r_0\Delta_0)=0.
 \end{equation*}
 We write 
 $$
\bea
 \sigma_0^2(r_0-1)-2r_0(r_0^2-a^2-r_0\Delta_0)&=(r_0-1)(r_0^2+a^2)^2-2r_0(r_0^2-a^2)\\
 & \phantom{=}\, +(2r_0^2-a^2\sin^2\theta(r_0-1))\Delta_0. 
\eea
$$
 We have
 \begin{eqnarray*}
 2r_0^2-a^2\sin^2\theta(r_0-1)\ge2r_0^2-a^2r_0=r_0(2r_0-a^2)>0,
 \end{eqnarray*}
and
$$
\bea
 (r_0-1)(r_0^2+a^2)^2-2r_0(r_0^2-a^2)&\ge 2r_0((r_0-1)(r_0^2+a^2)-(r_0^2-a^2))\\
 &=2r_0(r_0^3+a^2r_0-r_0^2-a^2-r_0^2+a^2)\\
 &= 2r_0^2(r^2_0+a^2-2r_0)=2r_0^2\Delta_0>0.
\eea 
$$
 Therefore $\xi\neq\frac{\sigma_0^2}{2ar_0}$ which proves the lemma. \qed

\begin{corollary}\label{corolaro}
 There exists $0<a_{1}\leq 1$ such that for $|a|M^{-1}<a_{1}$ the Killing vector fields $v_{\sH}$ and $v_{\sI}$ are future directed time-like at all points $(r, \theta, \varphi)\in \clopen{r_{+}, +\infty}\times \bS^{2}$ such that $r$ is a double zero of $R$.
  \end{corollary}
\proof We can assume without loss of generality $M= 1$ as explained at the beginning of the section. Let $f_{\sH/\sI}(a, r, \theta)= v_{\sH/\sI}\dual g v_{\sH/\sI}$.  We have  $f_{\sH/\sI}(0, 3, \theta)= - 1/3$, which implies that there exists $\epsilon, \delta>0$ such that $v_{\sH_{-}/ \sI_{-}}$ are time-like for $|a|\leq \epsilon$ and $|r-3|\leq \delta$. By applying Lemma \ref{lemilemo} we obtain that $v_{\sH}$ and $v_{\sI}$ are  time-like at all points $(r, \theta, \varphi)\in \clopen{r_{+}, +\infty}\times \bS^{2}$ such that $r$ is a double zero of $R$. They are clearly future directed since $-\nabla t$ is future directed.  \qed

\subsection{Null  geodesics in ${\rm M}_{\rm I}$}
We recall the classification of null geodesics according to the behavior of $r(s)$ for $s\to \infty$.  By \eqref{e0.0} the region $\{R(r)<0\}$ is not accessible to null geodesics. One can then discuss the possible connected components of $\{R(r)\geq 0\}$, depending of the values of the first integrals $E, L,Q$.

Another standard fact, see e.g.~\cite[Ch.~4]{N},  is that  $r(s)$ can reach a first order zero of $R$ only for some finite affine time $s_{0}$, and then $\dot{r}(s)$ changes sign at $s= s_{0}$.  
On the contrary $r(s)$ can reach a double zero $r_{0}$ of $R$ only for $s= \pm \infty$, or else $r(s)\equiv r_{0}$.   These facts follow easily from \eqref{e0.0}. 
The horizon $\{r= r_{+}\}$ can  be crossed only transversally, at finite affine time, see \cite[Sect.~4.4]{N}. 

We obtain the following classification for future directed null geodesics.   

In the tables below,   \emph{type} $[a, b]$ for example means that  $\gamma$ starts at $r= a$ and ends at $r=b$ for finite values of the affine parameter, while  \emph{type} $[a, b[$ means that $\gamma$ ends at $r=b$ for infinite value of the affine parameter. Furthermore, $r_{0}$ denotes some double zero of $R$, and $[r_0]$ corresponds to a geodesic for which $r(s)=r_0$ for all $s$. 

\[
\begin{tabular}{|c|c|}
  \hline
  \multicolumn{2}{|c|}{ Table 1.\  $E=0$}\\\hline
    Type& Constants \\ \hline
  \multirow{1}{*}{$[r_{+}\!\to \! r_{+}]$}  
    &$E=0,\, K>0,\, L\neq 0$ \\ \hline
     \multirow{1}{*}{$[r_{+}\!\to \!\infty[$ or $ ]\infty\!\to \! r_{+}]$}  
    &$E=0, \, K=0, \, L\neq 0$ (princ.~null) \\ \hline
         \end{tabular}\]
         \medskip
           \medskip  \medskip
\[
\begin{tabular}{|c|c|}
  \hline
  \multicolumn{2}{|c|}{ Table 2. \  $E\neq 0$}\\\hline
    Type& Constants \\ \hline
  \multirow{2}{*}{$[r_{+}\!\to \! r_{+}]$}  
    &$E\neq 0,\, Q>0,\, R(r_{+})>0$ \\
    &$E\neq 0,\, Q=0,\, L\neq aE,\, R(r_{+})>0$ \\ \hline
    \multirow{2}{*}{$[r_{+}\!\to \! \infty[$ or $ ]\infty\!\to \! r_{+}]$}  
    &$E\neq 0,\, Q\geq 0,\, R(r_{+})>0$ \\
    &$E\neq 0,\, Q< 0$ \\ \hline
    \multirow{2}{*}{$]\infty\!\to \! \infty[$}  
    &$E\neq 0,\, Q>0$ \\
    &$E\neq 0,\, Q=0,\, L\neq aE$ \\ \hline
    \multirow{2}{*}{$[r_{+}\!\to \! r_{0}[$ or $ ]r_{0}\!\to \! r_{+}]$}  
    &$E\neq 0,\, Q>0,\, R(r_{+})>0$ \\
    &$E\neq 0,\, Q=0,\, L\neq aE,\, R(r_{+})>0$ \\ \hline
    \multirow{1}{*}{$]r_{0}\!\to \!\infty[$ or $ ]\infty\to r_{0}[$}  
    &$E\neq 0,\, Q>0,\, R(r_{+})>0$ \\  \hline 
      \multirow{2}{*}{$[r_0]$}
        &$L=\frac{1}{a(r_0-M)}[M(r_0^2-a^2)-r_0\Delta(r_0)]E,$\\
        &$Q=\frac{r_0^3}{a^2(r_0-M)^2}[4M\Delta(r_0)-r_0(r_0-M)^2]E^2$\\[1.5mm] \hline
          \end{tabular}\medskip   \medskip               
          \] 
 To summarize the  tables above, let $\gamma$ a future directed null geodesic in ${\rm M}_{\rm I}$, parametrized by $s\in I$. We have  $I= \open{s_{-}, s_{+}}$ for 
 \[
\begin{array}{l}
 -\infty=s_{-}, \ s_{+}=+\infty\hbox{ if }\gamma\hbox{ is of type }]\infty\!\to \! \infty[,\ ]r_{0}\!\to \! \infty[,  \ ]\infty\!\to \!  r_{0}[,\\[2mm]
-\infty<s_{-}, \ s_{+}=+\infty\hbox{ if }\gamma\hbox{ is of type }[r_{+}\!\to \! \infty[, \ [r_{+}\!\to \!  r_{0}[,\\[2mm]
-\infty=s_{-}, \ s_{+}<+\infty\hbox{ if }\gamma\hbox{ is of type }]\infty\!\to \! r_{+}],\  ]r_{0}\!\to \! r_{+}]\\[2mm]
-\infty<s_{-}, \ s_{+}<+\infty\hbox{ if }\gamma\hbox{ is of type }[r_{+}\!\to \! r_{+}].
\end{array}
\] 

 \begin{lemma}\label{lemalemo}  \ben
 \item Let $\gamma$ be a future directed null geodesic. Then
 \begin{equation}
 \label{e0.08}
 \begin{array}{rl}
 i)&\lim_{s\to s_{\pm}}t(s)= \pm\infty, \\[2mm]
 ii)&\lim_{s\to s_{-}}t\kst(s)= -\infty\hbox{ if }\gamma \hbox{ starts at }r_{+}.
 \end{array}
 \end{equation}
 \item \label{secite}
 There exists $0<a_{0}\leq 1$ such that for $|a|M^{-1}<a_{0}$ any future directed null geodesic which does not start at $r_{+}$ nor at $\infty$ passes through a region where  the Killing vector fields $v_{\sH}$ and $v_{\sI}$ are both time-like. 
  \een
  \end{lemma}
  \proof
 Again we assume $M=1$.  Statement (2) follows from Corollary \ref{corolaro} and the tables above.  Let us now prove (1).
  
  Let us first check {\it ii)}. If $\gamma$ starts at $r_{+}$, then $\gamma$ enters ${\rm M}_{\rm I}$ from $\MII'$,  and since ${\rm M}_{\rm I}$ and $\MII'$ belong to a common $\stk {\rm K}$ patch we have $\lim_{s\to s_{-}}\stk t(s)$ finite, $\lim_{s\to s_{-}}r(s)= r_{+}$, so $\lim_{s\to s_{-}}t\kst(s)= -\infty$.

  Let us now prove {\it i)}. We first assume that $E\neq 0$.

  From \eqref{e0.0}, \eqref{e0.0b} we have
\beq
  \label{C.4.1}
t= \pm \int\frac{T(r, \theta)}{R^{\12}(r)}dr.
\eeq
{\it Case 1}: assume  that $\lim_{s\to s_{\pm}}r(s)= +\infty$.
We have:
\begin{equation}
\label{ez.1}
\frac{T(r, \theta)}{R^{\12}(r)}= \frac{E}{|E|}+ O(r^{-1}), \ r\to +\infty.
\end{equation}
Note that $E>0$ since $\gamma$ is future directed. Therefore  by \eqref{C.4.1} $\lim_{s\to s_{\pm}}t(s)= +\infty$.

{\it Case 2}: assume next that $\lim_{s\to s_{\pm}}r(s)=r_{+}$.  By Table 2 we have $R^{\12}(r_{+})= |P(r_{+})|\neq 0$
hence
 \begin{equation}
\label{ez.2}
\frac{T(r, \theta)}{R^{\12}(r)}=\frac{2r_{+}}{(r-r_{+})(r_{+}-r_{-})}+ O(1), \ r\to r_{+},
\end{equation}
which by \eqref{C.4.1} shows that $\lim_{s\to s_{\pm}}t(s)= +\infty$. 

{\it Case 3}: assume that $\lim_{s\to s_{\pm}}r(s)= r_{0}$, where $r_{0}\neq r_{+}$ is a double root of $R$.   By Lemma \ref{zbrodj}, $T(r_{0}, \theta)\neq 0$ for $\theta\in [0, \pi]$.  If $r(s)$ is not constant, then  \eqref{C.4.1} implies that $\lim_{s\to s_{\pm}}t(s)= +\infty$. 
If $r(s)\equiv r_0$ we use \eqref{e0.0b} to obtain again that    $\lim_{s\to s_{\pm}}t(s)= +\infty$. 
 
 Consider now the case $E= 0$. If $K>0$ and $L\neq 0$ then $\gamma$ is of type $[r_{+}\!\to \! r_{+}]$ and
 \begin{equation}
 \label{e0.06}
 \begin{array}{l}
 \rho^{-2}T=-\frac{2a^{2}Lr_{+}}{(r-r_{+})(r_{+}-r_{-})}+ O(1),\\[2mm]
 \rho^{-2}R^{\12}(r)= |aL|+ O(r-r_{+}),
 \end{array}\hbox{ when }r\to r_{+},\hbox{ uniformly in }\theta,
 \end{equation}
  Again, the assertion of the lemma is satisfied.  If $K=0, L\neq 0$, $\gamma$ is a principal null geodesic and {\it i)} is satisfied.  If $K=L=0$ $\gamma$ is not in ${\rm M}_{\rm I}$. \qed

 \subsection{Null geodesics in ${\rm M}_{\rm II}$}
 In ${\rm M}_{\rm II}$ we have other possibilities: a future directed null geodesic can originate at the 
 bifurcation sphere $S(r_{+})$ and terminate at the bifurcation sphere $S(r_{-})$. We obtain the following table of null geodesics in ${\rm M}_{\rm II}$:\medskip
 \[
\begin{tabular}{|c|c|}
  \hline
  \multicolumn{2}{|c|}{ Table 3.\  $E=0$}\\\hline
    Type& Constants \\ \hline
  \multirow{2}{*}{$[r_{\pm}\!\to \! r_{\mp}]$ }  
    &$E=0,\, K> 0,\, L\neq 0$ \\
   &$E=0,\, K=0,\, L\neq 0$ (princ.~null) \\ \hline
   $[S(r_{\pm})\!\to\! S(r_{\mp})]$& $E=0,\, K>0,\, L\neq 0$ \\ \hline
         \end{tabular}\]
         \medskip
           \medskip  
\[
\begin{tabular}{|c|c|}
  \hline
  \multicolumn{2}{|c|}{ Table 4. \  $E\neq 0$}\\\hline
    Type& Constants \\ \hline
  \multirow{2}{*}{$[r_{\pm}\!\to \! r_{\mp}]$ }  
    &$E\neq 0,\, Q\geq 0,\, R(r_{\pm})>0$ \\
    &$E\neq 0,\, Q<0,\,  R(r_{\pm})>0$ \\ \hline
    \multirow{2}{*}{$[r_{-}\!\to \! r_{-}]$ }  
    &$E\neq 0,\, Q> 0,\, R(r_{-})>0$ \\
    &$E\neq 0,\, Q= 0,\, L\neq aE,\, R(r_{-})>0$ \\ \hline
    \multirow{2}{*}{$[r_{+}\!\to \! r_{+}]$ }  
    &$E\neq 0,\, Q> 0,\, R(r_{+})>0$ \\
    &$E\neq 0,\, Q= 0,\, L\neq aE,\, R(r_{+})>0$ \\ \hline
           \multirow{2}{*}{$[r_{+}\!\to\! S(r_{-})]$ }  
    &$E\neq 0,\, Q> 0,\, R(r_{-})=0$ \\
    &$E\neq 0,\, Q=0,\,  L\neq aE,\,   R(r_{-})=0$ \\ \hline
      \multirow{2}{*}{$[S(r_{+})\!\to\! r_{-}]$ }  
    &$E\neq 0,\, Q> 0,\, R(r_{+})=0$ \\
    &$E\neq 0,\, Q=0,\, L\neq aE,\,   R(r_{+})=0$ \\ \hline

             \end{tabular}   \medskip
                        \medskip  
\]
 Let $\gamma$ be a future directed null geodesic in ${\rm M}_{\rm II}$, parametrized by $s\in I$. We have  $I= \open{s_{-}, s_{+}}$ for $-\infty<s_{-}<s_{+}<+\infty$.
 We have the following analogue of Lemma \ref{lemalemo}.
\begin{lemma}\label{lemalemo2}
 Let $\gamma$ be a future directed null geodesic in ${\rm M}_{\rm II}$. Then:
 \[
\begin{array}{rl}
1)&\lim_{s\to s_{-}}t\kst(s)= -\infty\hbox{ if }\gamma\hbox{ enters }\MII\hbox{ from }{\rm M}_{\rm I'}, \\[2mm]
2)&\lim_{s\to s_{-}}t\kst(s)= -\infty\hbox{ if }\gamma\hbox{ enters }\MII\hbox{ from }{\rm M}_{\rm II'}, \\[2mm]
3)&\lim_{s\to s_{-}}t\kst(s)\hbox{  finite }\hbox{ if }\gamma\hbox{ enters }\MII\hbox{ from }\MI, \\[2mm]
4)&\lim_{s\to s_{+}}t\kst(s)= +\infty\hbox{ if }\gamma\hbox{ leaves }\MII\hbox{ into }{\rm M}_{\rm III'}, \\[2mm]
5)&\lim_{s\to s_{+}}t\kst(s)= +\infty\hbox{ if }\gamma\hbox{ leaves }\MII\hbox{ into }{\rm M}_{\rm II'}, \\[2mm]
6)&\lim_{s\to s_{+}}t\kst(s) \hbox{ finite }\hbox{ if }\gamma\hbox{ leaves }\MII\hbox{ into }{\rm M}_{\rm III}. \\[2mm]
\end{array}
\]
\end{lemma}
\proof 
We recall that 
\begin{equation}
\label{zlobo}
t\kst= \stk t + 2x(r), \ \lim_{r\to r_{\pm}}x(r)= \mp\infty.
\end{equation}
The blocks ${\rm M}_{\rm I'}$, ${\rm M}_{\rm II}$ and ${\rm M}_{\rm III'}$ belong to a common $\stk {\rm K}$ patch. This implies that if 
$\gamma$ enters ${\rm M}_{\rm II}$ from ${\rm M}_{\rm I'}$ then $\lim_{s\to s_{-}}\stk t(s)$ finite, $\lim_{s\to s_{-}}r(s)= r_{+}$, and if $\gamma$ leaves ${\rm M}_{\rm II}$ into ${\rm M}_{\rm III'}$ then $\lim_{s\to s^{+}}\stk t(s)$ finite, $\lim_{s\to s^{+}}r(s)= r_{-}$. Using \eqref{zlobo} this implies 1) and 4). 

Since  ${\rm M}_{\rm I}$, ${\rm M}_{\rm II}$ and ${\rm M}_{\rm III}$  belong to a common ${\rm K}\kst$ patch we obtain  similarly 3) and 6). 

If $\gamma$ enters ${\rm M}_{\rm II}$ from ${\rm M}_{\rm II'}$, then $\gamma$ passes through the crossing sphere $S(r_{+})$, hence $\lim_{s\to s_{-}}V(s)=0$, which proves 2), using \eqref{e2}. Finally if $\gamma$ leaves ${\rm M}_{\rm II}$ into ${\rm M}_{\rm II'}$, then $\gamma$ passes through the other crossing sphere $S(r_{-})$. 
One uses then the   KBL coordinates used to construct the Kerr--Kruskal extension $D(r_{-})$, see \cite[Def.~3.4.5]{N}. The same argument as before shows that $\lim_{s\to s_{+}}t\kst(s)= +\infty$, which proves 5). \qed

\subsection{Cauchy surfaces}\label{ss:cauchy}
We now prove that several hypersurfaces are Cauchy surfaces in ${\rm M}_{\rm I}, \MIUII$  or ${\rm M}$. To prove that a set $S$ is a Cauchy surface in some spacetime $(M, g)$ we use the following facts. We recall that a set $S\subset M$ is called {\em achronal} if each time-like curve intersects $S$ at most once. 
 
\begin{theorem}\label{theotheo}
\ben
\item  A closed achronal set $S$ is a Cauchy surface iff each maximal null geodesic intersects $S$ and enters $I^{+}(S)$ and $I^{-}(S)$. 
\item   A closed connected non time-like hypersurface $S$ is achronal if $M\setminus S$ is disconnected.
\een 
\end{theorem}
Statement (1) can be found in \cite{Ger2}, \cite[Thm.~8.3.7]{W}, and statement (2) in \cite[Corr. 3.10.2]{N}.

We will use the following corollary of Theorem \ref{theotheo}.
\begin{corollary}\label{corocoro}
 Suppose $u\in \cinf(M)$ is such that: \ben
 \item $\nabla u$ is time-like,
 \item $\sup_{\gamma}u= +\infty$ and $\inf_{\gamma}u= -\infty$ for any maximal future directed null geodesic $\gamma$.
\een
Then the level sets $S_{T}= u^{-1}(\{T\})$ are Cauchy surfaces in $(M, g)$. 
\end{corollary}
\proof Since $\nabla u$ is time-like,    $S_{T}$ is space-like and achronal and clearly closed and connected. Possibly replacing $u$ by $-u$, we can assume that $\nabla u$ is future directed. This implies that $\{u(x)>T\}\subset I_{+}(S_{T})$ and $\{u(x)<T\}\subset I_{-}(S_{T})$, since the integral curves of $\nabla u$ are time-like future directed.  Using (2)  this shows that any maximal null geodesic intersects $S_{T}$ and enters $I^{\pm}(S_{T})$. \qed 

   \begin{proposition}\label{prop1}
 The surfaces $\Sigma_{T}= \{t=T\}$ are Cauchy surfaces in ${\rm M}_{\rm I}$.
\end{proposition}
\proof 
$g^{tt}<0$ in ${\rm M}_{\rm I}$, so $\nabla t$ is time-like in ${\rm M}_{\rm I}$. We apply then  Lemma \ref{lemalemo} {\it i)} and Corollary \ref{corocoro}. \qed
 
 \subsubsection{Auxiliary Cauchy surfaces in $\MI$}
 Recall that in \ref{pain-in-the-ass} we defined the family of hypersurfaces $\bar{\Sigma}_{T},\widetilde{\Sigma}_{T}$. We now introduce  families $\bar{\Sigma}_{T, n},\widetilde{\Sigma}_{T, n}$ of Cauchy surfaces in ${\rm M}_{\rm I}$ converging to $\bar{\Sigma}_{T},\widetilde{\Sigma}_{T}$ when $n\to \infty$.
 
 We start with defining $\bar{\Sigma}_{T, n}$.
 We set $x_{n}(r)= \min (r- 2\ln r, n)$ for $n\in \nn$ and 
 \[
 \bar{\Sigma}_{T, n}= \{t+ x_{n}(r)= T\}. 
 \]
 \begin{figure}[H]
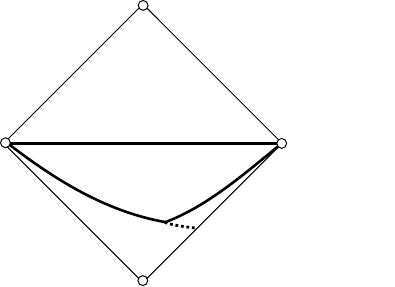\caption{The Cauchy surfaces $\bar{\Sigma}_{T, n}$.}\label{fig3}
\end{figure}
\begin{proposition}\label{prop2}
 The surfaces $\bar{\Sigma}_{T, n}$ for $T\ll -1, n\in \nn$ are Cauchy surfaces in ${\rm M}_{\rm I}$.
\end{proposition}
 \proof Let $u_{n}(t, r)= t+x_{n}(r)$. We have using \eqref{e5.1a}:
 \[
\bea
  - du_{n}\dual g^{-1}d u_{n}&\geq \dfrac{1}{\Delta\rho^{2}}((r^{2}+a^{2})^{2}- a^{2}\Delta- \Delta^{2}(1- 2r^{-1})^{2})\\
  &\geq \dfrac{1}{\Delta\rho^{2}}((r^{2}+a^{2})^{2}- a^{2}\Delta- \Delta^{2})\\
  &=\dfrac{1}{\Delta\rho^{2}}(2r(r^{2}+a^{2})+ (2r-a^{2})\Delta)>0\hbox{ for }r>r_{+},
\eea
 \]
 so $\nabla u_{n}$ is time-like in ${\rm M}_{\rm I}$. 
 For each $n$, $u_{n}(t, r)-t$ is bounded, so we conclude again by Lemma \ref{lemalemo} and Corollary \ref{corocoro}. \qeds

Let us now define $\widetilde{\Sigma}_{T, n}$. We set $\tilde{x}_{n}(r)= \min(-n, \tilde{x}(r))$ for $n\in \nn$, where $\tilde{x}(r)$ is defined in \eqref{matzno} and  
 \[
 \widetilde{\Sigma}_{T, n}= \{t+ \tilde{x}_{n}(r)= T\}. 
 \]
 \begin{figure}[H]
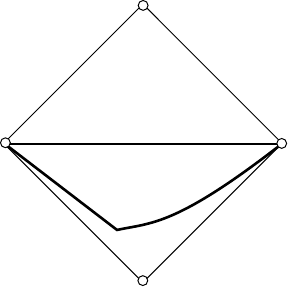\label{fig4-bis}\caption{The Cauchy surfaces $ \widetilde{\Sigma}_{T, n}$.}
\end{figure}

\begin{proposition}\label{prop3}
  The surfaces $\widetilde{\Sigma}_{T, n}$ for $T\ll -1, n\in \nn$ are Cauchy surfaces in ${\rm M}_{\rm I}$. 
\end{proposition}
\proof 
     Let $\tilde{u}_{n}(t, r)= t- \tilde{x}_{n}(r)$. We have
     \[
     \bea
     - d\tilde{u}_{n}\dual g^{-1}d \tilde{u}_{n}&\geq  \dfrac{1}{\Delta \rho^{2}}((r^{2}+ a^{2})^{2}- a^{2}\Delta- \Delta^{2}(\tilde{x}_{n}')^{2})\\
     &\geq  \dfrac{1}{\Delta \rho^{2}}((r^{2}+ a^{2})^{2}- a^{2}\Delta- \Delta^{2}(y')^{2})= \dfrac{1}{\rho^{2}},
     \eea
   \]
   where we recall that $y(r)$ is defined in \eqref{matznoni}. It follows that $\nabla \tilde{u}_{n}$ is time-like in ${\rm M}_{\rm I}$.For each $n$ $\tilde{u}_{n}(t, r)-t$ is bounded, and we complete the proof as for Proposition \ref{prop2}. \qed
     
     \subsubsection{Cauchy surfaces in $\MIUII$}\label{slip}
We construct a family of space-like Cauchy surfaces in $\MIUII$. 

It is easy to see that for $T\gg 1$ the equation $x(r)-r= -T$ has a unique solution $r_{T}$ in $]r_{+}, 2 r_{+}[$ with $r_{T}- r_{+}\in O(\e^{ -cT})$ for some $c>0$.
 We fix  a smooth decreasing function $\chi$ with  $\chi=1$ in  $]-\infty, r_{-}+\epsilon]$ and $\chi=0$ in  $]\12(r_{+}+ r_{-}), +\infty[$.  We define $v(r)$ by \[
 v'(r)= 1+ \chi(r)\frac{1}{r-r_{-}}, \ \ v(r_{+})= r_{+}
 \]
  so \begin{equation}
  \label{e.mizo2}
  v(r)=r\hbox{ in }\clopen{r_{+}, +\infty}, \ \ v(r)\to -\infty\hbox{ when }r\to r_{-}.
  \end{equation}
  We define the function $u_{T}$ on $\MIUII$ by
\[
u_{T}=\begin{cases}
t\kst-v(r)+T, \hbox{ for }r\leq r_{T},\\
t\ \hbox{ for }r_{T}<r.
\end{cases}
\]
We set
\[
Z_{T}\defeq \{u_{T}(x)=0\}\subset \MIUII.
\]
\begin{figure}[H]
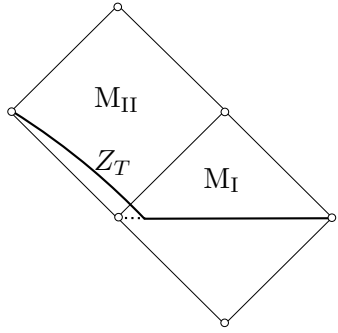\label{fig7}\caption{The Cauchy surfaces $Z_{T}$.}
\end{figure}
\begin{proposition}\label{pripoto}
   The surfaces $Z_{T}$ for $T\gg 1$ are Cauchy surfaces in $\MIUII$. 
\end{proposition}
\proof  In $\{r>r_{T}\}$ we have $du_{T}\dual g^{-1}du_{T}= g^{tt}<0$. In $\{r<r_{T}\}$  we have
\[
\bea
du_{T}\dual g^{-1}du_{T}&=  \frac{1}{\rho^{2}}\big(\Delta(r) (v')^{2}(r)- 2(r^{2}+a^{2})v'(r)+ a^{2}\sin^{2}\theta\big)\\
&\leq  \frac{1}{\rho^{2}}\big(\Delta(r) (v')^{2}(r)- 2(r^{2}+a^{2})v'(r)+ a^{2}\big).
\eea
\]
In $[r_{+}, +\infty[$ we have
\[
\Delta(r) (v')^{2}(r)- 2(r^{2}+a^{2})v'(r)+ a^{2}= \Delta(r)- 2(r^{2}+a^{2})+a^{2}= -r^{2}- 2r<0.
\]
 In $]r_{-}, r_{+}]$ we have $\Delta(r)\leq 0$ and $(v')^{2}(r)\geq v'(r)$ so
 \[
\bea
 \Delta (r)(v')^{2}(r)- 2(r^{2}+a^{2})v'(r)+ a^{2}&\leq (\Delta(r)- 2(r^{2}+a^{2}))v'(r)+ a^{2}\\ 
 &=(-r^{2}- 2r)v'+ a^{2}- a^{2}v'\leq -r^{2}- 2r<0,
\eea
 \]
 since $v'\geq 1$.
It follows that  $\nabla u_{T}$ is time-like in $\MIUII$. 
Let now $\gamma: J\ni s\mapsto x(s)$ be a future directed null geodesic in $\MIUII$.  
We claim that
\begin{equation}
\label{ziboto}
\sup_{\gamma}u_{T}= +\infty,
\end{equation}
\begin{equation}
\label{ziboti}
\inf_{\gamma}u_{T}= -\infty, 
\end{equation}
which using Corollary \ref{corocoro} will complete the proof.

Let us set $\gamma_{\rm I}= \gamma\cap {\rm M}_{\rm I}$,  $\gamma_{\rm II}= \gamma\cap {\rm M}_{\rm II}$.  

If $\gamma_{\rm I}=\gamma_{\rm II}= \emptyset$, then $\gamma$ is included in the horizon $\sH_{+}$ and $\gamma$ is a  rest photon, see \cite[Lem.~4.2.9]{N}. By \cite[Lem.~3.4.10]{N} we have $V(x(s))= s$ which implies that $\sup_{\gamma}t\kst= +\infty$, $\inf_{\gamma}t\kst= -\infty$ and hence \eqref{ziboto}, \eqref{ziboti}, since $v(r_{+})$ is finite. 

It remains to check \eqref{ziboto}, \eqref{ziboti} if $\gamma_{\rm I}$ or $\gamma_{\rm II}$ is not empty.
Let us first check \eqref{ziboto}. 

{\it Case I}: $\gamma_{\rm II}\neq \emptyset$.  Using Lemma \ref{lemalemo2} 4), 5), 6) and the fact that $\lim_{r\to r_{-}}v(r)= -\infty$, we obtain \eqref{ziboto}.

 {\it Case II}:   $\gamma_{\rm II}= \emptyset$, $\gamma_{\rm I}\neq \emptyset$. Let $]s_{-}, s_{+}[$ be the interval of affine parameter $s$ such that $x(s)\in \MI$.  Note that $\gamma_{\rm I}$ cannot end at $r_{+}$ otherwise $\gamma$ enters ${\rm M}_{\rm II}$. If $\gamma_{\rm I}$ ends at $\infty$ or some double root $r_{0}$, then by Lemma \ref{lemalemo} we have $\lim_{s\to s_{+}}t(s)=+\infty$ and since $r_{T}<r_{0}$ (at least for $T$ large enough), we have $u_{T}(x(s))= t(s)$ for $s$ close to $s_{+}$, which shows \eqref{ziboto}.

Let us now check \eqref{ziboti}.

{\it Case I}: $\gamma_{\rm I}\neq \emptyset$. 

If $\gamma_{\rm I}$ starts at $r_{+}$ then by Lemma \ref{lemalemo}, we have $\lim_{s\to s_{-}}t\kst= -\infty$, $\lim_{s\to s_{-}}r(s)= r_{+}$ which implies \eqref{ziboti}. 

If $\gamma_{\rm I}$ starts at $\infty$ or some double root $r_{0}$, then by Lemma \ref{lemalemo} we have $\lim_{s\to s_{-}}t(s)= -\infty$ and since $r_{T}<r_{0}$ for $T$ large enough, we have $u_{T}(x(s))= t(s)$ for $s$ close to $s_{-}$ hence \eqref{ziboti} holds also in this case.

{\it Case II}: $\gamma_{\rm I}= \emptyset$, $\gamma_{\rm II}\neq \emptyset$. We use Lemma \ref{lemalemo2} 1), 2) and the fact that $v(r_{+})$ is finite to obtain \eqref{ziboti}.\qed

     \subsubsection{Cauchy surfaces in ${\rm M}$}
     
\begin{proposition}\label{propitoli1}
 The  surface $\Sigma_{\rm M}\defeq \{U= V\}$ is a space-like Cauchy surface in ${\rm M}$.
\end{proposition}

\proof  Using \eqref{e2}, we obtain  first that
\beq\label{etoo.1}
\Sigma_{\rm M}= (\{t= 0\}\cap \MI) \cup(\{t=0\}\cap {\rm M}_{\rm I'})\cup S(r_{+}),
\eeq
 and that over $\Sigma_{\rm M}\cap \MI$ we have $d(U-V)= - 2 \kappa_{+}U dt$, while over $\Sigma_{\rm M}\cap {\rm M}_{\rm I'}$ $d(U-V)= 2\kappa_{+}U dt$. Since $g^{tt}<0$ on $\MI, {\rm M}_{\rm I'}$ we obtain that $d(U-V)\dual g^{-1}d(U-V)<0$ on $\Sigma_{\rm M}\setminus S(r_{+})$.  On $S(r_{+})$ we see directly that $d (U-V)\dual g^{-1}d(U-V)<0$, using \eqref{trif.5}. This implies that $\Sigma_{\rm M}$ is space-like, hence  achronal by Theorem \ref{theotheo} (2), since ${\rm M}\setminus \Sigma_{\rm M}$ is clearly disconnected.

Let us now $\gamma$ a future directed null geodesic in ${\rm M}$.  If $\gamma$ passes through $\MI$,  then by Proposition \ref{prop1} $\gamma$ crosses $\{t=0\}\cap \MI$  and enters $I^{\pm}(\{t=0\}\cap \MI)$, hence crosses $\Sigma_{\rm M}$ and enters $I^{\pm}(\Sigma_{\rm M})$. 
The similar conclusion  holds if $\gamma$ passes through ${\rm M}_{\rm I'}$. 

 Assume now that $\gamma$ passes through $\MII$.  If $\gamma$ enters $\MII$ from $\MI$ or ${\rm M}_{\rm I'}$ we argue as before. Otherwise  $\gamma$ enters $\MII$ through $S(r_{+})$.  Since $S(r_{+})$ is totally geodesic, $\gamma$ is transverse to $S(r_{+})$ and passes through $\MII$ and ${\rm M}_{\rm II'}$, hence enters $I^{\pm}(\Sigma_{\rm M})$.  
We argue similarly if $\gamma$ passes through $\MII'$. 

Finally if $\gamma$ passes through one horizon without entering any of the Boyer--Lindquist blocks,  then $\gamma$ is included in a horizon, for example in  $\sH_{R}= \{U=0\}$ and $\gamma$ is a  rest photon, see \cite[Lem.~4.2.9]{N}. By \cite[Lem.~3.4.10]{N} we have $V(x(s))= s$ so  $\gamma$ again crosses $\Sigma_{\rm M}$ and enters $I^{\pm}(\Sigma_{\rm M})$. Applying  Theorem \ref{theotheo} (1) we obtain that $\Sigma_{\MK}$ is a Cauchy surface. \qed

\section{Spectral projections in exponential coordinates}\label{appC}\init

\subsection{A one-dimensional lemma} For $\beta>0$, let $\chi^\pm_{\beta}(s)=(1+\e^{\mp\beta s})^{-1}$. Let $\one_{\rr^\pm}(s)$ be the characteristic function of $\pm \clopen {0,+\infty}$.

We consider the momentum operator $D_x = \i^{-1} \p_x$ acting in $L^2(\rr)$. Let us denote by
\[
\chi^\pm_{\infty}(D_x)\defeq \imath^{*} \circ \one_{\rr^\pm}(D_x) \circ \imath 
\]
the restriction of $\one_{\rr^\pm}(D_x)$ to $L^2(\rr^{+})$ using the canonical embedding $\imath: L^2(\rr^{+})\to L^2(\rr)$. We also consider  the selfadjoint generator of dilations 
\[
A= \12(x D_x + D_x x)=\i^{-1}(x\p_x + \12)
\]
defined  on  $L^2(\rr^+)$  by $\e^{\i sA}u(x)= \e^{s/2}u(\e^{s}x)$, $s\in \rr$. 

\begin{lemma}\label{mirlemma} On $L^2(\rr^{+})$ we have $\chi^\pm_\infty(D_x)= \chi^\pm_{2\pi}(A)$. 
\end{lemma}
\proof Using  the formula for the Fourier transform of $\one_{\rr^\pm}$ (see e.g.~\cite[Ch.~2.3, (22)]{gelfand}) one finds that the Schwartz kernel of $\chi^\pm_\infty(D_x)$ equals
\beq\label{eq:t0}
\chi^\pm_\infty(D_x)(x,y)= \pm \i (2\pi)^{-\12}  (x-y\pm \i 0)^{-1}.
\eeq
We will use the following convention for the Mellin transform on $C_{\rm c}^\infty(\rr^{+})$ and the inverse Mellin transform:
\beq\label{eq:mellin}
\bea
(\cM f)(\sigma) &= \frac{1}{\sqrt{2\pi}} \int_{\rr^{+}} x^{-\12-\i \sigma} f(x) dx,  \\ (\cM^{-1} g)(x) &= \frac{1}{\sqrt{2\pi}} \int_{\rr} x^{-\12+\i \sigma} g(\sigma) d\sigma.   
\eea
\eeq
It is well known that $\cM$ extends to a unitary map $\cM: L^2(\rr^{+})\to L^2(\rr)$, and the extension of $\cM^{-1}$ to $L^2(\rr)$ is the inverse of $\cM$. Another essential fact is that $\cM$ diagonalizes the generator of dilations, meaning that 
\beq\label{eq:t1}
\chi^\pm_{\beta}(A) = \cM^{-1}\circ \chi^\pm_{\beta}(\sigma) \circ \cM,
\eeq
where $\chi^\pm_{\beta}(\sigma)$ denotes the operator of multiplication by $\chi^\pm_{\beta}$. A brief computation using \eqref{eq:mellin} shows   that the Schwartz kernel of \eqref{eq:t1} equals
\beq\label{eq:t2}
\chi^\pm_{\beta}(A) (x,y) = \frac{1}{y} (\cM^{-1}\chi^\pm_{\beta})\big( \textstyle\frac{x}{y}\big).
\eeq
On the other hand, (as follows from e.g.~Formula 2.4 in \cite{mellintables}) 
\[
(\cM^{-1}\chi^\pm_{2\pi})(x) = \pm \i (2\pi)^{-\12}  (x-1\pm\i 0)^{-1}
\]
 in the sense of distributions. Plugging this into \eqref{eq:t2} and comparing with \eqref{eq:t0} yields the result. \qeds

Let us now rename the variable $x$ on $\rr^{+}$ to $U$, and let us consider the change of coordinates $U=\e^{-\kappa u}$ for some $\kappa>0$. This change of coordinates is implemented by a unitary map $\Phi: L^2(\rr^{+},dU)\to L^2(\rr,du)$, which satisfies
\[
\Phi\circ(-\kappa(U\p_{U}+\p_{U}U))= D_u \circ \Phi.
\]
From Lemma \ref{mirlemma} we obtain immediately
\beq
\chi^\pm_\infty(D_U)= \Phi^{-1} \circ \chi^\mp_{\beta}\big( D_u \big)\circ \Phi, \mbox{ with  } \beta=\frac{2\pi}{\kappa}. 
\eeq
This identity (together with its analogue for bosons) plays an important r\^ole in the description of the Unruh and Hawking effect.

\section{Wavefront sets and oscillatory test functions}\label{secapp1}\init

\subsection{Oscillatory test functions}
We  first recall  a well-known characterization of the wavefront set of a distribution using oscillatory test functions. The equivalence with other standard definitions, stated in Lemma \ref{lem:osc} below, follows from e.g.~\cite[Sect.~1.3]{duistermaat}.

Let $\Omega\subset \rr^n$ be an open set. For $x\in\Omega$, $q=(y,\eta)\in T^*\Omega\setminus\zero$ and $\chi\in C_{\rm c}^\infty(\Omega)$  we denote
\beq\label{eq:modes}
v_q^\lambda (x)\defeq \chi(x) \e^{\i\lambda(x-y)\cdot \eta}, \ \ \lambda\geq 1.
\eeq
We then extend the definition to manifolds by chart diffeomorphism pullback.  We will say that a function  $v_q^\lambda$
 of this form is an  \emph{oscillatory test function at} $q_0=(x_0,\xi_0)$ if $v_q^\lambda (x_0)\neq 0$. Oscillatory test functions can be used to give the following elementary characterization of the wavefront set. 
 
 Let $(\cdot|\cdot)_M$ be the $L^2(M, d\mu)$ pairing associated to some arbitrary smooth density $d\mu$. 
 
\begin{lemma}\label{lem:osc}  Let $u\in\cD'(M)$ and $(x_0,\xi_0)\in T^*M\setminus\zero$. Then $(x_0,\xi_0)\notin \wf(u)$ iff there exists an oscillatory test function $v_q^\lambda$ at $q_0$ such that for all $N\in \nn$, 
\beq\label{eosc1}
|(v^\lambda_q | u)_M| \leq C_N \lambda^{-N}, \ \lambda\geq 1, 
\eeq
uniformly for $q$ in a neighborhood of $(x_0,\xi_0)$ in $T^*M\setminus \zero$.
\end{lemma}

This fact  extends in a straightforward way to distributional sections of a hermitian vector bundle.

A very simple, but useful observation is that microlocal regularity can be tested with more general oscillatory functions.

\begin{definition}\label{def:osc2} We say that $w^\lambda_q$ is a \emph{generalized oscillatory test function at} $q_0=(x_0,\xi_0)\in T^*M\setminus\zero$ if it is of the form $w^\lambda_q=A^* v^\lambda_q$, where $A\in \Psi^0(M)$ is properly supported and elliptic at $q_0$, and $v^\lambda_q$ is an oscillatory test function at $q_0$.
\end{definition}

\begin{lemma}\label{lem:osc3} Let $u\in\cD'(M)$ and $q_0\in T^*M\setminus\zero$. Then $q_0\notin \wf(u)$ iff there exists a generalized oscillatory test function $w_q^\lambda$ at $q_0$ such that for all $N\in \nn$, 
$$
|(w^\lambda_q | u)_M| \leq C_N \lambda^{-N}, \ \lambda\geq 1, 
$$
uniformly for $q$ in a neighborhood of $q_0$ in $T^*M\setminus \zero$.
\end{lemma}
\proof Let $A$ and $v^\lambda_q$  be as in Definition \ref{def:osc2}.  Since $A$ is elliptic at $q_0$, $q_0\in \wf(u)$ iff  $q_0\in \wf(Au)$. We also have $(w^\lambda_q | u)_M=(v^\lambda_q | Au)_M$, so the assertion follows from Lemma \ref{lem:osc} applied to $Au$. 
\qeds

The next lemma shows that the oscillatory function $w_q^\lambda$ (and the neighborhood of $q_0$ where the estimate holds) can be chosen in a uniform way if $u$ varies in some set.

\begin{lemma}\label{lem:osc4} Let $\cX\subset \cD'(M)$ and let $\Gamma\subset T^*M\setminus\zero$ be closed. Then $\wf(u)\subset\Gamma$ for all $u\in\cX$ iff for all non-zero $q_0\in T^*M\setminus\Gamma$ there exists a generalized oscillatory test function $w_q^\lambda$ at $q_0$ such that for all $u\in\cX$ and $N\in \nn$, 
\beq\label{eosc3}
|(w^\lambda_q | u)_M| \leq C_{u,N} \lambda^{-N}, \ \lambda\geq 1,
\eeq
uniformly for $q$ in a neighborhood of $q_0$ in $T^*M\setminus \zero$. 
\end{lemma}
\proof Suppose that $\wf(u)\subset\Gamma$ for all $u\in\cX$.  Let $q_0\in T^*M\setminus\Gamma$.  Let $A\in\Psi(M)$ be properly supported and such that $\wf'(A)\cap \Gamma=\emptyset$ and $A$ is elliptic at $q_0$.  Then $Au\in \cinf(M)$ for all $u\in\cX$. Thus, if  $v_q^\lambda$  is an arbitrary oscillatory test function at $q_0$, then for all $u\in\cX$, 
$$
|(v^\lambda_q | Au)_M| \leq C_{u,N} \lambda^{-N}, \ \lambda\geq 1, \ N\in\nn,
$$
uniformly in $q$. Thus, the  assertion \eqref{eosc3} follows by setting $w^\lambda_q=A^*v^\lambda_q$.

The opposite direction trivially follows from Lemma \ref{lem:osc3}.
\qed
\subsection{Wavefront set estimates}\label{ss:wfe}
The  proposition below provides a refinement of the general strategy used by Moretti in \cite{Mo2} and in subsequent works \cite{DMP1,DMP2,radiative,characteristic} to estimate wavefront sets of solutions in terms of their traces on a null hypersurface. Strictly speaking, we will use its generalization to distributional sections of vector bundles, which is immediate.

\begin{proposition}\label{prop-key}
 Let $(M, g)$ be an oriented and time oriented Lorentzian manifold of dimension $n$, and let $S\subset M$ be a null hypersurface equipped with a smooth density $dm$. For $u\in \cE'(S)$  we define $\delta_{S}\otimes u\in \cE'(M)$ by:
 \[
\int_{M} (\delta_{S}\otimes u)\varphi \dVol_{g}\defeq \int_{S} u \varphi \diff m, \ \ \varphi\in \coinf(M).
\]
Let also $X$ be a vector field on $M$, tangent to $S$, null, future directed on $S$ and 
suppose $G\in \cD'(M\times M)$ satisfies $\WF(G)'\subset \{(q, q') \,:\, q\sim q'\}$. 
  Then for any $u\in \cE'(M)$ one has the implication:
\[
\WF(u)\subset\{(y, \eta)\in T^{*}S\setminus \zero \,:\, \pm \eta\dual X(y)\geq 0\}\Rightarrow \WF(G(\delta_{S}\otimes u))\cap \pi^{-1}(M\setminus S)\subset \cN^{\pm}.
\]
\end{proposition}
\proof  Since $\WF (G)'_{M}= \emptyset$  (i.e.~$\WF (G)'$ has no points of the form $(x_1,0,x_2,\xi_2)$), we can apply $G$ to $\delta_{S}\otimes u$ and we have  (see e.g.~\cite[7.2.7]{G}) 
\beq\label{e11.3}
\WF (G(\delta_{S}\otimes u))\subset \WF(G)'(\WF (\delta_{S}\otimes u)).
\eeq
Denoting by $i: S\to M$ the canonical injection, we have:
\beq\label{e11.2}
\WF(\delta_{S}\otimes u)\subset (i^{*})^{-1}(\WF (u))\cup N^{*}S,
\eeq
where $N^{*}S= \{(x, \xi)\in T^{*}M\setminus \zero \,:\,  x\in S, \ \xi_{| T_{x}S}=0\}$ is the conormal bundle to $S$. 

Let now $(x_{1}, \xi_{1})\in \WF( G(\delta_{S}\otimes u))$ with $x_{1}\not \in S$. By \eqref{e11.3} there exists $(x_0,\xi_{0})\in \WF(\delta_{S}\otimes u)$ such that $(x_{1}, \xi_{1})\sim (x_0,\xi_{0})$. 
 Since $g_{| T_{x_{0}}S}$ is positive semi-definite with kernel $\rr X(x_{0})$, we can  find $L\subset T_{x_{0}}S$ space-like with $T_{x_{0}}S= L\oplus \rr X(x_{0})$. The orthogonal $L^{\perp}$ is time-like and $2$-dimensional, hence contains two null lines, $\rr X(x_{0})$ and $\rr v$ for $v\in T_{x_{0}}M$ transverse to $S$. We can assume that $X(x_{0})\dual g(x_{0})v=-1$ and $v$ is future directed. 

We fix a basis $(w_{1}, \dots, w_{n-2})$ of $L$ and denote by $x= (y_{1}, y_{2}, y')$, $y'\in \rr^{n-2}$ the coordinates in the basis $(v, X(x_{0}),w_{1}, \dots , w_{n-2})$ of $T_{x_{0}}M$. We have then
\[
x\dual g(x_{0})x= - 2 y_{1}y_{2}+ y'\dual h y', \hbox{ where } h>0,
\]
and consequently, for $\xi_0= (\eta_{1}, \eta_{2}, \eta')$ expressed in dual coordinates, we have
\beq\label{xixi}
\xi_0\dual g(x_{0})^{-1}\xi_0= - 2 \eta_{1}\eta_{2}+ \eta'\dual h^{-1}\eta'.
\eeq
Since $(x_0,\xi_0)\in\cN$, we have  $\xi_{0} \dual g^{-1}(x_{0})\xi_{0}=0$. On the other hand,  from \eqref{e11.2} either $(x_{0},\eta_{2}, \eta')\in \WF(u)$ or $\eta_{2}= \eta'= 0$, i.e.~$(x_{0}, \xi_{0})\in N^{*}S$. 
Since $h$ is positive definite and the l.h.s.~of \eqref{xixi} vanishes,  $\eta_{2}=0$ implies $\eta'=0$. Therefore we have $\xi_{0}=(\eta_{1}, \eta_{2}, \eta')$ with  either
\beq\label{e11.4}
-2 \eta_{1}\eta_{2}+ \eta' \dual h \eta'=0, \ \ (x_{0}, \eta_{2}, \eta')\in \WF(u), \ \ \eta_{2}\neq 0,
\eeq
 or $(x_{0}, \xi_{0})\in N^{*}S$. Let us first consider the second case. The fact that $S$ is null is equivalent to $N^{*}S\subset \cN$, which using the fact that $N^{*}S$ is a Lagrangian submanifold of $T^{*}M$ implies that $N^{*}S$ is invariant under the bicharacteristic flow. Therefore the null bicharacteristic from $(x_0,\xi_0)$ stays in $N^{*}S$, hence above $S$, and thus cannot reach  the point $(x_{1}, \xi_{1})$ which is above $M\setminus S$.
 
 Let us now consider the first case.  Since by assumption $\WF(u)\subset\{(y, \eta)\in T^{*}S\setminus \zero \,:\, \pm \eta\dual X(y)\geq 0\}$ we deduce from \eqref{e11.4} that $\pm \eta_{2}>0$ and $\pm \eta_{1}= \eta_{2}^{-1} \eta'\dual h \eta'>0$. Therefore $(x_{0}, \xi_{0})\in \cN^{\pm}$ hence $(x_{1}, \xi_{1})\in \cN^{\pm}$ since $(x_{1}, \xi_{1})\sim (x_{0}, \xi_{0})$. This completes the proof of the proposition. \qeds

\medskip

\end{document}